%% file: HSM_ArXiv.tex
\documentclass[]{scrartcl} 
\usepackage[table]{xcolor}
\usepackage{pifont}
\newcommand{\cmark}{\ding{52}}
\usepackage{graphicx} 
\usepackage[round]{natbib}  
\usepackage{caption} 
\usepackage[utf8]{inputenc} 
\usepackage[T1]{fontenc}    
\usepackage{url}            
\usepackage{bm}       
\usepackage{amsfonts}       
\usepackage{nicefrac}       
\usepackage{amsmath}      
\usepackage{graphicx}
\usepackage{placeins}
\usepackage{comment}
\usepackage{bm}
\usepackage{bbm}
\usepackage[ruled, vlined]{algorithm2e}
\usepackage{booktabs} 
\usepackage[T1]{fontenc}
\usepackage[utf8]{inputenc}
\usepackage{authblk}
\usepackage[parfill]{parskip}
\usepackage[utf8]{inputenc} 
\usepackage[T1]{fontenc}    
\usepackage{url}            
\usepackage{siunitx}
\usepackage[colorlinks=true,linkcolor=blue,citecolor=blue]{hyperref}

\usepackage{pifont}
\usepackage{microtype}      
\usepackage{amsmath}      
\usepackage{graphicx}
\usepackage{placeins}
\usepackage{booktabs} 
\usepackage[switch]{lineno} 
\usepackage{caption}
\usepackage{geometry}
\title{A Horseshoe mixture model for Bayesian screening with an application to light sheet fluorescence microscopy in brain imaging}

\date{}
\author[1]{Francesco Denti}
\author[2,3]{Ricardo Azevedo}
\author[2,3]{Chelsie Lo}
\author[3]{Damian Wheeler}
\author[2,3,4]{Sunil P. Gandhi}
\author[5]{Michele Guindani}
\author[6]{Babak Shahbaba\footnote{\texttt{babaks@uci.edu}}}
\affil[1]{Department of Statistics, Università Cattolica del Sacro Cuore, Milan, Italy}
\affil[2]{Department of Neurobiology and Behavior, University of California, Irvine}
\affil[3]{Translucence Biosystems, Inc, Irvine}
\affil[4]{Center for the Neurobiology of Learning and Memory, University of California, Irvine}
\affil[5]{Department of Biostatistcs, University of California, Los Angeles}
\affil[6]{Department of Statistics, University of California, Irvine}

\begin{document}
\maketitle
	\begin{abstract}
	\input{HSM_ABSTRACT_R2}\	\textbf{Keywords:}
		Bayesian inference,
		Variable selection,
		Mixture models,
		Neuroscience	
	\end{abstract}

\input{HSM_MAIN_R2}
\bibliographystyle{plainnat} 
\bibliography{HSM_ArXiv}       


\clearpage
\appendix
\input{Suppmatt_body}

\end{document}

%% file: HSM_ABSTRACT_R2.tex
In this paper,  we focus on identifying differentially activated brain regions using a light sheet fluorescence microscopy -- a recently developed technique for whole-brain imaging. 
Most existing statistical methods solve this problem by partitioning the brain regions into two classes: significantly and non-significantly activated. However, for the brain imaging problem at the center of our study, such binary grouping may provide overly simplistic discoveries by filtering out weak but important signals, that are typically adulterated by the noise present in the data. To overcome this limitation, we introduce a new Bayesian approach that allows classifying the brain regions into several tiers with varying degrees of \emph{relevance}. Our approach is based on a combination of shrinkage priors -- widely used in regression and multiple hypothesis testing problems -- and mixture models -- commonly used in model-based clustering. 
In contrast to the existing regularizing prior distributions, which use either the spike-and-slab prior or continuous scale mixtures, our class of priors is based on a \emph{discrete mixture of continuous scale mixtures} and devises a cluster-shrinkage version of the Horseshoe prior. 
As a result, our approach provides a more general setting for Bayesian sparse estimation, drastically reduces the number of shrinkage parameters needed, and creates a framework for sharing information across units of interest. We show that this approach leads to more biologically meaningful and interpretable results in our brain imaging problem, since it allows the discrimination between active and inactive regions, while at the same time ranking the discoveries into clusters representing tiers of similar importance.


%% file: HSM_MAIN_R2.tex

\section{Introduction}
\label{experiment_description}

A central goal of many neuroscience studies is to detect regional patterns of brain activation associated with an activity, preferably at cellular resolution. A recent strategy to accomplish this goal involves using thin-section microscopy. This technique allows to detect immediate-early gene (IEG) activation, that is, the coordinate activation of genes for which the transcription is fast in response to external stimuli. IEG activation is thus closely related to changes in neurons' activity \citep{sheng1990regulation}. By using fluorescent antibodies for labeling IEG proteins along with advanced optical tissue clearing techniques and light sheet fluorescence microscopy (LSFM) we can obtain high resolution, three-dimensional snapshots of activity in individual neurons across the entire brain \citep{Richardson2015, Renier2016}.  Using a specific IEG, the Neuronal Per-Arnt-Sim Domain Protein 4 \citep[Npsa4 -][]{Lin2008,Sun2016}, our goal in this paper is to detect differentially activated brain regions in response to light exposure.
Statistical methods for assessing regional differences in activity across the whole brain using IEGs are currently in their infancy. The screening procedure proposed in this paper is a first step to improve statistical inference for quickly emerging  high-content imaging techniques such as LSFM.

Existing statistical methods for multiple hypothesis testing and variable selection typically group the individual estimates across the regions into two classes: \emph{significant} and \emph{non-significant}. This approach, however, oversimplifies the overall objective of such studies as the noise in the data may affect the discovery process.
    In particular, by using arbitrary cutoffs, the binary partition can also dismiss (i.e., classify as non-significant) many weak but biologically relevant signals. The limitations imposed by a dichotomous, symmetric screening are well-known, and  proposals to improve the decision problem date back at least to \citet{Tukey1993}. 
   In a recent report, the U.S. National Academies recommended the consideration of alternatives to binary decision rules (e.g., to reject or not to reject a null hypothesis) as one way to improve the replicability of scientific results \citep{NAP25303}. See also \citet{Wasserstein2019} and \citet{Shane2019} for more discussion of this concept.
    Here, we propose an alternative to classical binary discrimination, with a  method that can partition the potential findings into multiple tiers with varying degrees of \emph{relevance} -- a term we use instead of significance to distinguish our approach from other hypothesis testing methods \citep[see, for similar usage in Bayesian variable selection,][]{TadesseVannucci2021}. 
By allowing the sharing of information across the different regularization profiles and shrinking the noise to zero, our proposed model can better discriminate between signal and noise. Furthermore, this approach allows scientists to rank and classify brain regions without resorting to arbitrary cutoffs or pre-specifying the grouping. Thus, investigators can identify interesting activation pathways to consider in their follow-up studies. To achieve these goals, our method combines shrinkage priors with mixture models.

\subsection{One- and Two-group based screening}
 
Screening procedures play a central role in many statistical inference problems involving high throughput scientific studies. Whether presented as a multiple comparisons problem within a hypothesis testing framework or a variable selection problem within a regression framework, they typically involve inference regarding a set of $n$ parameters, say $\bm{\beta}=\{\beta_i\}_{i=1}^n$.
In a Bayesian framework, many methodologies have been proposed based on \emph{regularization} -- or \emph{shrinkage} -- of these parameters by using either the spike-and-slab (two-group) models \citep{Mitchell1988,McCulloch1993} or the continuous scale mixture (one-group) models \citep{Polson2012}.  

The first approach treats the prior over the parameters as a discrete mixture of a point mass at 0 (or a distribution centered at zero with low variance) and a ``flat'' distribution with large variance. This way, the resulting model-based clustering can discriminate between relevant and irrelevant units. 
\citet{Rockova2016} have recently proposed an extension of the Bayesian Lasso \citep{Park2008}, called the spike-and-slab Lasso, where the two competing densities are assumed to be from the Laplace family.

The second approach 
places hierarchical priors on the scale parameter of a given kernel distribution, typically Gaussian \citep[see][ for a review]{Bhadra2019}. The scale parameter is often decoupled into the product of global (i.e., shared across all the regression coefficients) and local shrinkage parameters (i.e., specific to each unit). 
This framework includes the Bayesian Lasso \citep{Park2008}, the Normal-Gamma \citep{Griffin2010}, the Horseshoe \citep{Carvalho2010}, the Horseshoe+, and the Dirichlet-Laplace \citep{Bhattacharya2015} priors, all based on Gaussian kernels. 
Due to their continuous shrinkage profile, the selection between relevant and irrelevant variables needs to be done through post-hoc analysis, usually by thresholding a proxy of the posterior probability $\mathbb{P}\left[\beta_i\neq 0 |\text{data}\right]$. 

In this paper, we propose a discrete mixture of continuous scale mixtures that bridges the gap between those two alternatives and provides a unified framework.
As carefully highlighted in \citet{Hahn2015}, the idea of adopting mixtures to model the scale parameters can be traced back to the seminal paper by \citet{Ishwaran2005}, where the authors discuss it within the context of bimodal mixtures. By building on that idea, our contribution allows combining the regularization effect typical of continuous shrinkage priors while inducing a grouping of the coefficients similarly to the spike-and-slab case. In our application to thin-section microscopy, our approach leads to automated model-based detection of groups of brain regions driven by different sparsity levels, imposing an adaptive regularization within each group. With our model, we can also rank the discoveries into blocks of increasing relevance, facilitating the interpretation of the results. 
The discrete mixture also greatly reduces the complexity of the model, avoiding the usual specification of a local shrinkage parameter for each variable and enabling at the same time sharing of information across the parameters.
From a multiple comparison perspective, the induced clustering goes beyond the classical ``significant vs. non-significant'' paradigm, and allows to capture signals that would be otherwise lost within the canonical binary framework. 
In summary, our approach (i) provides a more general setting for Bayesian sparse estimation without resorting to arbitrary cutoffs, (ii) drastically reduces the number of shrinkage parameters needed, and (iii) creates a framework for sharing information across units of interest without pre-specifying any grouping of the units. The combination of model-based clustering and shrinkage is important for our application since it allows the discrimination between a group of inactive regions (whose effect is aggressively shrunk to zero) and a group of active ones. Moreover, the group of discoveries is partitioned into tiers characterized by a similar amount of signal, providing neuroscientists with a ranking that can be invaluable for prioritizing further investigation. 

Our approach is related to several other methods using mixture models to improve the efficacy of the variable selection and shrinkage processes and models for hypothesis testing. 
Our proposed method is also related to -- but different than -- the scale mixture of Gaussian distributions for relevance determination of \citet{Shahbaba2013} and the Dirichlet-$t$ distribution of \citet{Finegold2011,Finegold2014}. We further elaborate on the connections between our model and the literature in Section A of the Supplementary Material \citep{Denti2023}. 


In the next section, we describe our study and the pre-processing steps required for preparing the raw data for analysis. We also present some preliminary results based on commonly-used methods whose limitations led to the development of our model. We introduce our methodology in Section \ref{Sec::model} and the derivation of the corresponding posterior inference in Section \ref{Sec::PostInf}. Section \ref{sec::applicationBM} is devoted to applying our method to the whole-brain imaging data (discussed above) using light sheet fluorescence microscopy to detect degrees of activation across brain regions. Then, in Section \ref{Sec::SimuStud}, we evaluate our model and confirm its validity using several simulation studies.
Finally, in Section \ref{Sec::discussion}, we summarize the advantages and the shortcomings of our proposed method and discuss future directions.

\section{Thin-section microscopy: experimental setup, pre-processing pipeline, and preliminary results}
\label{sect::preproc}
\begin{figure}[tbh!]
    \centering
    \includegraphics[width=\linewidth]{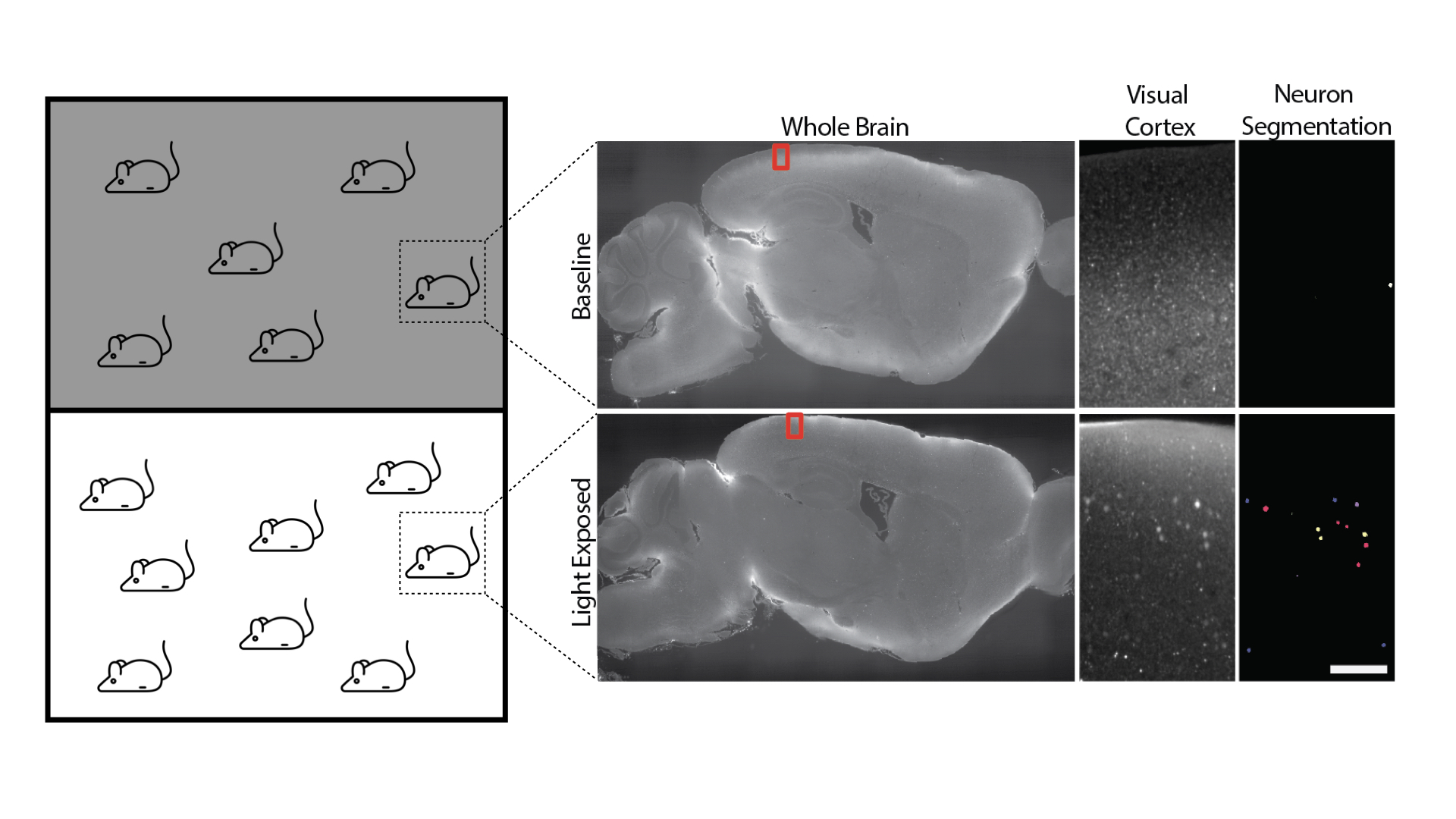}
    \caption{Visual representation of the experiment setting and images obtained with the LSFM technology. For both rows, the brain areas highlighted in the rectangles are reported in the right four panels to show the high level of resolution achievable with the LSFM.}
    \label{fig:experiment}
\end{figure}

Figure \ref{fig:experiment} shows a visual representation of the experimental setting in our case study, along with sample images obtained from two representative mice. More specifically, fourteen mice were individually housed in the dark for 24 hours to establish baseline visual activity. Mice were then transferred into a new cage exposed to ambient light. The brains of six mice were examined 0-15 minutes after light exposure to serve as the baseline group. The brains of another eight mice were examined 30-120 minutes after light exposure, within the window of Npas4 protein up-regulation \citep{Ramamoorthi2011}. Equal numbers of left and right hemispheres were sampled.  The goal was to assess differences in brain activation by comparing the baseline and light-exposed groups. We expect that light exposure induces widespread, visually evoked activity  in terms of fluorescence intensity.
Through this experiment, we measured the location of almost 300,000 active neurons within a common three-dimensional reference space and extracted their {intensity} and {volume} with remarkable precision.
 The neurons are classified into regions according to the \emph{Allen Brain Atlas} \citep{Sunkin2013}, the anatomic reference atlas commonly used in studies involving brains structures of mice.  

\begin{figure}[t!]
    \centering
    \includegraphics[width = \linewidth,trim ={4cm 0 4cm 0},clip]{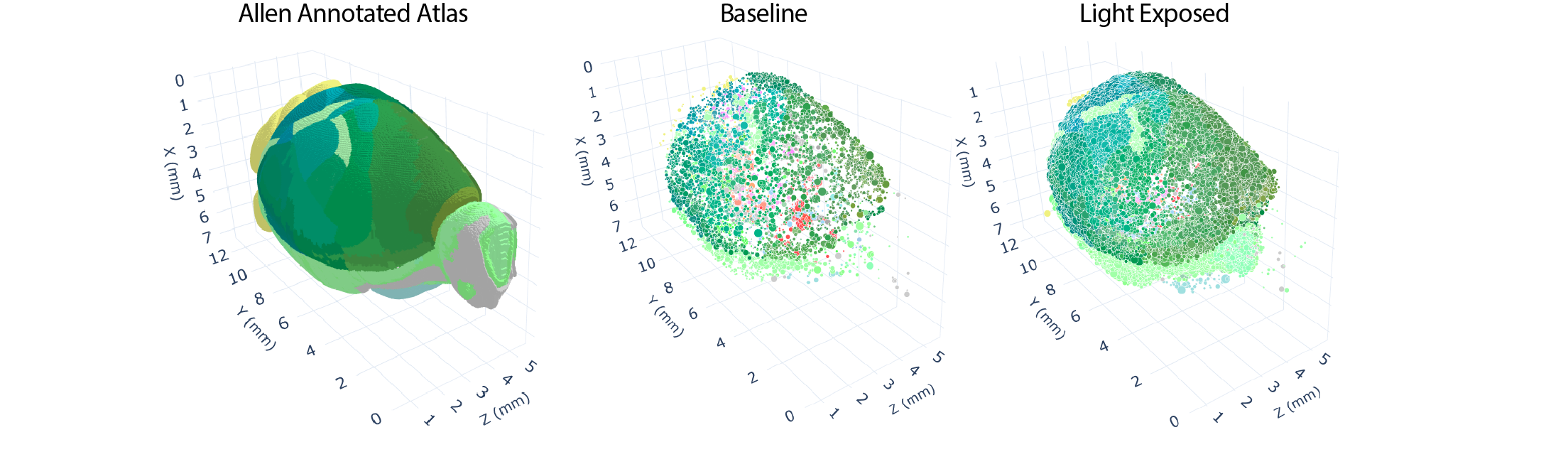}
    \caption{Comparison between detected Npas4 expressing neurons in brains of two representative mice exposed to different experimental conditions (Allen annotated atlas- left, baseline- middle, light-exposed- right). The points represent the detected neurons. The size of each dot corresponds to the neuron's volume. As we can see, the activated neuron count is higher in the light-exposed group of mice.}
    \label{fig:brains}
\end{figure}

\begin{figure}[t!]
    \centering
    \includegraphics[width = \linewidth]{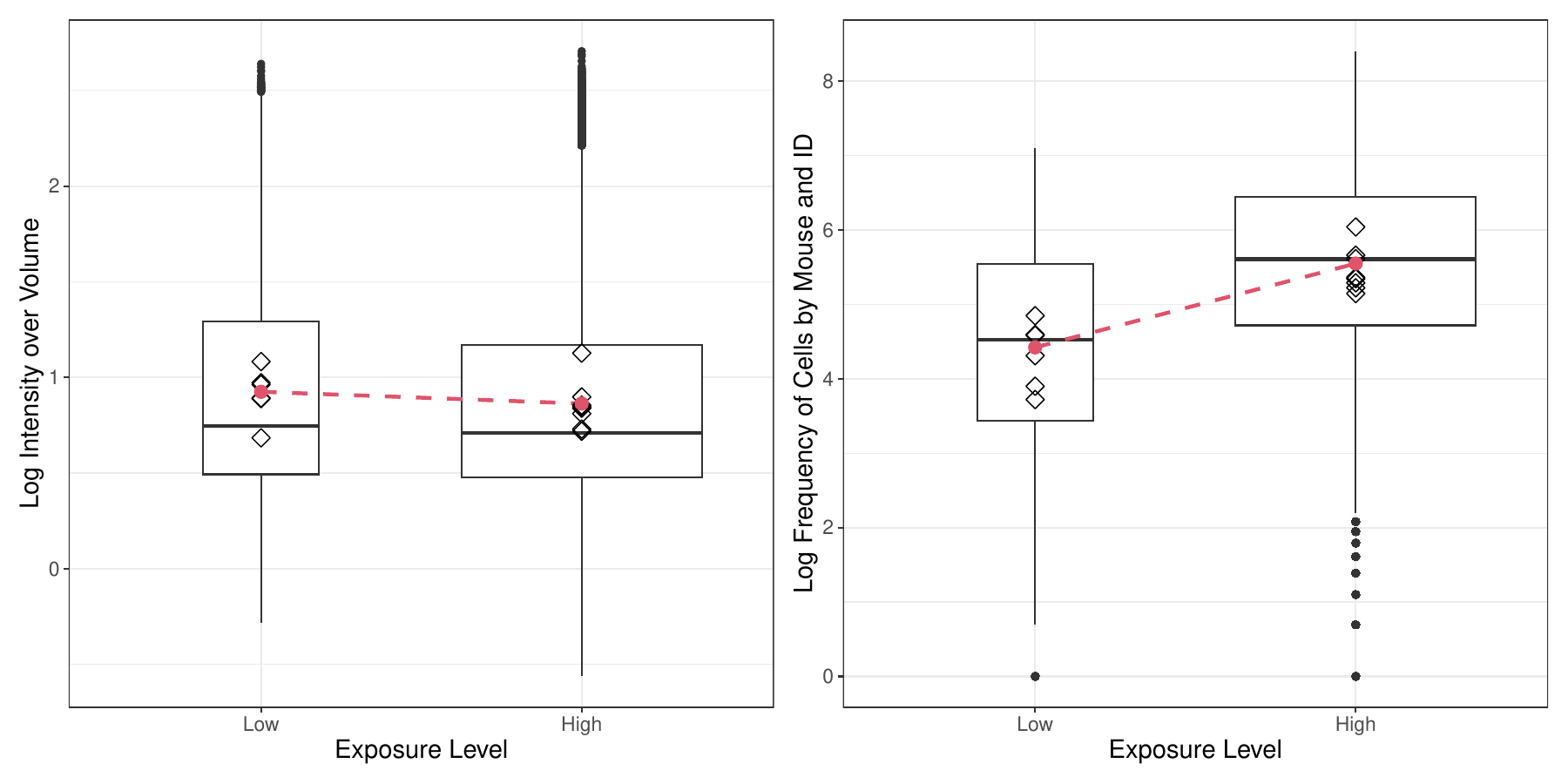}
    \caption{Boxplots representing the distributions of $\log(\texttt{iov})$ and the log-frequency of cells in each region and mouse stratified by level of exposure. The widths of the boxplots are proportional to the square-roots of the number of observations under the two 
    experimental conditions (baseline $\approx$ 55,000 neurons; light-exposed $\approx$ 235,000 neurons). The diamonds represent the mean for each mouse, while the dashed lines connect the overall means across the two subpopulations.}
    \label{fig:box1}
\end{figure}

Figure \ref{fig:brains} displays the three-dimensional images of brain cells measured in two representative mice under the two different experimental conditions: baseline and light-exposed. 
The intensity per unit of volume, \texttt{iov}, is the primary variable of interest in this study.  However, before starting our analysis, Figure \ref{fig:brains} reveals an important feature of the data: the frequency of observed neurons is strongly affected by the light exposure level.  This effect can also be observed in Figure \ref{fig:box1}, which reports two boxplots comparing the distributions of the logarithm of \texttt{iov} (left panel) and the logarithmic frequency of cells detected in the brain regions of each mouse (right panel), under the two experimental conditions. The log scale is chosen to enhance the visual representation. Basing the entire analysis on \texttt{iov} alone would be insufficient and could lead to misleading results as no clear difference emerges between the two experimental groups. However, the right panel suggests a positive association between the exposure level and the number of activated cells.  In other words, a proper definition of ``activation'' needs to incorporate the number of detected cells.  Therefore, we will base our analysis on a score derived as a combination of frequency and intensity.

To compare the regions under the two different exposure levels, we need to adjust for the effects of possible confounders. 
 An important source of information provided by our data is the multi-resolution, hierarchical organization of the brain regions. Each neuron is assigned to a terminal region, and different terminal regions are connected to a shared, higher-level parent region. This mechanism goes on until all the regions are assigned to a common region, called \emph{root}.
We aim to remove the potential distortion in the intensity given by specific mouse effects and the possible influence of parent areas (i.e., the closest ancestors). In fact, certain areas may have higher intensity because of the dimension and overall intensity of their parents, which in turn may blur the activation measures. 
Therefore, we regress the variable \texttt{iov} on all possible interactions between the mouse identifiers and the ancestor identifiers. We denote the resulting residual for each neuron as $r_{i,c}$, highlighting the membership of the cell $c$ to the brain region $i$. Let $m_i$ indicate the number of cells found in brain region $i$, with $i=1,\ldots,n$.
To take into account the frequency distribution of the neurons, we multiply $r_{i,c}$ by the density of neurons per unit of parent volume.  This way, we obtain a new variable of interest: $ \tilde{r}_{i,c} = r_{i,c} \times m^*_{i}/\textrm{Vol}^*_{i}$, where with $m^*_{i}$ and $\textrm{Vol}^*_{i}$ we indicate the frequency of cells and the volume of the parent of region $i$, respectively.  
Finally, we retain all brain regions with at least 15 ($m_i\geq 15$) neurons, leaving $n=281$ regions for our analysis.

 In a typical analysis, neuroscientists would consider a standard two-sided Welch t-test to detect differential activation of brain regions, comparing the averages of the vector $\tilde{\bm{r}}_i=\left(r_{1,i},\ldots,r_{m_i,i}\right)$ under the baseline vs. light-exposed conditions.  In the following, we show how how this typical approach may fail to identify important regions of interest. We obtain the t-statistics $\bm{t}=\{t_i\}_{i=1}^{n}$, the degrees of freedom estimated by the Welch–Satterthwaite equation $\bm{d}=\{d_i\}_{i=1}^{n}$, and the corresponding p-values $\bm{p}=\{p_i\}_{i=1}^{n}$ for each brain region. 
The p-values are post-processed following \citet[BH, ][]{Benjamini1995} and thresholded at 5\% to detect the activated regions. This result provides a first benchmark for later comparisons. We also use Efron’s empirical Bayes two-group model \citep{Efron2007}: local false discover rate (lFDR).
To do so, we first transform the t-statistics to z-scores: $z_i = \Phi^{-1} \left( F_{T_{d_i}} (t_i)\right) \: \forall i$, where $\Phi$ and $F_{T_{d}}$ denotes the cumulative distribution function (c.d.f.) of the standard normal distribution and a Student-t distribution with $d$ degrees of freedom, respectively. Then, we threshold the resulting local false discovery rate at 0.20, as suggested in the literature. 

Within this setting, the BH discovers 142 regions. In contrast, the lFDR method flags only 38 brain regions as important, missing many pertinent regions known to be associated with the visual task. 
On the one hand, such a difference in the results suggests that some brain regions may be active but show weaker signals than others. These regions are the ones that are likely missed by lFDR. 
On the other hand,  it is known that the BH method struggles in cases where the z-scores distribution departs from the theoretical null. This rigidity may explain the large number of regions identified as significant, potentially due to false discoveries. Nonetheless, the discrepancy between the numbers of findings of the two methods highlights the shortcomings of the classical binary hypothesis partition (e.g., significant vs. non-significant). Indeed, the model we present in the next section can provide more insights by ranking the signals into several tiers with varying degrees of relevance, which identify several levels of biological importance. As shown in Section \ref{sec::applicationBM}, this ranking allows scientists to examine groups of brain regions from the highest degree of relevance to the lowest degree without setting an arbitrary cutoff. More details about the discoveries can be found in Section G of the Supplementary Material.

\section{Methodology: a discrete mixture of continuous scale mixtures}
\label{Sec::model}
For analyzing the whole-brain imaging data (and potentially similar high-throughput studies), we propose a novel discrete mixture models to cluster the brain regions into several tiers of varying relevance with respect to their activation levels. More specifically, we consider the following model
	\begin{equation}
		\boldsymbol{z} = \bm{\beta}+\boldsymbol{\varepsilon}, \quad \quad \boldsymbol{\varepsilon}\sim \mathcal{N}_n(\bm{0},\bm{\Sigma}),
		\label{EQ:Like}
	\end{equation}
where $\boldsymbol{z}=\{z_i\}_{i=1}^n$ is an outcome vector (e.g., z-scores)  of length $n$,   $\bm{\beta}=\{\beta_i\}_{i=1}^n$ is the mean vector, and $\boldsymbol{\varepsilon}$ is the noise term. We assume homoscedastic and uncorrelated errors, i.e. $\bm{\Sigma}=\sigma^2\mathbb{I}_n$, for simplicity; this assumption seems to hold for our data, but our approach also can be readily generalized for more complex structures. In what follows, $\mathcal{N}_k(\bm{a},\bm{A})$ indicates a multivariate Normal distribution of dimension $k$ with mean vector $\bm{a}$, covariance matrix $\bm{A}$, and density function $\phi_k(\bm{a},\bm{A})$. 
In the univariate case, we let $\mathcal{N}_1\equiv \mathcal{N}$ and $\phi_1\equiv\phi$.

Our main focus is the specification of a suitable prior distributions for the coefficients $\bm{\beta}$. In the usual global-local shrinkage parameter models \citep{Polson2012}, the regression coefficients are assumed to be distributed as a continuous scale mixture of Gaussian distributions, i.e. $\beta_i |\tau, \bm{\lambda}_n,\sigma^2 \sim \mathcal{N}(0, \sigma^2\cdot\tau^2\cdot \lambda^2_{i}) \:\: \forall i=1,\ldots,n$, with $\lambda_i$ assumed to be stochastic. Here, $\tau \in \mathbb{R}^+$ denotes a \emph{global} shrinkage parameter, while the vector $\bm{\lambda}_n=\{\lambda_i\}_{i=1}^n,\: \lambda_{i} \in \mathbb{R}^+$ contains all the \emph{local} shrinkage parameters. Conditioning on the variance of the data, $\sigma^2$, guarantees a unimodal posterior \citep{Park2008}. 

We extend this framework and consider a discrete mixture of continuous scale mixtures of Gaussians.  As a result, the large number of local shrinkage parameters is substituted by a more parsimonious set of $L$ \emph{mixture component} shrinkage parameters. More specifically, we assume
\begin{equation}
	\beta_i | \tau,\bm{\lambda}_L,\bm{\pi},\sigma^2 \sim \sum_{l=1}^{L} \pi_l\:\phi(\beta_i; 0, \sigma^2\cdot\tau^2\cdot \lambda^2_{l}), \quad\quad  i=1,\ldots,n,
	\label{EQ::HSM}
\end{equation}
where $\bm{\pi}$ is the $L$-dimensional vector of mixture weights, and the elements of the vector $\bm{\lambda}_L=\{\lambda_l\}_{l=1}^L$ assume the role of \emph{mixture component} shrinkage parameters. The specification in \eqref{EQ::HSM} is very general and encompasses many known models. In particular, when $L=2$ and $\lambda_1\approx0$, we recover the continuous spike-and-slab framework of \citet{George1997}, while when $L=n$ and $\pi_l=\delta_i(l)\:\:\forall l,i$ (i.e., inducing $n$ different singleton clusters) we recover the continuous shrinkage framework. In our application, the mixture model allows to identify signals characterized by similar levels of shrinkage.

The mean estimation scenario is often considered for hypothesis testing, where the task is to identify the test statistics that depart from the standard Gaussian distribution specified under the null hypothesis (e.g., $H_{0,i}:\beta_i=0$). 
Adopting the classical global-local shrinkage prior for $\bm{\beta}$ to induce sparsity and setting $\sigma^2=1$, one can easily show that $z_i|\lambda_i,\tau \sim \mathcal{N}(0,1+\tau^2\lambda_i^2)$.
In our discrete mixture of continuous scale mixture model, the induced sampling distribution is itself a mixture:
\begin{equation}
	z_i|\tau,\bm{\lambda}_L,\bm{\pi} \sim \sum_{l=1}^{L}\pi_l\: \phi(z_i;0,1+\tau^2\lambda_l^2).
	\label{mult}
\end{equation}
In the multiple comparison setting, we can see $\bm{z}$ as a vector of $n$ properly standardized test statistics corresponding to $n$ different null hypotheses. Thus, model \eqref{mult} can be interpreted as a \emph{multi-group} extension of the classical two-group model \citep{Efron2007}. This connection is crucial since it reveals the limitations of well-established multiple hypothesis testing methods when applied to our neuroscience data, as highlighted in Section \ref{sect::preproc}. See Section B of the Supplementary Material for the derivation of \eqref{mult}.

The interpretation of \eqref{mult} as a multi-group version of the model presented in \citet{Efron2007} provides an additional justification for the use of continuous scale mixture of Gaussians. Without loss of generality, let us assume that the first mixture component is characterized by the smallest scale parameter $\lambda_{(min)}=\min_{l} \lambda_l$. One can impose this constraint \emph{a priori} or identify the mixture component with the smallest scale parameter after model estimation. Whenever the product $\tau\lambda_{(min)}\approx 0$, the corresponding mixture component can be interpreted as the null distribution, resembling the theoretical standard Gaussian. At the same time, the product $\tau \lambda_{(min)}$ is allowed to be different from zero to reflect a departure from the theoretical null, leading to the estimation of the so-called \emph{empirical} null, which could capture, for example, unexplained correlations among brain regions \citep{Efron2004}. The remaining mixture components  describe the alternative distribution, which can be decomposed into degrees of relevance according to the magnitude of the remaining parameters, $\bm{\lambda}\setminus \lambda_{(min)}$.  

Finally, we highlight that our proposal can be extended to a more generic regression problem. The linear regression case can be obtained by simply substituting $\bm{X}\tilde{\bm{\beta}}$ as the mean term of model \eqref{EQ:Like},
where $\bm{X}$ is a $n\times p$ covariate matrix and $\tilde{\bm{\beta}}=\{\tilde{\beta}_j\}_{j=1}^p$ the corresponding regression coefficients. We explore the performance of our prior specification in such scenario with a simulation study, reported in Section \ref{Sec::SimuStud}.

\subsection{Mixture and shrinkage: the Horseshoe Mix}
Whether we are adopting our model to perform variable selection or hypothesis testing, we need to specify prior distributions for the remaining parameters to complete the Bayesian specification. In addition, we can also specify a distribution for the global shrinkage parameter $\tau$.
A common choice for the prior distribution of the error variance $\sigma^2$ is the Jeffreys prior $\pi(\sigma^2)\propto 1/\sigma^2$.
The prior distribution for the weights changes if we assume a countable or uncountable number of mixture components. If we assume $L$ to be finite, we can simply set $\bm{\pi}\sim {Dirichlet}(a_1,\ldots,a_L)$. Notice that even $L>p$ is a viable option since one has to distinguish between mixture components and active components, i.e. the actual clusters found in the dataset. See \citet{Malsiner-Walli2016} for more discussion on the use of sparse finite mixture (SFM) models. Setting the hyperparameters $a_l=\epsilon\:\: \forall l$ with $\epsilon$ small ($\leq0.05$) allows the model to parsimoniously select the number of active components needed to describe the data. Another option is to specify a nonparametric model via a Dirichlet Process (DP) mixture model: 
\begin{equation}
	\beta_i|\tau,\bm{\lambda}_\infty,\sigma^2 \sim \mathcal{N}(0, \tau^2\sigma^2\lambda_i^2), \quad \lambda_i|G \sim G,\quad G\sim DP(\alpha,H),
	\label{EQ:DP}
\end{equation}
where $DP(\alpha,H)$ indicates a Dirichlet Process with concentration parameter $\alpha$ and base measure $H$. Adopting the Stick Breaking (SB) representation of \citet{sethuraman94}, model \eqref{EQ:DP} becomes
	\begin{equation}
		\beta_i|\tau,\bm{\lambda}_\infty,\sigma^2,\bm{\pi}\sim \sum_{l=1}^{+\infty} \pi_l \, \phi(\beta_i; 0, \sigma^2\cdot\tau^2\cdot\lambda_l^2), \quad   \lambda_l\sim H, \quad \bm{\pi}\sim SB(\alpha),
		\label{EQ:DPSB}
	\end{equation}
where the weights $\bm{\pi}$ are defined as $\pi_1=u_1$, $\pi_l=u_l\prod_{q<l}(1-u_q)$ for $l>1$ and $u_l\sim Beta(1,\alpha)$ for $l\geq 1$.

We have introduced multiple mixture specifications (both parametric and nonparametric) to present a general working framework that can be adapted beyond our specific application. Depending on the problem at hand, specific priors can be used to incorporate our domain knowledge about the possible number of tiers. The Bayesian nonparametric approach is preferable if the number of clusters ($L$) is expected to increase with the number of tests (i.e., regions). In our application, a higher resolution brain atlas would lead to a larger number of tests and possibly the identification of new activation profiles of brain sub-regions. In contrast, using a sparse finite mixture implies that the number of clusters has the upper bound $L$. Nevertheless, as we will show in the simulation study of Section \ref{Sec::SimuStud}, the two approaches achieve very similar results if $L$ is set to a sufficiently large number to ensure that many superfluous mixture components are not assigned any observation \textit{a posteriori}. This rule of thumb is based on the posterior behavior of overfitted mixtures \citep{Rousseau2011}. In our experience, the two methods usually provide similar results for all practical purposes when $L>30$.

To summarize,  the introduction of mixture component shrinkage parameters is beneficial for several reasons. This specification can improve the effectiveness of the regularization with respect to common global-local scale mixtures models. A discrete mixture allows the model to use a relatively small number of shrinkage parameters to borrow information across all the units and self-adapt to the different degrees of sparsity characterizing subsets of the coefficients.  In our application, this feature would help compound the signal in each tier and thus differentiate between pure noise -- effectively shrunk to zero -- and weak signals. 
Also, the model-based clustering nature of our approach enables the ranking of groups of coefficients into several \emph{shrinkage profiles}, improving on commonly-used binary solutions (i.e., significant vs. non-significant) by providing more flexibility and insight for decision making.

In what follows, we will adopt a Half-Cauchy prior for the mixture component shrinkage parameters: $\lambda_l\sim \mathcal{C}^+(0,1)$, $\forall l$.  The Half-Cauchy has been successfully employed in sparse mean estimation tasks, and its aggressive shrinkage property is ideal for our discovery problem. Henceforth, we refer to this model as Horseshoe Mix (HSM), in the spirit of the Horseshoe (HS) prior introduced by \citet{Carvalho2010}.

Finally, we point out that although the two models involve similar distributions, our model is fundamentally different from the Dirichlet-Laplace (DL) prior of \citet{Bhattacharya2015}. Under the DL prior, the conditional distribution of each coefficient  is $\beta_j|\pi^{*}_j,\lambda_j,\tau \sim \mathcal{N}\left(0, \lambda_j^2\pi^{*^2}_j\tau^2 \right),$ where $\bm{\pi}^*\sim Dirichlet(a_1,\ldots,a_n)$.
Thus, the model by \citet{Bhattacharya2015} assumes that the Dirichlet random vector $\bm{\pi}^*$ lies on the $n-1$-dimensional simplex, i.e. its dimension is tied to the sample size. In our model, the vector of mixture weights has only $L$ entries. More importantly, our likelihood is the convex combination of $L$ different kernels, which is different from the single-parameter kernel structure assumed under the DL distribution.

\subsection{Mixture component and cluster shrinkage} 

Consider now the Normal mean estimation framework, and define $\kappa_i=1/(1+\tau^2\lambda^2_i)\in (0,1)$. It follows that  $\mathbb{E}\left[\beta_i|z_i\right] = (1-\mathbb{E}\left[\kappa_i|z_i\right])\cdot z_i$, and $\mathbb{E}\left[\beta_i|\lambda_i,\tau, z_i\right] = \frac{\tau^2\lambda^2_i}{1+\tau^2\lambda^2_i}\cdot z_i$,
where $\kappa_i$ is known as the \emph{shrinkage factor} for observation $i$, which can be interpreted as a proxy of the complement of the posterior probability of relevance in the two-group model \citep{Carvalho2010}. It is interesting to see how these key quantities change under our model specification. For the conditional model, the posterior expected values of the coefficients become 
	\begin{equation}
		\begin{aligned}	\mathbb{E}\left[\beta_i|\bm{\pi},\bm{z}\right] &= \sum_{l=1}^L\mathbb{E}\left[ r_{l}(z_i)(1-\kappa^*_l)|\bm{z}\right]\cdot z_i,\\	\mathbb{E}\left[\beta_i|\tau,\bm{\lambda}_L,\bm{\pi},\bm{z}\right] &= \left(\sum_{l=1}^{L} r_{l}(z_i)(1-\kappa^*_l)\right)\cdot z_i = (1-\tilde{\kappa}_{j})\cdot z_i,
		\end{aligned}
		\label{A1}
	\end{equation}
where $r_{l}(z_i)=\frac{\pi_l \phi(z_i;0,1+{\tau^2}\lambda_l^2)}{\sum_{l=1}^{L}\pi_l \phi(z_i;0,1+{\tau^2}\lambda_l^2)}$. See Section B of the Supplementary Material for the derivation of \eqref{A1}. Here, we distinguish between the \emph{mixture component shrinkage factors} (MCSF - one for every mixture component) defined as $\kappa_l^*=1/(1+\tau^2\lambda^2_l)$ and the \emph{cluster shrinkage factors} (CSF - one for every coefficient) $\tilde{\kappa}_{j}=\sum_{l=1}^{L} r_{l}(z_i) \kappa_l^*$. Each CSF is a function of a convex combination of the $L$ MCSFs and directly controls the amount of shrinkage that affects each parameter $\beta_i$. Simultaneously, the weights of the convex combination depend on the components of the marginal sampling distribution $\phi(z_i;0,1+{\tau^2}\lambda_l^2)$. 
It becomes clear how the model structure takes advantage of the the sharing of statistical strength across parameters.  Indeed, the posterior mean for $\beta_i$ is the result of two effects. Given its mixture nature, the shrinkage is affected by all the other mixture components parameters through information sharing. However, since the mixture is driven by weights that directly depend on each data point's contribution to the marginal likelihood, we retain an observation-specific effect in the shrinkage process.
These simultaneous effects help the estimating procedure to place more emphasis on shrinkage profiles that better describe the data points in $\bm{z}$.

\section{Posterior Inference}
\label{Sec::PostInf}
To conduct posterior inference, we rely on Markov Chain Monte Carlo (MCMC) algorithms because the posterior distribution is not directly available in closed form. To simplify posterior simulation, we augment model \eqref{EQ::HSM} with the latent membership labels $\bm{\zeta}=\{\zeta_i\}_{i=1}^p$, where $\zeta_i\in \{1,\ldots,L\}$, linking each coefficient with a cluster; i.e., $\zeta_i=l$ if the $i$-th coefficients has been assigned to the $l$-th cluster. We obtain
\begin{equation}
	\beta_i | \tau,\bm{\lambda}_L,\zeta_i,\sigma^2  \sim \mathcal{N}(0, \sigma^2\cdot\tau^2\cdot \lambda^2_{\zeta_i}),
	\: \quad \zeta_i|\bm{\pi} \sim \sum_{l= 1}^L \pi_l \delta_l(\cdot).
	\label{EQ::zetas}
\end{equation}
Once the auxiliary membership labels are introduced in the model, it is straightforward to derive the full conditional for the corresponding Gibbs sampler. Both the global and the mixture component shrinkage parameters can be efficiently sampled following a parameter augmentation strategy \citep{Makalic2016} or via slice sampler \citep[as in the Supplementary Material of][]{Polson2014}. The details of the Gibbs sampler are deferred to Section C of the Supplementary Material.  In Section D of the Supplementary Material, we also comment on additional insights that the data augmentation procedure \eqref{EQ::zetas} provides about the model.

\subsection{Postprocessing of the results} 
Once the posterior samples have been collected, we can estimate the cluster-shrinkage factors from the membership labels. We map each coefficient $\beta_i$ to the assigned local shrinkage parameter via $\zeta_i$, constructing the vector $\left(\lambda_{\zeta_1},\ldots,\lambda_{\zeta_p}\right)$. It is then straightforward to compute $\hat{\tilde{\kappa}}_i={1}/(1+\tau^2\lambda^2_{\zeta_i})$.
One of the main advantages of our model is that once the MCMC samples of size $T$ are collected, it allows the estimation of the best partition that groups the coefficients into classes of similar magnitude. Let $\bm{\zeta}^{(t)}=\{\zeta_1^{(t)},\ldots,\zeta_p^{(t)}\}$ be the realization of the membership labels at iteration $t=1,\ldots,T$. With this information, we can estimate the Posterior Probability Coclustering (PPC) matrix, whose entries are defined as $\widehat{PPC}_{i,i'} = \sum_{t=1}^T\mathbbm{1}_{\left(\zeta^{(t)}_i=\zeta^{(t)}_{i'}\right)}/T,$ for $i,i'=1,\ldots,n.$ In other words, $\widehat{PPC}_{i,i'}$ estimates the proportion of times that coefficients $i$ and $i'$ have been assigned to the same cluster along the MCMC iterations. Hierarchical clustering can be applied directly to the $\widehat{PPC}$ matrix for fast solutions as in \citet{Medvedovic2004}. The choice of the number of tiers can be driven by the simultaneous inspection of the dendrogram obtained from the hierarchical clustering approach and exogenous knowledge from domain experts. When the latter is unavailable, we recommend thresholding the resulting dendrogram using a moderate value of potential tiers (e.g., ranging between 2 and 6) and avoid partitions with clusters containing only a negligible fraction of the observations. 
We elaborate more on this point in Section H of the Supplementary Material.

The resulting partition is easy to interpret. The HSM prior allows for a model-based clustering driven by the cluster-shrinkage parameter vector $\bm{\lambda}_L$. Therefore, the clusters in the solution specified by the optimal partition $\hat{\bm{\zeta}}$ can be described as classes of different magnitude. Therefore, we can explicitly identify the subgroup of coefficients characterized by the smallest magnitude that can be deemed as irrelevant, similarly to the null component in the two-group model. In a linear regression framework, this means that we are able to identify the set of indices that indicate the least relevant covariates, say $\mathcal{B}_0=\{i \in \{1,\ldots,n\}: \beta_i=0 \}$, inducing a variable selection solution. Moreover, the model also allows the classification of the remaining parameters into subsets of different magnitudes, yielding an interpretable ranking. 
 
In the next section, we apply the HSM model to the light-sheet fluorescence microscopy data presented in Section \ref{sect::preproc}. We emphasize that, despite being tailored to the differential activation detection problem, our HSM model represents a viable alternative to shrinkage priors in a wide range of problems.  In Section \ref{Sec::SimuStud}, we compare HSM to well-established and state-of-the-art methods for variable selection and multiple hypothesis testing.

\section{Application -- segmenting brain regions into activation tiers} \label{sec::applicationBM}

 In Section \ref{sect::preproc}, we presented the pre-processing steps along with the results obtained from lFDR (38 discoveries) and BH procedures (142 discoveries). 

To present additional benchmarks, here we estimate the posterior mean of the vector of z-statistics using the Spike and Slab (SnS) and Horseshoe (HS) models, after proper centering. Under the former model, we deem a region as relevant if its inclusion probability over the MCMC samples is over 5\%, given the high level of sparsity induced by the SnS model in our data. Under the latter model, we select a brain region as significant if the credible set for the corresponding mean does not contain zero \citep{VanderPas2017}.

Next, we apply the HSM model directly to the centered z-scores:
\begin{equation}
    z_i|\beta_i,\sigma^2 \sim \mathcal{N}(\beta_i,\sigma^2),\quad\quad \beta_i|\lambda,\tau,\sigma\sim \sum_{l\geq 1} \pi_l\: \phi(0,\lambda^2_l\:\tau^2\:\sigma^2), \quad i=1,\ldots,n.
    \label{HSM_zeta}
\end{equation}
As mentioned previously, the expression in \eqref{HSM_zeta} can be regarded as a multi-group model in a multiple hypothesis testing framework. Within this setting, we will interpret the component characterized by the lowest variance as representative of the null distribution. 
In contrast, the other components, which are ranked in increasing order, represent different degrees of relevance.
To fit model \eqref{HSM_zeta}, we use a Bayesian nonparametric approach with a DP stick--breaking representation over the mixture weights, and adopt an Inverse Gamma distribution for $\tau^2$. 
A sparse finite mixture would also suffice, as the experiments we conduct in Section \ref{Sec::SimuStud} show that the two specifications appear to provide similar results. The hyperprior on $\tau^2$ was chosen to ensure good mixing of the global shrinkage parameter. 
We ran 10,000 iterations as burn-in period and used the next 10,000 samples for inference. Then, we postprocessed the resulting posterior coclustering matrix with the Medvedovic approach. Although the flexibility of our model allows estimating the number of tiers through inspection of the non-empty components of the posterior distribution, for practical and inferential purposes a choice often needs to be made post-MCMC. For this application, the inspection of the postprocessing results and the insight of our collaborators led to partitioning the z-scores into four tiers of relevance ranging from no activation ({Tier 4}) to clear activation ({Tier 1}).

\begin{figure}[t!]
    \centering
    \includegraphics[width = .8\columnwidth]{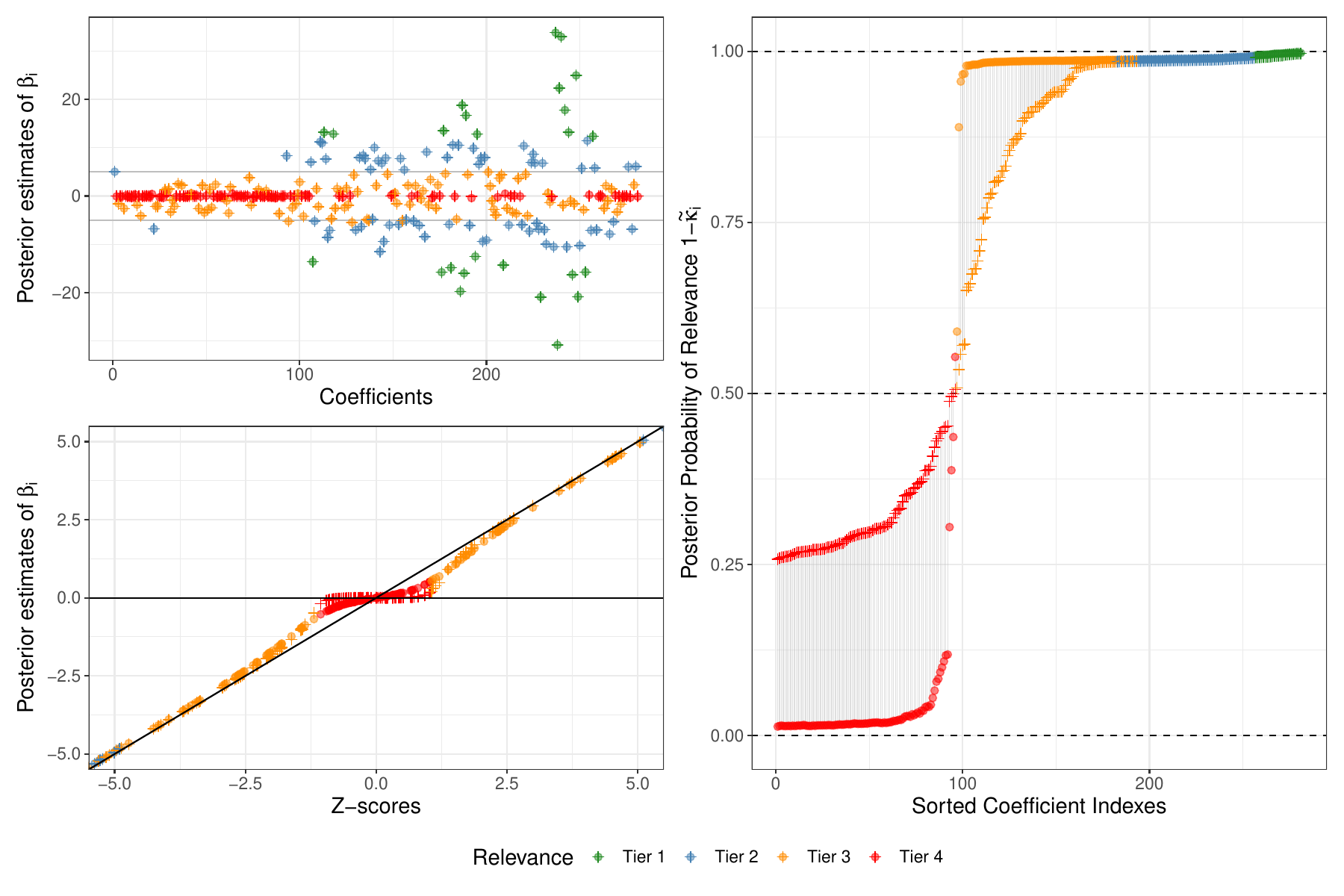}
    \caption{All the panels show posterior means (circles) and posterior medians (crosses) for different quantities. Top left panel: estimates for $\bm{\beta}$ stratified according to the retrieved segmentation. 
    Bottom left panel: posterior estimates for $\bm{\beta}$ plotted against the z-scores. The plot is cropped between $(-5,5)$ on both axes to show the shrinkage induced on the z-scores belonging to the low tiers of relevance. 
Right panel: posterior probability of relevance, approximated as the complement to one of the cluster shrinkage factors ${\tilde \kappa}_i$, linked with a gray vertical line to highlight the variability in the posterior distributions.}
    \label{fig:coef_prob}
\end{figure}

Figure \ref{fig:coef_prob} presents the posterior means (circles) and posterior medians (crosses) for different quantities. The elements in both panels are represented according to the tier to which they are assigned. The top-left panel shows the estimated coefficients. We can see how the model groups the scores according to their magnitude. A scatter plot of the z-scores vs. the posterior estimates is displayed in the bottom-left panel. The axes are cropped to showcase the shrinking effect of the HSM model on the z-scores for {Tiers 3} and {4}. Finally, the right panel presents the posterior probabilities of relevance $1-\bar{\kappa}_i$, for $i=1,\ldots,n$. This plot helps interpret the tiers of relevance: we notice the shift from {Tier 4} to {Tier 3} in both posterior estimates occurring around 0.5. Therefore, our method can be seen as an extension of the two-group model, automatically detecting the null group. 
Moreover, after filtering out the irrelevant units, it partitions the remaining ones into different sets with increasing levels of importance, capturing more information from the z-scores.

\begin{figure}[th]
    \centering
    \includegraphics[width = \linewidth]{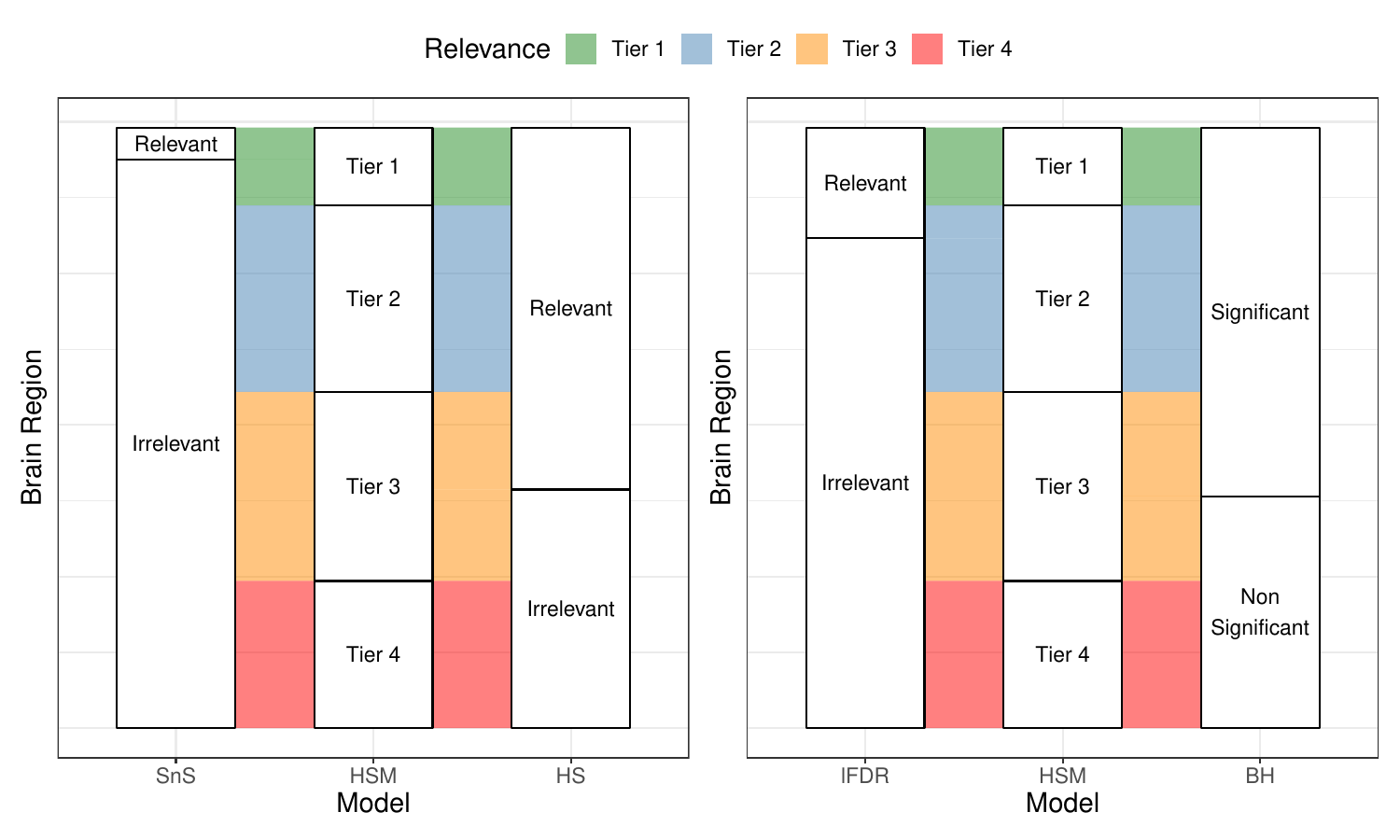}
    \caption{Alluvial plots displaying how the partitions of the 281 regions obtained with different models (SnS, HS, lFDR, and BH) relate with the HSM partition into tiers.}
    \label{fig:alluv}
\end{figure}

We now compare the results obtained by the five alternative methods discussed in this paper: BH, lFDR, HS, SnS, and HSM. Figure \ref{fig:alluv} juxtaposes the different results with an alluvial plot. Each column represents a model, and the horizontal lines (the brain regions) display how HSM tiers are associated with the results of the other models. A contingency table is also provided in Section F of the Supplementary Material. The SnS and lFDR methods are the most conservative, detecting only 9 and 38 regions, respectively. All these selected regions are part of HSM's top two tiers. The results from HS and BH are also similar: the two methods detect 138 and 142 regions, respectively. The HSM model places 25 regions in the top tier, 68 regions in the second tier, 49 in the third, and deems 96 regions as irrelevant. 

We next sought to identify the biological relevance of these findings. We expect that introduction of animals to light will drive Npas4 expression in neurons within the laminar sub-regions (e.g., layers) of different visual cortex areas \citep{Hubener2003, andermann2011functional}. Previous studies have shown that neurons in the primary visual (V1) area of the cortex respond to light exposure by expressing Npas4 mRNA \citep{hrvatin2018single}. Our results align with the literature, capturing the activation of the V1 laminae due to light exposure in terms of increased Npas4 protein expression.  Other cortex regions are expected to exhibit visually evoked activity, such as the lateral, posteromedial, anterolateral, and anteromedial visual areas. According to our results, all of these regions show Npas4 expression following light exposure \citep{andermann2011functional}. However, across the twenty brain laminae comprised by these regions, five of them fall into Tiers 3 and 4, reflecting their lack of activation. We report the list of the two top-tier areas in Table \ref{tier1}. The complete list of findings for the four models we considered is reported in Section G of the Supplementary Material. From that list, we can appreciate how the HSM model provides a more articulate solution, mediating between the more conservative lFDR and SnS methods and the numerous discoveries of the BH and Horseshoe models.

The tiering results obtained from our method allow us to stratify our findings by the level of activation without resorting to successive manual and arbitrary p-value cutoffs. We can utilize this approach to identify laminar activity patterns. For example, neurons in layer 2/3 of the primary visual cortex are known to exhibit lower activity than those in other V1 laminae \citep{niell2010modulation}. HSM identifies this by placing layer 2/3 in tier 2, while all other V1 laminae are placed in tier 1. Interestingly,  layers 2/3 of different visual cortex areas are assigned one tier below the other laminae, suggesting lower activity in layer 2/3 may be a common feature throughout the visual cortex. To our knowledge, these results are novel and have not been previously reported in mice. Hence, these findings warrant further investigation. These results illustrate the additional insights provided by HSM when applied to high throughput studies.

\input{Table_regions}

\section{Numerical experiments validating the HSM model} 
\label{Sec::SimuStud}

 In this section, we illustrate our method and evaluate its performance, in terms of generic mean estimation and linear regression, to establish its competitiveness with commonly-used and the-state-of-the-art statistical methodologies. 

\subsection{Illustrative example} 
\begin{figure}[t!]
	\centering
	\includegraphics[width=\columnwidth]{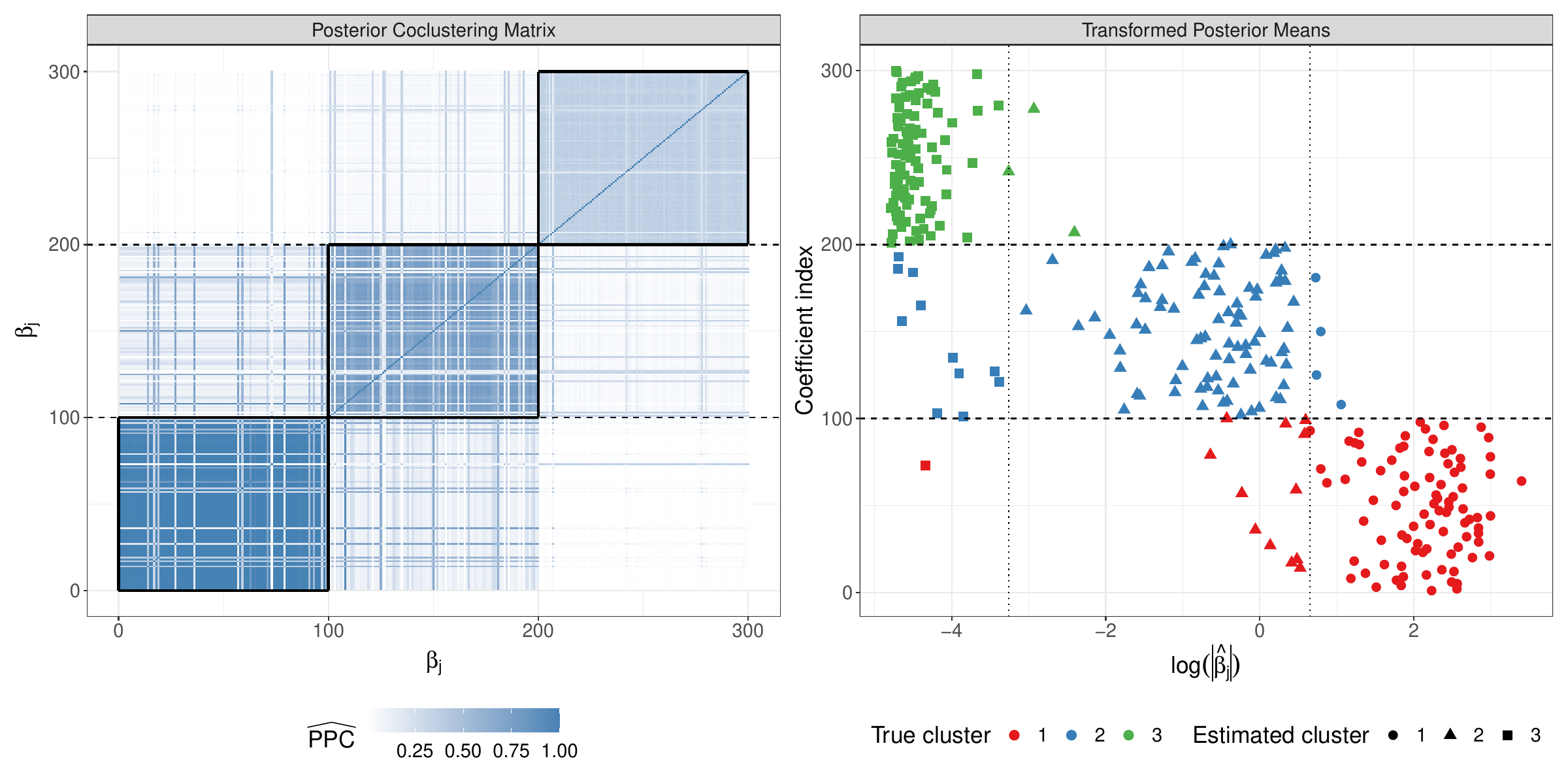}
	\caption{Left panel: the estimated posterior coclustering matrix where the actual clusters are superimposed with solid lines. Right panel: a scatterplot presenting the estimated posterior mean for each coefficient transformed as $\hat{\lambda}_i\tau\sigma$. The horizontal lines separate the true clusters while the vertical lines highlight the estimated partition. Note how the magnitude of the coefficients leads the estimation of the clusters.}
	\label{fig:HSvsHSM}
\end{figure}

As a simple example, consider a sample of 500 observations generated from a linear regression model with true vector of coefficients $\bm{\beta}$ composed of 100 zeros and 200 realizations generated in equal proportions from two Normal distributions centered around zero with variances $100$ and $1$, respectively. The error noise is set to $\sigma^2=0.5$. Given this dataset, we estimate the HSM model with a nonparametric specification of the mixture weights and fix $\tau^2=0.001$.
The two panels of Figure \ref{fig:HSvsHSM} show the estimated posterior coclustering matrix $\widehat{PPC}$ (left) and the posterior mean for each $\beta_i$ (right), transformed as $\log{|\hat{\beta_i}|}$ to emphasize the differences in terms of magnitude. On top of the PPC matrix, we highlighted three blocks representing the true clusters present in the data, which are also represented by the different shapes in the right plot. From the right panel, we can see that the model can effectively group the parameters in terms of magnitude. The accuracy of the classification is 0.89, with a Adjusted Rand Index of 0.72.

\subsection{Performance in mean estimation}

Next, we investigate the performance of the HSM model in terms of mean estimation. To this end, we generate random vectors from a multivariate Gaussian distribution with mean $\bm{\beta}$. The elements in $\bm{\beta}=\{\beta_i\}_{i=1}^n$ are organized into three different blocks: $\beta^{(1)}_{i}\sim \mathcal{N}(0,100)$ for $i=1,\ldots,q$, $\beta^{(2)}_{i}\sim \mathcal{N}(0,1)$ for $i=q+1,\ldots,2q,$ and $\beta^{(3)}_{i}\sim \delta_0$ for $i=2q+1,\ldots,n.$
We consider four scenarios (S1-S4), varying according to the values assumed by $q$ and $n$. Specifically, for S1 and S2 we set $n=500$, while for S3 and S4 $n=1000$. Moreover, we set $q=50$ for S1 and S3, while $q=100$ for S2 and S4. 
We compare the results obtained from three models: HSM, Horseshoe (HS), and Spike and Slab (SnS). Results for the latter two models were obtained via the R packages \texttt{horseshoe} (v0.2.0) 
and \texttt{BoomSpikeSlab} (v1.2.5), 
respectively. We consider six different specifications for the HSM model by varying mixture type (sparse finite mixture, SFM, and Dirichlet process mixture, BNP - in all cases, we set $L=50$) and distribution for the global shrinkage parameter (fixed, Inverse Gamma, and half-Cauchy). We summarize the specifications and the corresponding acronyms in Table \ref{tab:simstud_setting}. We also consider two specifications for the HS model, where $\tau\sim \mathcal{C}^+(0,1)$ (HS1) or $\tau \sim \delta_{0.0001}$ (HS2).

\begin{table}[ht!]
    \centering
    \begin{tabular}{lcccccc}
    \toprule
    & HSM1 & HSM2 & HSM3 & HSM4 & HSM5 & HSM6 \\
    \midrule     
    Mixture & SFM & SFM & SMF & BNP & BNP & BNP \\
    Global shrinkage parameter & 0.0001   & $\tau^2\sim IG$ &  $\tau\sim \mathcal{C}^+$& 0.0001   & $\tau^2\sim IG$ &  $\tau\sim \mathcal{C}^+$\\
    \bottomrule     
    \end{tabular}
    \caption{Summary of the different HSM model specifications used in our first simulation experiment.}
    \label{tab:simstud_setting}
\end{table}

\begin{figure}[t!]
    \centering
    \includegraphics[width =\linewidth]{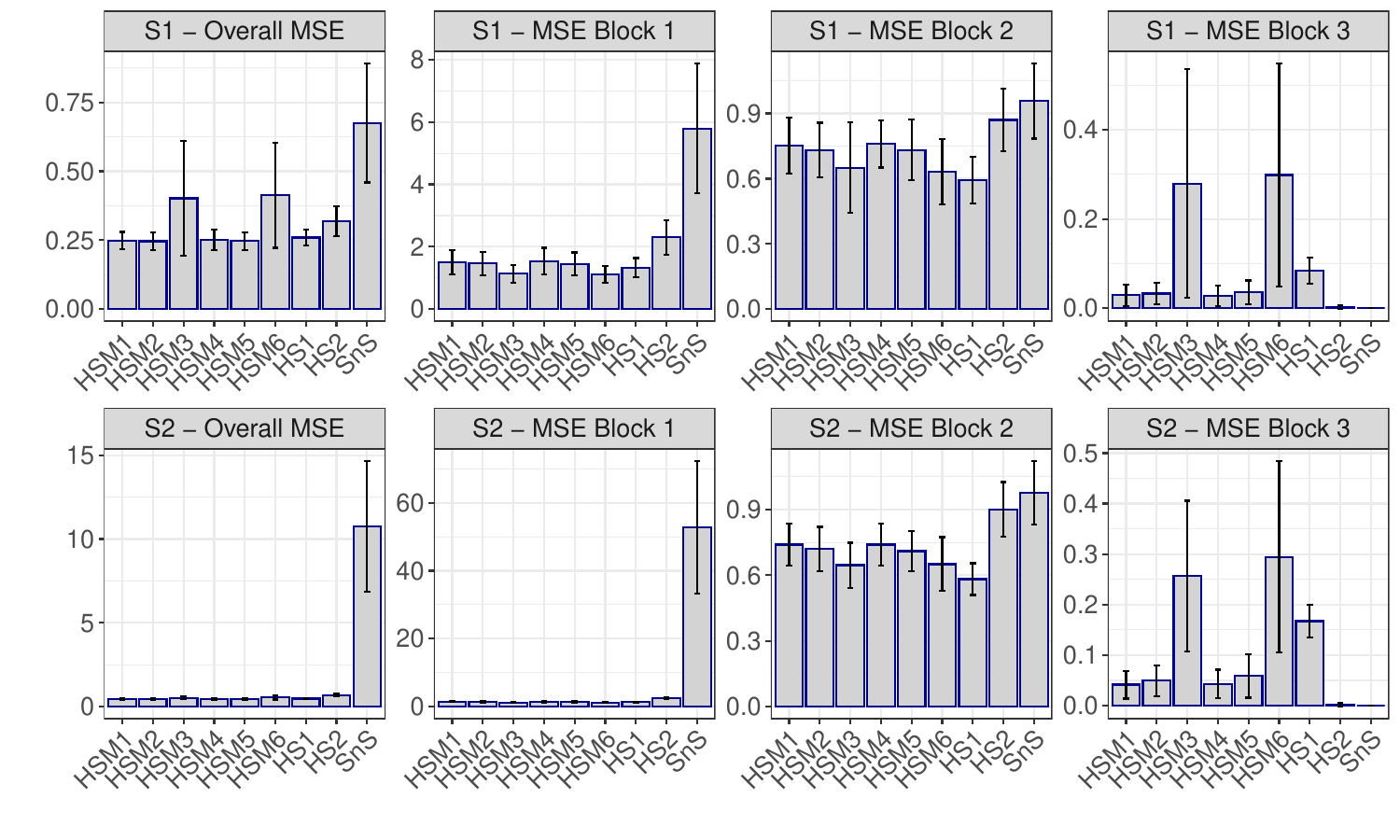}
    \caption{Bar plots of the average overall and stratified MSEs, and corresponding standard errors (error bars) for simulation scenarios S1 and S2.}
    \label{fig:extab:S1}
\end{figure}

To quantify the performance of the models, for each dataset we compute the mean squared error between the posterior mean $\hat{\bm{\beta}}_k$ and the ground truth, defined as $\textrm{MSE}(\bm{\beta}_k,\hat{\bm{\beta}}_k) = \sum_{i=1}^{n}(\bm{\beta}_{i,k}-\hat{\bm{\beta}}_{i,k})^2/n$. We also stratify the same quantity across the three different parameter blocks to understand which magnitude group contributes the most to the error. All the results are averaged over 30 replicates. For each replicate, we ran 10,000 MCMC iterations, discarding the first half as burn-in. The outcome of the first and second scenarios are displayed in the bar plots with error bars (representing the standard errors) in Figure \ref{fig:extab:S1}. The table containing the values used to draw the bar plots is reported in Section F of the Supplementary Material. The same table also includes the results for S3 and S4, for which the MSE values are very similar to the ones we display here. Lastly, in Section E of the Supplementary Material, we also include an alternative version of Figure \ref{fig:extab:S1} without the SnS output to ease the visual comparison of the remaining models. 

The HSM obtains very competitive results both in terms of overall and stratified MSE. Overall, the HSM specifications with constant or Inverse-Gamma global shrinkage parameters attain low MSE combined with small standard errors across the replicates. Instead, whenever a Half-Cauchy prior is used, the average overall MSE increases, but so does the variability of the results. The same happens for the HS model. 
Also, we do not observe any clear result that favors one mixture type over the other. 
The stratification of the results into the parameter blocks shows an interesting trade-off between the precision of the parameter estimation across the different magnitudes. As expected, the SnS perfectly captures the true zeros, but it tends to over-shrink the non-zero $\bm{\beta}$'s. In the cases where $\tau \sim \mathcal{C}^+$ (HSM3 and HSM6), the model better captures the non-zero parameters, while the remaining specifications perform very good regularization. The same rationale applies to the HS models.

\subsection{Performance in the estimation of regression coefficients}

Here, we consider a linear regression framework with $n$ observations and $p$ covariates, and compare the estimation performance of the HSM model with both well-established and recent Bayesian shrinkage models including Bayesian Lasso, Horseshoe and Horseshoe+. To estimate these models under different regularizing prior specifications, we use the \texttt{R} package \texttt{bayesreg} (v1.2.0).  

Our experiment consists of three scenarios, characterized by different values of the ratio $n/p$, describing the proportion between the sample size and the number of variables. Specifically, we consider the following three ratios: $n/p \in \{ (500,250)=2 , (500,500)=1, (500,750)=0.667 \}$. Under each scenario, we generate $K=30$ datasets as follows. We first sample $n$ independent observations from a multivariate Gaussian as $X_{i,k}\sim\mathcal{N}_p\left(0,\mathbb{I}_p\right)$, $i=1,\ldots,n$, creating the design matrix $\bm{X}_k$, $k=1,\ldots,K$. Then, we sample the regression coefficients $\bm{\beta}_k$ organized in three different blocks: $\beta^{(1)}_{j_1,k}\sim \mathcal{N}(0,100)$ for $j_1=1,\ldots,100$, $\beta^{(2)}_{j_2,k}\sim \mathcal{N}(0,1)$ for $j_2=1,\ldots,100,$ and $\beta^{(3)}_{j_3,k}\sim \delta_0$ for $j_3=1,\ldots,p-200.$
That is, for a fixed number of covariates $p>100$, we generate $100$ coefficients of high magnitude ($\sigma^{(1)}=10$), $100$ coefficients of low magnitude ($\sigma^{(2)}=1$), and $p-100$ coefficients identically equal to zero. Finally, we set $y_k=5 + X_k\bm{\beta}_k+\bm{\epsilon}_k$, with $\bm{\epsilon}_k\sim\mathcal{N}_n\left(0,\mathbb{I}_n\right)$.

For the mixture weights, we adopt a sparse mixture specification using $L=50$ mixture components and $a=0.05$, while we fix $\tau^2 = 0.0001$. As in the previous case, we ran 5,000 iterations as burn-in period and retained 5,000 iterations (thinning every 5 steps) for posterior inference. 

The average and standard errors of the overall MSE obtained by each model over the 30 replicates are reported in Table \ref{tab:overall}. Each row corresponds to a simulation scenario. In general, all models obtain very good performance. Indeed, in the first scenario, the average MSEs obtained by different models are very similar. As the number of covariates increases, we see how the sharing of information across the HSM parameters leads to lower MSE, followed, in order, by the Horseshoe+, the Horseshoe, and the Bayesian Lasso.

\begin{table}[t!]
    \centering
\begin{tabular}{lcccccccc}
\toprule
Scenario & \multicolumn{2}{c}{HSM} &
\multicolumn{2}{c}{HS} &
\multicolumn{2}{c}{HS+} &
\multicolumn{2}{c}{Lasso}\\
\midrule
1 & 0.0031 & (4e-04) & 0.0035 &(4e-04) & 0.0033  &(4e-04) & 0.0039  &(0.0005)\\
2 & 0.0019 &(3e-04) & 0.0044 &(6e-04) & 0.0031  &(4e-04) & 0.0287 &(0.0059)\\
3& 0.0014  &(3e-04) & 0.0039 &(6e-04) & 0.0028 &(5e-04) & 0.3075 &(0.0707)\\

\bottomrule
\end{tabular}
\caption{Overall average MSE (and relative standard errors) for the four different models under the three simulation scenarios. }
\label{tab:overall}
\end{table}

\begin{figure}[t!]
    \centering
    \includegraphics[width=\linewidth]{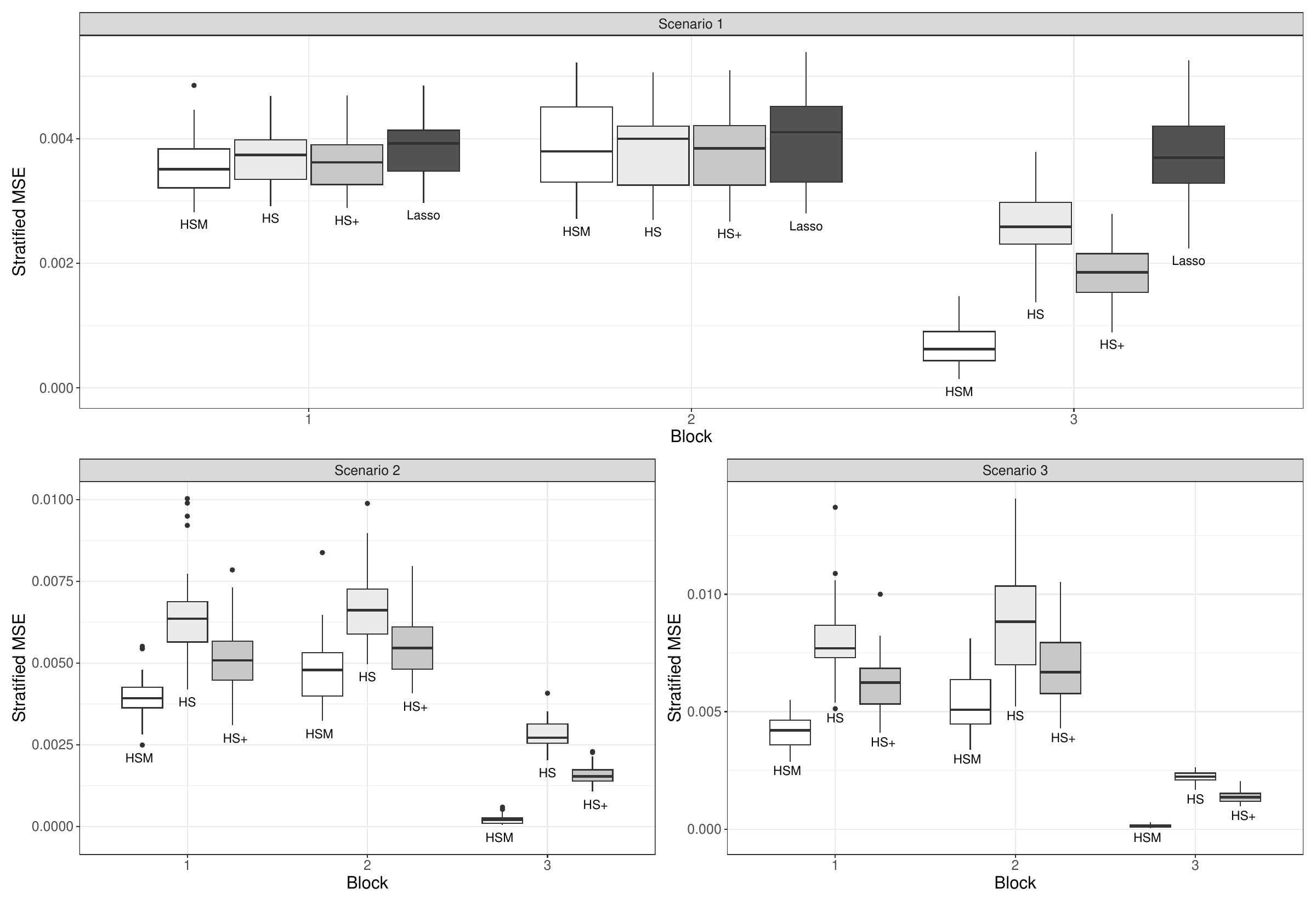}
    \caption{Boxplots of the stratified MSE obtained over 30 replicates by the different models. Each panel corresponds to a simulation scenario, with boxplots organized by blocks of parameters. The results obtained with the Lasso are omitted in the center and bottom panels to ease the visual comparison.}
    \label{fig:boxplot_reg}
\end{figure}

We decompose the overall MSE replicates into different magnitude blocks and report the results in Figure \ref{fig:boxplot_reg}. The boxplots describe the distributions of the results obtained over 30 replicates. Each panel depicts a simulation scenario. In the center and bottom panels, we removed the Bayesian Lasso's results to facilitate the visual comparison since its MSE is much larger than the ones obtained with HSM, Horseshoe, and Horseshoe+. The complete figure is reported in Section E of the Supplementary Material.
First, we notice how the HSM model obtains better performance in block 3, regardless of the scenario. Here, the gain in MSE reflects the ability of the model to target and shrink the true zeros effectively. Second, it is interesting to see how HSM obtains lower MSE than its competitors as the ratio $n/p$ decreases. While the boxplots for blocks 1 and 2 in the top panel are almost equivalent, MSE gains start to manifest in the other two panels.

\section{Discussion}
\label{Sec::discussion}

 In this paper, we have developed a novel class of priors that for multiple hypothesis testing and variable selection problems. Our proposed method groups the units of interests and their corresponding parameters into several tiers with varying degrees of relevance. This feature was proved to be particularly valuable for the specific application discussed in this paper: discovering differentially activated brain regions from data collected via a state-of-the-art brain imaging technology.   Specifically, our approach involves adopting a discrete mixture model, reminiscent of the two-group models, where each mixture component is itself a continuous scale mixture distribution. We then assumed half-Cauchy priors for the shrinkage parameters, mimicking the Horseshoe model. This way, we can retain the strong shrinkage properties of the continuous mixtures while performing model-based clustering typical of the discrete mixtures. Furthermore, the clustering detects irrelevant units and potentially segments the relevant ones into tiered classes, according to each coefficient's magnitude.
 With respect to the specific application discussed in this paper, the enhanced stratification of coefficients induced by our model detects important regions that were left out by more conservative methods. Further, these regions are ranked into groups of varying importance, avoiding arbitrary decisions in screening. Notably, our model ranks in Tier 2 regions that are related to the visual task, such as the primary visual area (layer 2/3) and various posterolateral visual areas, missed by the lFDR and SnS models. 
Combining the two discrete and continuous mixtures for shrinkage shows promising results, especially in targeting and regularizing the null coefficients via the clustering-induced shrinkage structure. We have showcased the potential of our approach using simulation studies.

The results presented in this paper can pave the way for many future research directions. For example, the data analysis can be enriched by including the hierarchical structure of brain regions -- each brain region can be divided into progressively smaller subregions. In our preprocessing workflow, we only considered the relationship between the areas at the highest resolution and their parents to account for potential correlations across regions. However, including information regarding the whole-brain structure could add substantial information since it is usually unclear at which hierarchical level a relevant differential effect emerges. Therefore, we are currently developing a two-group model that incorporates such information directly in the Bayesian model, studying how the activation probability is partitioned across the regions' ancestors.
From a methodological point of view, one can consider different continuous scale mixture types to improve the HSM model. For example, a possibility would be to take into consideration a Laplace distribution generalizing the model in \citet{Rockova2016}, or more refined horseshoe distributions such as the horseshoe-like distribution \citep{Bhadra2019a}, or regularized horseshoe \citep{Piironen2017}. 
Moreover, the prior specification for the global shrinkage parameter, $\tau$, could be improved in the context of our clustering approach, for example, by using the method of \citet{Piironen2017_tau}. Inspired by the idea of splitting the error rate introduced by \citet{Tukey1993}, it may be possible to develop a mixture of shrinkage priors that behaves asymmetrically around the origin. The approach would estimate the probability that each z-score is assigned to a particular relevance tier while allowing for different tail behaviors in over- and under-expressed brain regions.  
Lastly, the scalability of our method could also be improved using more efficient MCMC alternatives, such as the two algorithms for horseshoe estimation recently proposed in \citet{Johndrow2020}. Alternatively, one can adopt an approximate inference method such as mean-field variational Bayes \citep{Neville2014}.

%% file: Table_regions.tex
\begin{table}[ht!]
\centering
\resizebox{.9\columnwidth}{!}{%
\renewcommand{\arraystretch}{.55}
\footnotesize
\begin{tabular}{ll}
\toprule
Tier 1 & Tier 2\\
\midrule
Agranular insular area, ventral part: layer 1 & Agranular insular area, dorsal part: layer 1,  6a\\

Anterior cingulate area, dorsal part: layer 1,  5 & Anterior cingulate area, ventral part: layer 1,  2/3,  5\\

Anterolateral visual area: layer 1 & Anterior olfactory nucleus\\

Dorsal auditory area: layer 1,  2/3,  5,  6a & Anteromedial visual area: layer 1\\

Ectorhinal area: lyer 2/3 & Central amygdalar nucleus\\

Lateral visual area: layer 1,  5 & Ectorhinal area: layer 1,  5,  6a\\

Posteromedial visual area,  4,  5 & Fiber tracts\\

Postpiriform transition area & Hippocampal formation\\

Primary auditory area: layer 2/3,  5 & Infralimbic area: layer 1,  5\\

Primary visual area: layer 1,  4,  5,  6a & Main olfactory bulb\\

Subiculum & Nucleus accumbens\\

Taenia tecta & Orbital area, medial part: layer 1,  5\\

Temporal association areas: layer 2/3,  6a & Orbital area, ventrolateral part: layer 1\\

Ventral auditory area: layer 2/3 & Parasubiculum\\

 $\cdot$ & Perirhinal area: layer 1,  2/3,  5,  6a\\
 $\cdot$ & Piriform area\\
 $\cdot$ & Piriform-amygdalar area\\
 $\cdot$ & Posterior auditory area: layer 2/3\\
 $\cdot$ & Posterolateral visual area: layer 2/3,  5\\
 $\cdot$ & posteromedial visual area: layer 1\\
 $\cdot$ & Postsubiculum\\
 $\cdot$ & Prelimbic area: layer 1,  5\\
 $\cdot$ & Primary auditory area: layer 1,  6a\\
 $\cdot$ & Primary motor area: layer 1,  5\\
 $\cdot$ & Primary somat. area, barrel field: layer 1,  2/3,  5,  6a\\
 $\cdot$ & Primary somat. area, lower limb: layer 2/3,  5\\
 $\cdot$ & Primary somat. area, mouth: layer 2/3\\
 $\cdot$ & Primary somat. area, nose,  2/3\\
 $\cdot$ & Primary somat. area, trunk: layer 2/3,  4,  5\\
 $\cdot$ & Primary somat. area, upper limb: layer 1,  2/3\\
 $\cdot$ & Primary visual area,  2/3\\
 $\cdot$ & Retrosplenial area, lateral agranular part,  1\\
 $\cdot$ & Retrosplenial area, lateral agranular part: layer 5\\
 $\cdot$ & Retrosplenial area, ventral part: layer 1,  6a\\
 $\cdot$ & Secondary motor area: layer 2/3,  5,  6a\\
 $\cdot$ & Suppl. somatosensory area: layer 1,  5,  6a\\
 $\cdot$ & Temporal association areas: layer 1,  5\\
 $\cdot$ & Third ventricle\\
 $\cdot$ & Unlabeled\\
 $\cdot$ & Ventral auditory area: layer 5,  6a\\
\bottomrule
\end{tabular}
}
\caption{Lists of brain regions assigned to \texttt{Tier 1} and \texttt{Tier 2} of activation by the HSM model.}
\label{tier1}
\end{table}
\renewcommand{\arraystretch}{1}

%% file: Suppmatt_body.tex
\title{Supplementary Material}
\maketitle
\section{Connections with relevant literature}

\subsection{The link between the Horseshoe Mix prior and Bayesian robust modeling}

To provide an additional perspective, we can relate our regularization prior to the ``classical'' vs. ``alternative'' paradigm discussed by \citet{Finegold2011} in terms of robust modeling. Suppose a generic random variable $\bm{Y}$ is distributed according to a continuous scale mixture of Normals. In that case, it can be equivalently represented as $\bm{Y}=\bm{X}\rho$, where $\bm{X}$ has the multivariate standard Gaussian distribution and $\rho$ is a random scale parameter. 
It is well-known that, if $\rho\sim Gamma(\nu/2,\nu/2)$, then $\bm{Y}$ is distributed as a multivariate Student-t.
In the ``classical'' case, $\rho$ is univariate and shared across all the coordinates of $\bm{X}=\{X_j\}_{j=1}^p$. This case is equivalent to a regularization model with a unique, random global shrinkage parameter $\rho=\tau$. 
The opposite situation, where each entry $X_j$ is paired with a unique scale $\rho_j$, is referred to as the ``alternative'' multivariate Student-t. This case corresponds to the classical global-local shrinkage models with deterministic $\bar{\tau}$, where $\rho_j=\bar{\tau}\lambda_j$. In \citet{Finegold2014}, the authors propose a third distribution, Dirichlet-t, assuming $\tau_j\sim G$ and $G\sim DP(\alpha, Gamma(\nu/2,\nu/2))$. This last option is directly linked with our proposal.

\begin{figure}[th!]
	\centering
	\includegraphics[scale=.32]{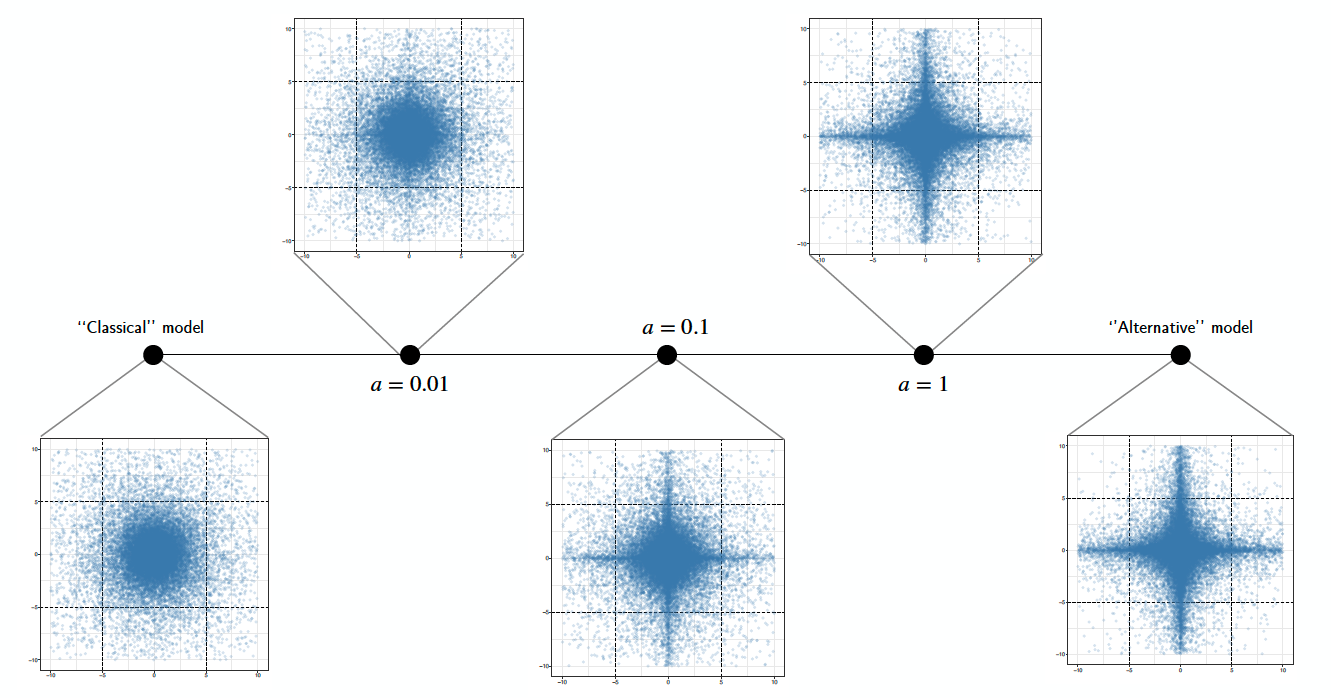}
	\caption{Parametric Dirichlet-HS realizations induced by different specification of the Dirichlet distribution adopted for $L=10$ mixture weights.}
	\label{fig:ALCL}
\end{figure}

Following their definition, once we assume a fixed global shrinkage parameter $\tau=\bar{\tau}$ and $\lambda_l\sim \mathcal{C}^+$, both models (2) and (4) in the main paper can be seen as the parametric and nonparametric versions of a novel Dirichlet-HS distribution, respectively. \citet{Shahbaba2014} suggested to use this structure in the context of robust modeling, given the appealing properties of the Horseshoe distribution. This argument can be generalized to Dirichlet-$\rho$ distributions, defined by considering different specifications for the scale parameter $\rho$. Notice that, despite similar names, these distributions are essentially different from the one proposed by \citet{Bhattacharya2015}.
Even in this context, the gain is twofold. First, we obtain a distribution that is more flexible than the ``classic'' one. Second, we allow sharing of statistical strength across the elements of the random vector even without assuming a distribution for $\tau$, for which special care is needed to carry out posterior simulation \citep{Piironen2017}. Moreover, we gain in terms of computational properties by greatly reducing the number of parameters.
\citet{Finegold2014} highlight that the structure of a distribution such as the generic Dirichlet-$\rho$ interpolates between the two extreme model specifications (``classical'' vs. ``alternative''). 
We add that, even in the simple case of a parametric mixture, model (2) in the main paper gives us direct control of where the resulting distribution takes place between the two extremes by tweaking the prior over the mixture weights. 
Figure \ref{fig:ALCL} provides an example with the Dirichlet-HS distribution with $L=10$. 
The overfitted mixture case ($a\leq 0.05$) is the closest one to the ``classical'' model, which can be recovered for $a\rightarrow 0$. As $a$ increases, if $L$ is large enough, the likelihood of sampling distinct values of $\rho_j$ for all $j$ increases, and therefore we get closer to the ``alternative'' model.

\subsection{Additional relevant literature}
\citet{MacLehose2010} proposed to model regression coefficients as binary mixtures between a traditional double exponential centered in zero (Bayesian Lasso) and a double exponential with non-zero location parameter. They employ DPs for both the location and scale parameters. This model was extended by \citet{Yang2011}, who adopted mixtures of Bayesian elastic-nets \citep{Li2010}.
In contrast, our approach focuses only on the modeling of variances, fixing the center of the mixture kernel distributions to 0. This choice allows to exploit the properties of the continuous scale mixture family. 
Recently, \citet{Ding2020} explored the effects of the combination between shrinkage priors and a covariate dependent DP mixtures. The authors jointly model the conditional response distributions and the covariates with a nonparametric mixture, inducing a partition over the observations.  We instead consider fixed covariates and employ the mixture to cluster the regression coefficients into shrinkage profiles.

\section{Marginal likelihood and posterior means under the HSM prior}
\subsection*{Marginal Likelihood}
Let $\sigma^2=1$ without loss of generality. Then, the sampling distribution for $z_i$, $i=1,\ldots,n$ after marginalizing out the parameter vector $\bm{\beta}$ is obtained as:
\begin{align*}
    \pi(z_i|\tau,\bm{\lambda},\bm{\pi})&= \int \pi(z_i|\bm{\beta})
    \pi(\bm{\beta}|\tau,\bm{\lambda},\bm{\pi}) d\bm{\beta}\\&= \int \pi(z_i|\beta_i)
    \pi(\beta_i|\tau,\bm{\lambda},\bm{\pi}) d\beta_i\\
    &=
    \int \frac{1}{\sqrt{2\pi}}e^{-\frac{(z_i-\beta_i)^2}{2}}
    \left( \sum_{l=1}^{L} \pi_l \frac{1}{\sqrt{2\pi\tau^2\lambda_l^2}}e^{-\frac{\beta_i^2}{2\tau^2\lambda^2_l}}  \right) d\beta_i\\
    &=
    \sum_{l=1}^{L}\pi_l     \frac{1}{2\pi\sqrt{\tau^2\lambda_l^2}}\int e^{-\frac{(z_i-\beta_i)^2}{2}
    -\frac{\beta_i^2}{2\tau^2\lambda^2_l}}  d\beta_i\\
    &=
    \sum_{l=1}^{L}\pi_l\frac{e^{-\frac{z_i^2}{2}}}{2\pi\sqrt{\tau^2\lambda_l^2}} \int e^{-0.5\left(\frac{1+\tau^2\lambda^2_l}{\tau^2\lambda^2_l}\beta_i^2 
    -2\beta_iz_i \right)}  d\beta_i\\
    &=\sum_{l=1}^{L}\pi_l \frac{1}{2\pi\sqrt{\tau^2\lambda_l^2}}\sqrt{\frac{\tau^2\lambda_l^2}{1+\tau^2\lambda_l^2}}e^{-\frac{z_i^2}{2}\left( 1-\frac{\tau^2\lambda_l^2}{1+\tau^2\lambda_l^2}\right)} \\
    &= \sum_{l=1}^{L} \frac{\pi_l}{\sqrt{2\pi(1+\tau^2\lambda_l^2)}}e^{-\frac{z_i^2}{2(1+\tau^2\lambda_l^2)}}.
\end{align*}
Therefore, $z_i|\tau,\bm{\lambda},\bm{\pi}\sim \sum_{l=1}^L \pi_l \mathcal{N}(0, 1+\tau^2\lambda^2_l)$. If $\sigma^2$ is supposed to be stochastic, we would obtain $z_i|\tau,\bm{\lambda},\bm{\pi},\sigma^2\sim \sum_{l=1}^L \pi_l \mathcal{N}(0, \sigma^2(1+\tau^2\lambda^2_l))$.
\subsection*{Posterior mean}
Again, let us suppose $\sigma^2=1$. We can derive the posterior mean of the parameter $\beta_i$ as follows.
\begin{align*}
    \mathbb{E}\left[\beta_i|\tau,\bm{\lambda},\bm{\pi},z_i\right] &= \int\beta_i \frac{\pi(z_i,\beta_i,\tau,\bm{\lambda},\bm{\pi})}{\pi(z_i,\tau,\bm{\lambda},\bm{\pi})}d\beta_i =
    \frac{\pi(\tau,\bm{\lambda},\bm{\pi})}{\pi(z_i,\tau,\bm{\lambda},\bm{\pi})}\int\beta_i \pi(z_i|\beta_i)\pi(\beta_i|\tau,\bm{\lambda})d\beta_i\\
    &= \frac{1}{\pi(z_i|\tau,\bm{\lambda},\bm{\pi})}
    \int\beta_i \pi(z_i|\beta_i)\pi(\beta_i|\tau,\bm{\lambda})d\beta_i\\
    &=\sum_{l=1}^L\pi_l\frac{e^{-\frac{z_i^2}{2}\left(1-\frac{\tau^2\lambda_l^2}{1+\tau^2\lambda_l^2}\right)} }{\pi(z_i|\tau,\bm{\lambda},\bm{\pi})\sqrt{2\pi(1+\tau^2\lambda^2)}}
    \int\beta_i \phi\left(\beta_i;z_i\frac{\tau^2\lambda_l^2}{1+\tau^2\lambda_l^2},\frac{\tau^2\lambda_l^2}{1+\tau^2\lambda_l^2}\right) d\beta_i\\
    &=\sum_{l=1}^L\pi_l\frac{e^{-\frac{z_i^2}{2(1+\tau^2\lambda_l^2)}} }{\pi(z_i|\tau,\bm{\lambda},\bm{\pi})\sqrt{2\pi(1+\tau^2\lambda^2_l)}}z_i\frac{\tau^2\lambda_l^2}{1+\tau^2\lambda_l^2}\\
    &=\sum_{l=1}^L\pi_l\frac{\phi(z_i;0,1+\tau^2\lambda^2_l) }{\pi(z_i|\tau,\bm{\lambda},\bm{\pi})}z_i(1-\kappa^*_l)\\
    &=\frac{\sum_{l=1}^L\pi_l\phi(z_i;0,1+\tau^2\lambda^2_l) (1-\kappa^*_l)}{\pi(z_i|\tau,\bm{\lambda},\bm{\pi})}z_i\\
    &=\frac{\sum_{l=1}^Lr_l(z_i) (1-\kappa^*_l)}{\sum_{l=1}^Lr_l(z_i)}z_i.\\
\end{align*}
where $1-\kappa^*_k=\frac{\tau^2\lambda_l^2}{1+\tau^2\lambda_l^2}$ and $r_l(z_i)=\pi_l\phi(z_i;0,1+\tau^2\lambda^2_l)$.

\clearpage

\section{Gibbs Sampler for the HSM model}

We report the algorithm for the Gibbs sampler used in our analyses. Here, we suppose $\tau \sim \mathcal{C}^+$. The algorithm can be straightforwardly modified in case one assumes $\tau^2 \sim IG$. The presented Gibbs sampler is devised for the linear regression framework with $n$ observations and $p$ covariates. We denote the target variable with $\bm{y}=\{y_i\}_{i=1}^n$ and the covariate matrix with $\bm{X}$. Therefore, we consider $\tilde{\bm{\beta}}=\{\tilde{\beta}\}_{j=1}^p$. By simply setting $\bm{X} = \mathbb{I}_n$, one recovers the mean estimation framework. At each iteration of the Gibbs sampler, we simulate from the following full conditionals:
\begin{enumerate}

     \item Let $\bm{\Lambda}_{\zeta}=\tau^2\cdot diag(\lambda^2_{\zeta_1},\ldots,\lambda^2_{\zeta_p})$\; 
     Sample $$\tilde{\bm{\beta}}\sim \mathcal{N}_p\left( A^{-1}\bm{X}'\bm{y}, \sigma^2 A^{-1}\right),$$ where $A=(\bm{X}'\bm{X}+\boldsymbol{\Lambda}_{\zeta})$. To efficiently sample from this distribution we follow \citet{Makalic2016} and employ the algorithm of \citet{Havard2001} when $p/n\leq 2$, while we use \citet{Bhattacharya2016} otherwise.
     
    \item Sample $\zeta_j$ according to $$\pi(\zeta_j=l) \propto \pi_l \cdot \phi(\tilde{\beta}_j;0;\sigma^2\cdot\tau^2\cdot\lambda^2_{l}).$$ Then, compute $n_l = \#\{j:\zeta_j=l\} $ for $l=1,\ldots,L$.

    \item Introduce the auxiliary variable $u_{\lambda_l}$.
    Let $t_l=1/\lambda^2_l$. Then, sample $u_{\lambda_l} \sim \mathcal{U}\left(0,1/(1+t_l)\right)$ and $$t_l \sim G\left(\frac{(n_l+1)}{2},\frac{\sum_{j:\zeta_j=l} \tilde{\beta}_l^2}{2\tau^2\sigma^2}\right)\mathbbm{1}_{t_l\in \left[0,1/u_{\lambda_l}-1\right]}.$$
     
    \item Introduce the auxiliary variable $u_{\tau}$. Let $t^*=1/\tau^2$. Then, sample $u_{\tau} \sim \mathcal{U}\left(0,1/(1+t^*)\right)$ and $$t^* \sim G\left(\frac{(p+1)}{2},\frac{\sum_{j} \tilde{\beta}_j^2/\lambda_ {\zeta_j}^2}{2\sigma^2}\right)\mathbbm{1}_{t^*\in \left[0,1/u_{\tau}-1\right]}.$$
     
    \item Sample the error variance $$\sigma^2 \sim IG\left(\frac{n+p}{2},\frac{\sum_i (y_i-X_i\tilde{\beta})^2+\sum_j\tilde{\beta}^2_{j}/\tau\lambda_{\zeta_j}}{2}\right).$$\;
     
    \item Sample the mixture weights $\boldsymbol{\pi}$ from $Dir(a_1+n_1,\ldots,a_J+n_L)$ for finite mixture models; for nonparametric mixtures,  use the corresponding step for the stick-breaking construction from blocked Gibbs sampler of \citet{Ishwaran2001}.  \;
\end{enumerate}

\clearpage

\section{Membership labels and conditional posterior distribution interpretation}

\begin{figure}[th!]
	\centering
	\includegraphics[scale=.25]{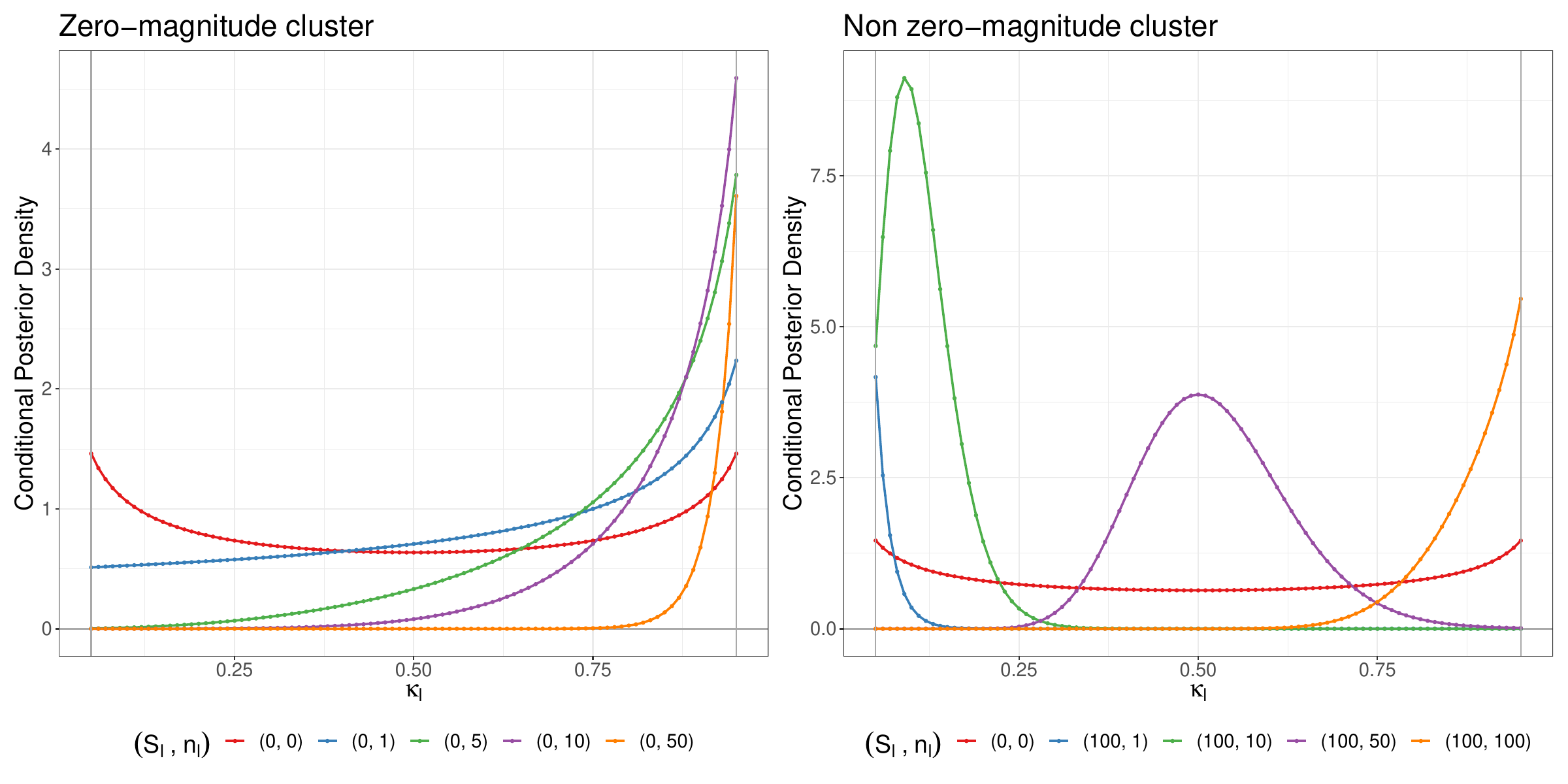}
	\caption{Posterior densities of the MCSF $\kappa^*_l $, changing according to the number of coefficients assigned to a specific mixture component and their magnitudes.}
	\label{fig:Behavior}
\end{figure}

The data augmentation we adopted in is not only useful to conduct feasible posterior inference, but also to provide more insights regarding the behavior of MCSFs. Indeed, we can derive the conditional posterior distribution for the $l$-th MCSF $\kappa^*_l$ under the Horseshoe Mix prior, obtaining
	\begin{equation}
		\begin{aligned}
			\label{eq::condPost}
			p(\kappa^*_l | \boldsymbol{\zeta}, \bm{z}, \tau^2, \sigma^2) \propto& \frac{(\kappa^*_l)^{-\frac{1}{2}}(1-\kappa^*_l)^{-\frac{1}{2}} }{\tau^2 \kappa^*_l +1 -\kappa_l^*}(\kappa_l^*)^{\frac{n_l}{2}}\cdot\exp\left[-\frac{\kappa^*_l}{2\sigma^2}\sum_{j:z_i=l} z_i^2\right].
		\end{aligned}
	\end{equation}
It is crucial to note how all the observations that are grouped in the $l$-th cluster explicitly contribute to the conditional posterior distribution of $\kappa^*_l$. Without loss of generality, we set $\sigma^2=\tau^2=1$. Moreover, define $S_l=\sum_{j:z_i=l} z_i^2$. Then, the distribution in \eqref{eq::condPost} simplifies into $p(\kappa^*_l | \boldsymbol{\zeta}, \bm{z}) \propto (\kappa^*_l)^{\frac{n_l-1}{2}}(1-\kappa^*_l)^{-\frac{1}{2}} \exp\left[-\frac{\kappa^*_l}{2}S_l\right]$. Whenever $L=p$, $n_l=1$, and $S_l=z_i^2$, we recover the the Horseshoe model.\\
We illustrate the behavior of this distribution in the two panels of Figure \ref{fig:Behavior}, where we report different shapes of the conditional posterior density functions for different combinations of $(S_l, n_l)$. We expect our model to group parameters characterized by similar magnitude, so the left panel shows what happens when $S_l=0$ and $n_l$ grow. Ideally, the more observations with 0 magnitude are assigned to the same cluster, the stronger is the shrinkage effect of the MCSF $\kappa^*_l$, concentrating all the mass around 1. In contrast, high values of $S_l$ and low values of $n_l$ will result in small $\kappa^*_l$, as we can see in the right panel. In other words, MCSF is lower when the relative cluster is comprised of few observations of great magnitude. In Section D of the Supplementary Material, we report a diagram that depicts this behavior, which explains how the sharing of information is exploited by our model to tune the amount of shrinkage according to the observed data.

\begin{figure}[h!]
    \centering
    \includegraphics[scale=.4]{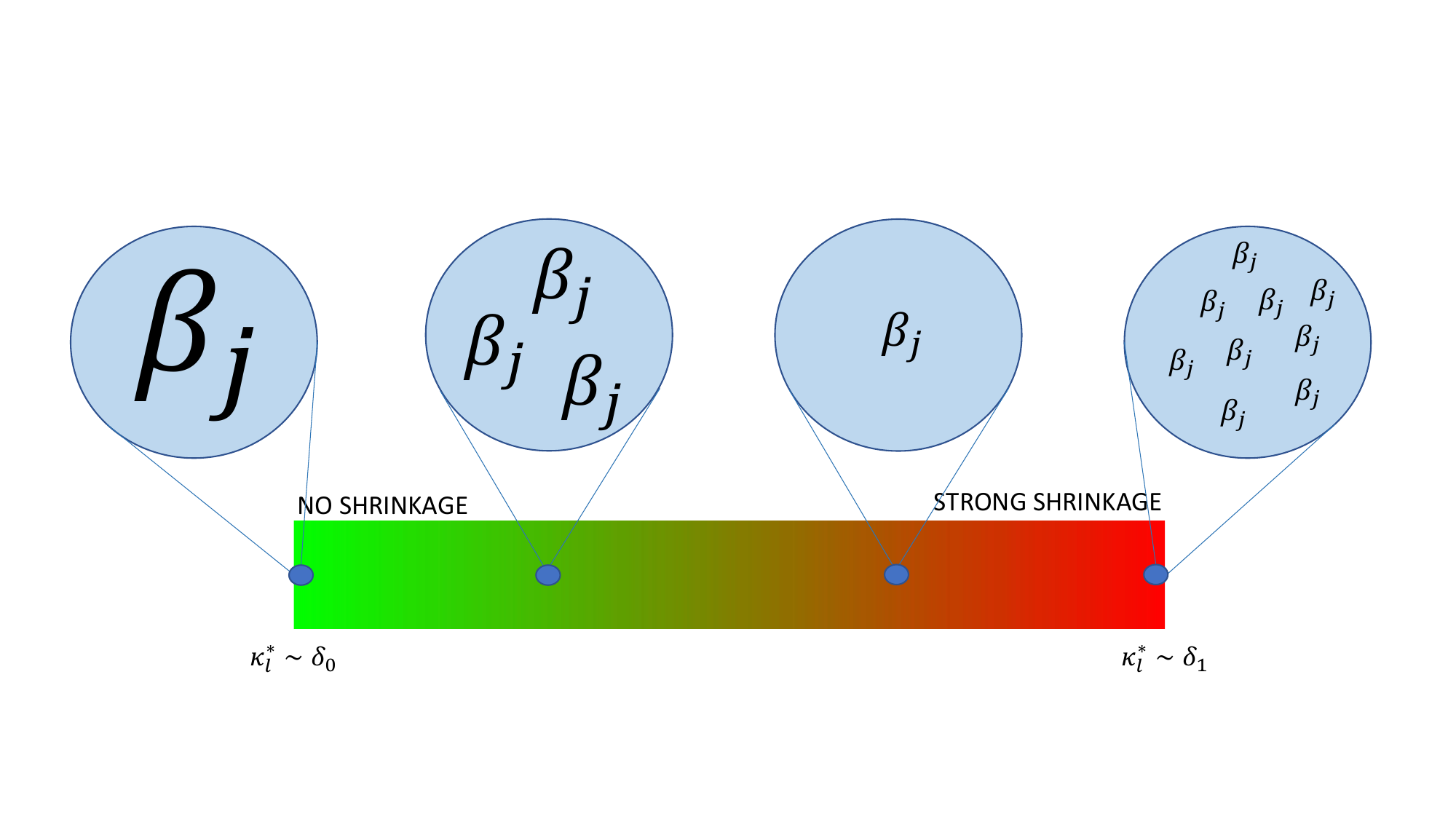}
    \caption{A visual depiction of the shrinkage induced by the HSM model. Clusters containing numerous coefficients with overall low magnitude are regularized the most (right end of the scale). On the contrary, cluster containing few, big coefficients suffer less regularization (left end of the scale).}
    \label{fig:scale}
\end{figure}

\FloatBarrier

\section{Additional Figures}

\begin{figure}[h!]
    \centering
    \includegraphics[scale=.5]{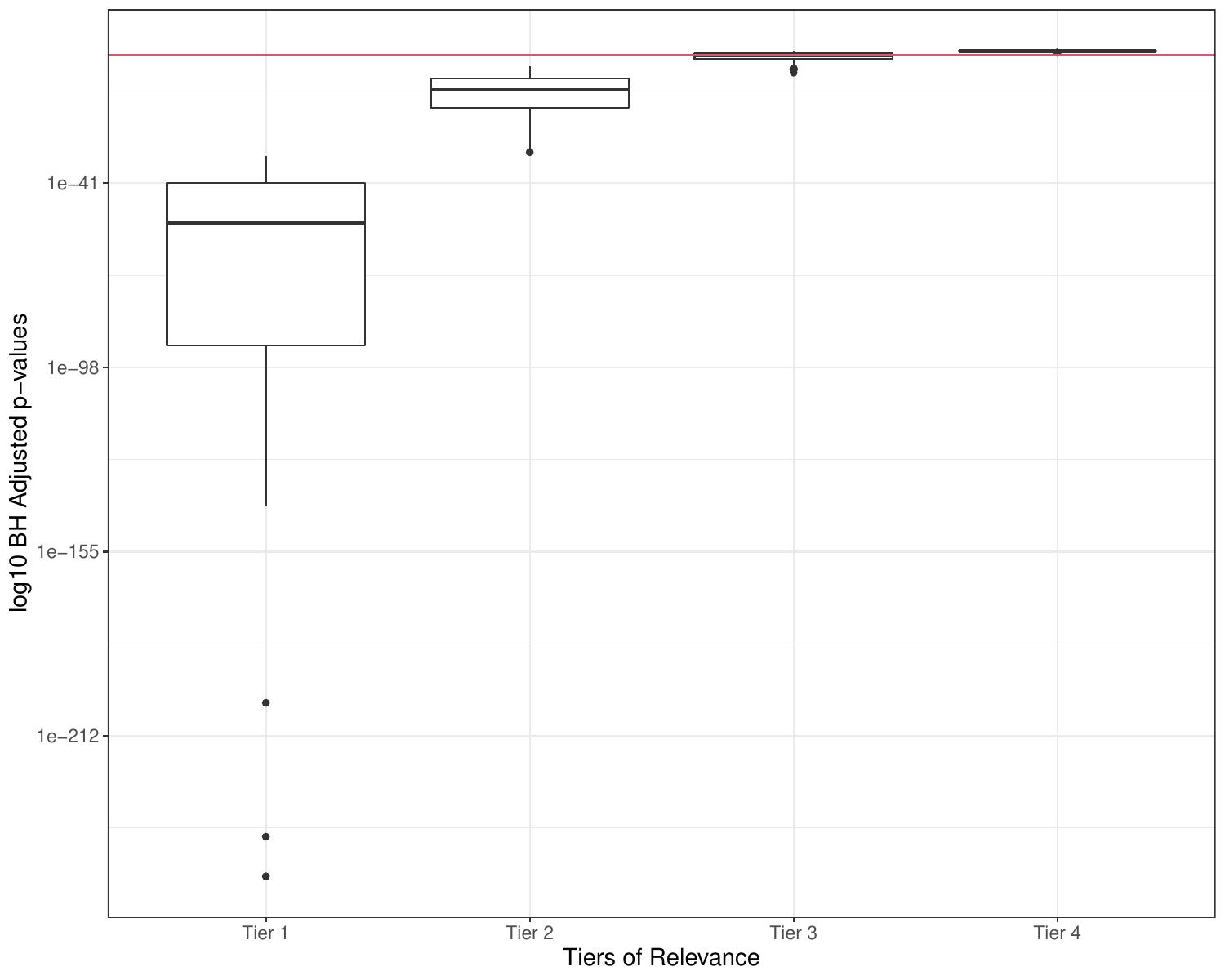}
    \caption{Comparison between the boxplots of the adjusted p-values (according to the \texttt{BH} procedure), stratified into tiers of relevance. The horizontal red lines depicts the cutoff value $0.05$. The y-axis is displayed in $\log 10$ scale.}
    \label{fig:boxplots_app}
\end{figure}
\begin{figure}[h!]
    \centering
    \includegraphics[width = \linewidth]{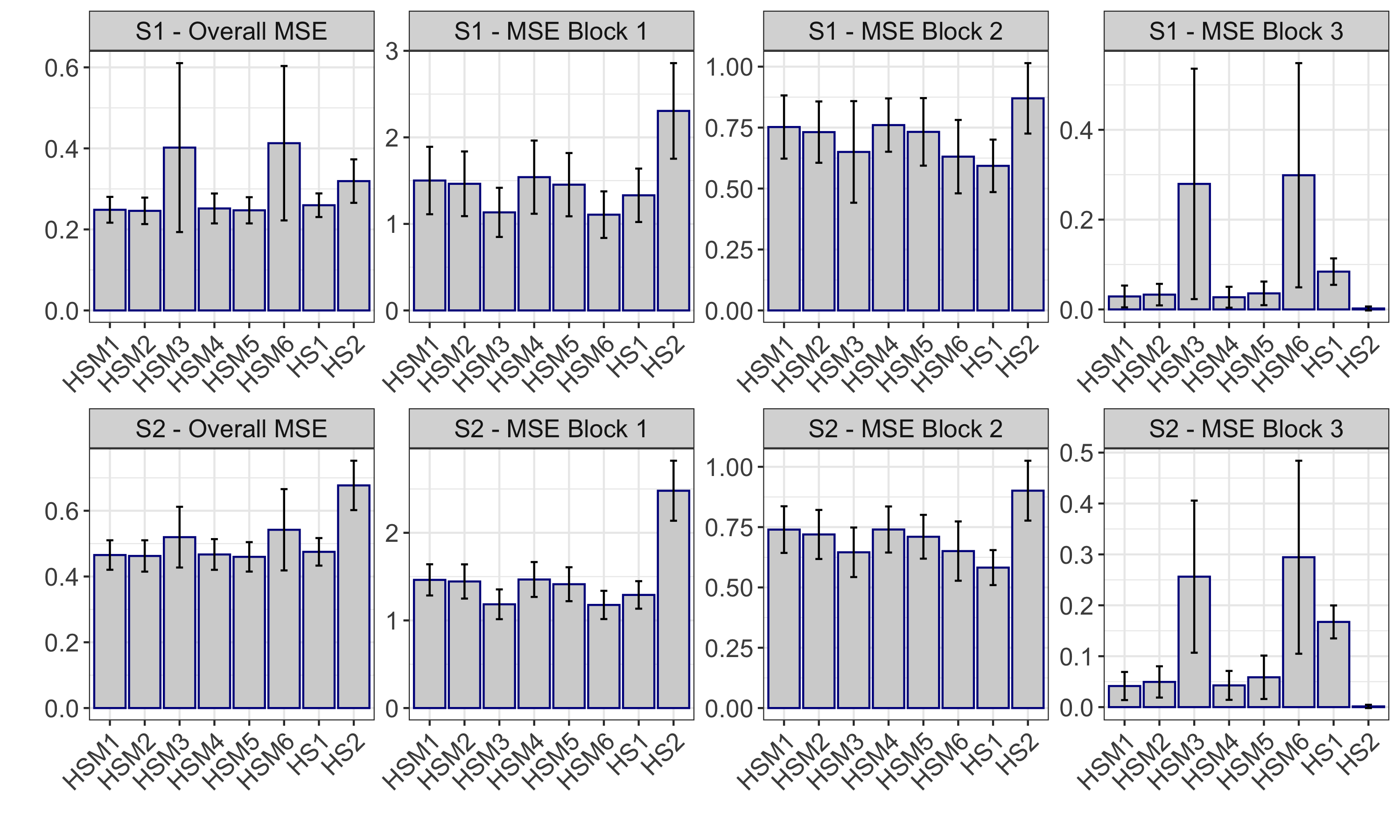}
    \caption{Bar plots of the average overall and stratified MSEs and error bars representing the corresponding standard errors for simulation scenarios S1 and S2. The plot is based on the values reported in Table \ref{tab:S2}. Here, the bars relative to the Spike and Slab model have been removed to ease the visual comparison of the remaining models.}
    \label{fig:scale2}
\end{figure}

\begin{figure}[h!]
    \centering
    \includegraphics[width=.9\linewidth]{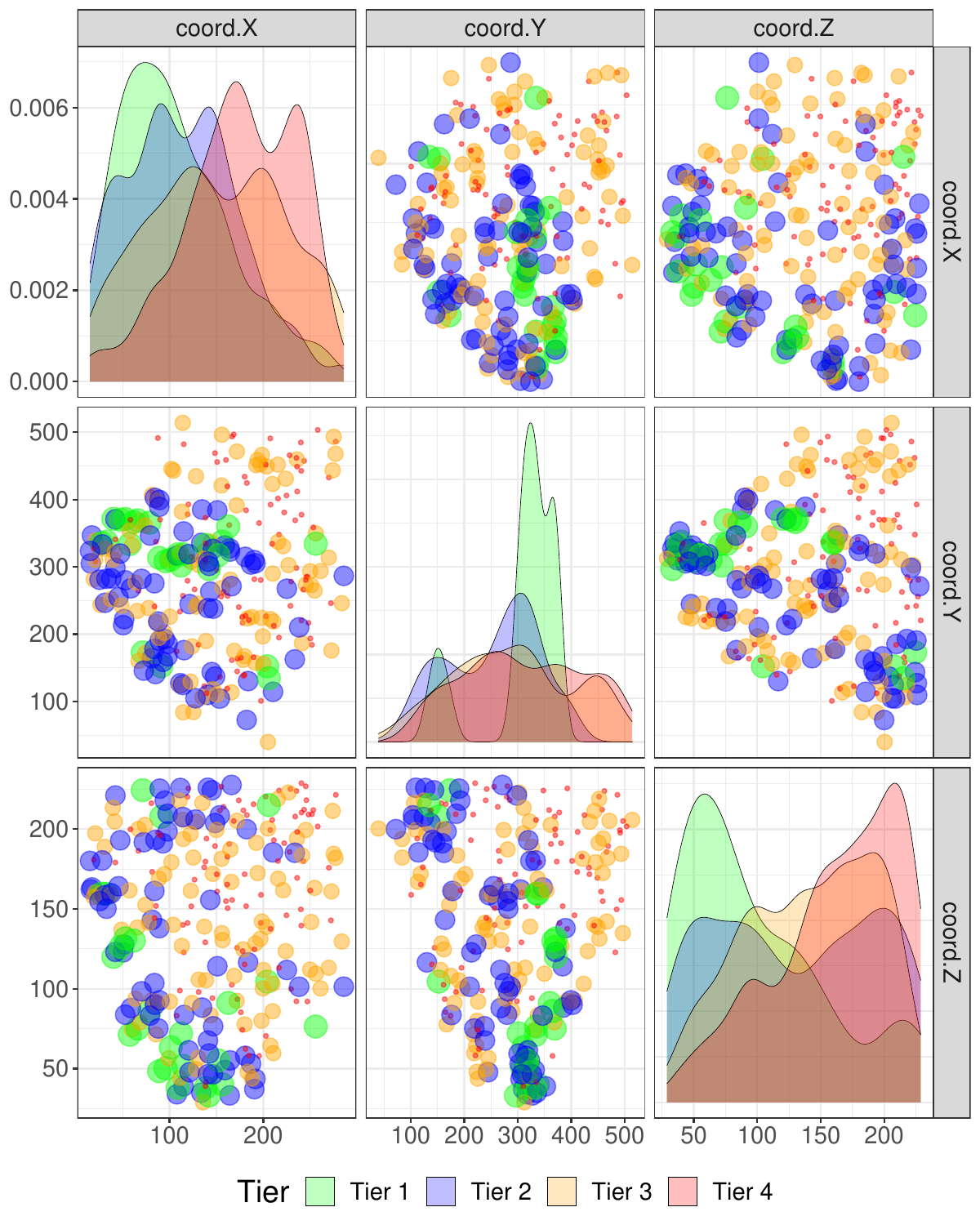}
    \caption{Pairwise scatterplots representing the projection of the 3D spatial coordinates. Each point indicates the spatial coordinates of the centroid computed over the recorded neurons for each brain region of interest. The colors and sizes reflect the different relevance tiers.}
    \label{fig:spatial}
\end{figure}

\clearpage
\FloatBarrier
\section{Additional Tables}
\FloatBarrier
\begin{table}[ht!]
\footnotesize
\centering
\begin{tabular}{llcccccccc}
\toprule
&Model & \multicolumn{2}{c}{Overall MSE} & \multicolumn{2}{c}{MSE Block 1} & \multicolumn{2}{c}{MSE Block 2} & \multicolumn{2}{c}{MSE Block 3} \\
\cmidrule(lr{1em}){1-2} \cmidrule(lr{1em}){3-4} \cmidrule(lr{1em}){5-6} \cmidrule(lr{1em}){7-8}   \cmidrule(lr{1em}){9-10}       
                    & HSM1 & 0.2485 & (0.0318) & 1.5012 & (0.3897) & 0.7523 & (0.1297) & 0.0289 & (0.0243)\\
                    & HSM2 & 0.2458 & (0.0327) & 1.4634 & (0.3737) & 0.7314 & (0.1257) & 0.0330 & (0.0240)\\
                    & HSM3 & 0.4020 & (0.2085) & 1.1331 & (0.2841) & 0.6501 & (0.2084) & 0.2795 & (0.2566)\\
                    & HSM4 & 0.2518 & (0.0369) & 1.5402 & (0.4228) & 0.7603 & (0.1093) & 0.0271 & (0.0233)\\
S1          & HSM5 & 0.2472 & (0.0323) & 1.4534 & (0.3655) & 0.7324 & (0.1385) & 0.0357 & (0.0263)\\
                    & HSM6 & 0.4128 & (0.1906) & 1.1067 & (0.2691) & 0.6308 & (0.1505) & 0.2988 & (0.2497)\\
                    & HS1  & 0.2597 & (0.0292) & 1.3304 & (0.3087) & 0.5931 & (0.1077) & 0.0842 & (0.0294)\\
                    & HS2  & 0.3192 & (0.0537) & 2.3054 & (0.5524) & 0.8699 & (0.1445) & 0.0021 & (0.0044)\\
                    & SnS   & 0.6746 & (0.2156) & 5.7892 & (2.0793) & 0.9568 & (0.1723) & 0.0000 & (0.0000)\\
\midrule
                    & HSM1 & 0.4653 & (0.0448) & 1.4627 & (0.1780) & 0.7397 & (0.0967) & 0.0414 & (0.0276)\\
                    & HSM2 & 0.4625 & (0.0477) & 1.4445 & (0.1952) & 0.7196 & (0.1017) & 0.0495 & (0.0307)\\
                    & HSM3 & 0.5197 & (0.0923) & 1.1839 & (0.1699) & 0.6456 & (0.1027) & 0.2563 & (0.1494)\\
                    & HSM4 & 0.4670 & (0.0467) & 1.4670 & (0.1991) & 0.7401 & (0.0955) & 0.0427 & (0.0284)\\
S2                    & HSM5 & 0.4598 & (0.0448) & 1.4132 & (0.1937) & 0.7100 & (0.0908) & 0.0586 & (0.0427)\\
                    & HSM6 & 0.5421 & (0.1236) & 1.1766 & (0.1612) & 0.6507 & (0.1229) & 0.2945 & (0.1896)\\
                    & HS1  & 0.4750 & (0.0419) & 1.2909 & (0.1570) & 0.5822 & (0.0725) & 0.1674 & (0.0324)\\
                    & HS2  & 0.6770 & (0.0751) & 2.4799 & (0.3425) & 0.9011 & (0.1241) & 0.0014 & (0.0032)\\
                    & SnS   & 10.7556 & (3.8950) & 52.8020 & (19.4701) & 0.9760 & (0.1454) & 0.0000 & (0.0000)\\
\midrule

&HSM1 & 0.1354 & (0.0181) & 1.6302 & (0.4099) & 0.7935 & (0.1290) & 0.0158 & (0.0105)\\
&HSM2 & 0.1355 & (0.0189) & 1.5953 & (0.4205) & 0.7811 & (0.1333) & 0.0186 & (0.0159)\\
&HSM3 & 0.1941 & (0.0740) & 1.2932 & (0.2906) & 0.6131 & (0.1264) & 0.1097 & (0.0892)\\
&HSM4 & 0.1354 & (0.0183) & 1.6453 & (0.4248) & 0.8050 & (0.1336) & 0.0144 & (0.0106)\\
S3 & HSM5 & 0.1381 & (0.0212) & 1.6015 & (0.4256) & 0.7787 & (0.1412) & 0.0212 & (0.0222)\\
&HSM6 & 0.3280 & (0.2627) & 1.2097 & (0.3586) & 0.6411 & (0.1472) & 0.2616 & (0.3019)\\
&HS1  & 0.1430 & (0.0147) & 1.4772 & (0.3506) & 0.6578 & (0.1091) & 0.0403 & (0.0140)\\
&HS2  & 0.1594 & (0.0266) & 2.2825 & (0.5586) & 0.8705 & (0.1464) & 0.0020 & (0.0029)\\
&SnS   & 0.2869 & (0.0758) & 4.7830 & (1.4444) & 0.9555 & (0.1722) & 0.0000 & (0.0000)\\
\midrule
&HSM1 & 0.2565 & (0.0258) & 1.5647 & (0.2085) & 0.7724 & (0.1167) & 0.0284 & (0.0146)\\
&HSM2 & 0.2580 & (0.0248) & 1.5747 & (0.1890) & 0.7824 & (0.1096) & 0.0279 & (0.0120)\\
&HSM3 & 0.2959 & (0.0706) & 1.3435 & (0.1700) & 0.6485 & (0.0992) & 0.1209 & (0.0934)\\
&HSM4 & 0.2593 & (0.0240) & 1.5941 & (0.1953) & 0.7831 & (0.1228) & 0.0270 & (0.0154)\\
S4 &HSM5 & 0.2576 & (0.0237) & 1.5846 & (0.2022) & 0.7823 & (0.1043) & 0.0261 & (0.0128)\\
&HSM6 & 0.4130 & (0.1816) & 1.2571 & (0.2007) & 0.6693 & (0.1045) & 0.2755 & (0.2375)\\
&HS1  & 0.2664 & (0.0227) & 1.4657 & (0.1706) & 0.6314 & (0.0857) & 0.0708 & (0.0153)\\
&HS2  & 0.3369 & (0.0358) & 2.4533 & (0.3234) & 0.9026 & (0.1242) & 0.0017 & (0.0024)\\
&SnS   & 0.7971 & (0.1898) & 6.9954 & (1.9543) & 0.9759 & (0.1453) & 0.0000 & (0.0000)\\
\bottomrule
\end{tabular}
\caption{Average overall and stratified MSEs and corresponding standard errors (between parentheses) for simulation scenarios S1-S4. Each row represents a different model specification. }
\label{tab:S2}
\end{table}

\begin{table}[ht!]
    \centering
    \begin{tabular}{lcccc|lcccc}
    \toprule
        HSM Tier & 1 & 2 & 3 &4 & HSM Tier & 1 & 2 & 3 &4 \\
        \midrule
        BH - \texttt{Non Significant}  & 0  & 0 & 43 & 96 & HS - \texttt{Irrelevant}  & 0  & 0 & 47 & 96 \\
        BH - \texttt{Significant} & 25 & 68 & 49 & 0 & HS - \texttt{Relevant} & 25 & 68 & 45 & 0 \\
        \midrule
        lFDR - \texttt{Irrelevant}   & 0 & 55 & 92 & 96 & SnS - \texttt{Irrelevant}   & 16 & 68 & 92 & 96\\
        lFDR - \texttt{Relevant}  & 25 & 13 &0 &0 & SnS - \texttt{Relevant}  & 9 & 0 &0 &0\\
        
    \bottomrule
    \end{tabular}
    \caption{Contingency matrix comparing the findings in the light sheet fluorescence microscopy dataset obtained with HS, BH, SnS, and lFDR methods versus vs. the HSM allocation in tiers of relevance.}
    \label{tab:frequency}
\end{table}
\clearpage

\FloatBarrier
\section{Complete list of findings}

\begin{table}[h!]
\centering
\resizebox{\linewidth}{!}{
\begin{tabular}[t]{lccccc}
\toprule
Brain region & HS & BH & lFDR & SnS &HSM\\
\midrule
\cellcolor{gray!6}{Agranular insular area, ventral part, layer 1} & \cellcolor{gray!6}{\cmark} & \cellcolor{gray!6}{\cmark} & \cellcolor{gray!6}{\cmark} & \cellcolor{gray!6}{$\cdot$} & \cellcolor{gray!6}{Tier 1}\\
Anterior cingulate area, dorsal part, layer 1 &  \cmark &  \cmark &  \cmark & $\cdot$ & Tier 1\\
\cellcolor{gray!6}{Anterior cingulate area, dorsal part, layer 5} & \cellcolor{gray!6}{\cmark} & \cellcolor{gray!6}{\cmark} & \cellcolor{gray!6}{\cmark} & \cellcolor{gray!6}{$\cdot$} & \cellcolor{gray!6}{Tier 1}\\
Anterolateral visual area, layer 1 &  \cmark &  \cmark &  \cmark & $\cdot$ & Tier 1\\
\cellcolor{gray!6}{Dorsal auditory area, layer 1} & \cellcolor{gray!6}{\cmark} & \cellcolor{gray!6}{\cmark} & \cellcolor{gray!6}{\cmark} & \cellcolor{gray!6}{$\cdot$} & \cellcolor{gray!6}{Tier 1}\\
Dorsal auditory area, layer 2/3 &  \cmark &  \cmark &  \cmark &  \cmark & Tier 1\\
\cellcolor{gray!6}{Dorsal auditory area, layer 5} & \cellcolor{gray!6}{\cmark} & \cellcolor{gray!6}{\cmark} & \cellcolor{gray!6}{\cmark} & \cellcolor{gray!6}{$\cdot$} & \cellcolor{gray!6}{Tier 1}\\
Dorsal auditory area, layer 6a &  \cmark &  \cmark &  \cmark & $\cdot$ & Tier 1\\
\cellcolor{gray!6}{Ectorhinal area, layer 2/3} & \cellcolor{gray!6}{\cmark} & \cellcolor{gray!6}{\cmark} & \cellcolor{gray!6}{\cmark} & \cellcolor{gray!6}{\cmark} & \cellcolor{gray!6}{Tier 1}\\
Lateral visual area, layer 1 &  \cmark &  \cmark &  \cmark &  \cmark & Tier 1\\
\cellcolor{gray!6}{Lateral visual area, layer 5} & \cellcolor{gray!6}{\cmark} & \cellcolor{gray!6}{\cmark} & \cellcolor{gray!6}{\cmark} & \cellcolor{gray!6}{\cmark} & \cellcolor{gray!6}{Tier 1}\\
Posteromedial visual area, layer 4 &  \cmark &  \cmark &  \cmark & $\cdot$ & Tier 1\\
\cellcolor{gray!6}{Posteromedial visual area, layer 5} & \cellcolor{gray!6}{\cmark} & \cellcolor{gray!6}{\cmark} & \cellcolor{gray!6}{\cmark} & \cellcolor{gray!6}{\cmark} & \cellcolor{gray!6}{Tier 1}\\
Postpiriform transition area &  \cmark &  \cmark &  \cmark & $\cdot$ & Tier 1\\
\cellcolor{gray!6}{Primary auditory area, layer 2/3} & \cellcolor{gray!6}{\cmark} & \cellcolor{gray!6}{\cmark} & \cellcolor{gray!6}{\cmark} & \cellcolor{gray!6}{$\cdot$} & \cellcolor{gray!6}{Tier 1}\\
Primary auditory area, layer 5 &  \cmark &  \cmark &  \cmark & $\cdot$ & Tier 1\\
\cellcolor{gray!6}{Primary visual area, layer 1} & \cellcolor{gray!6}{\cmark} & \cellcolor{gray!6}{\cmark} & \cellcolor{gray!6}{\cmark} & \cellcolor{gray!6}{\cmark} & \cellcolor{gray!6}{Tier 1}\\
Primary visual area, layer 4 &  \cmark &  \cmark &  \cmark &  \cmark & Tier 1\\
\cellcolor{gray!6}{Primary visual area, layer 5} & \cellcolor{gray!6}{\cmark} & \cellcolor{gray!6}{\cmark} & \cellcolor{gray!6}{\cmark} & \cellcolor{gray!6}{\cmark} & \cellcolor{gray!6}{Tier 1}\\
Primary visual area, layer 6a &  \cmark &  \cmark &  \cmark &  \cmark & Tier 1\\
\cellcolor{gray!6}{Subiculum} & \cellcolor{gray!6}{\cmark} & \cellcolor{gray!6}{\cmark} & \cellcolor{gray!6}{\cmark} & \cellcolor{gray!6}{$\cdot$} & \cellcolor{gray!6}{Tier 1}\\
Taenia tecta &  \cmark &  \cmark &  \cmark & $\cdot$ & Tier 1\\
\cellcolor{gray!6}{Temporal association areas, layer 2/3} & \cellcolor{gray!6}{\cmark} & \cellcolor{gray!6}{\cmark} & \cellcolor{gray!6}{\cmark} & \cellcolor{gray!6}{$\cdot$} & \cellcolor{gray!6}{Tier 1}\\
Temporal association areas, layer 6a &  \cmark &  \cmark &  \cmark & $\cdot$ & Tier 1\\
\cellcolor{gray!6}{Ventral auditory area, layer 2/3} & \cellcolor{gray!6}{\cmark} & \cellcolor{gray!6}{\cmark} & \cellcolor{gray!6}{\cmark} & \cellcolor{gray!6}{$\cdot$} & \cellcolor{gray!6}{Tier 1}\\
\bottomrule
\end{tabular}}
\end{table}
\clearpage

\begin{table}
\centering
\resizebox{\linewidth}{!}{
\begin{tabular}[t]{lccccc}
\toprule
Brain region & HS & BH & lFDR & SnS & HSM\\
\midrule
Agranular insular area, dorsal part, layer 1 & \cmark & \cmark & $\cdot$ & $\cdot$ & Tier 2\\
\cellcolor{gray!6}{Anterior cingulate area, dorsal part, layer 6a} & \cellcolor{gray!6}{\cmark} & \cellcolor{gray!6}{\cmark} & \cellcolor{gray!6}{\cmark} & \cellcolor{gray!6}{$\cdot$} & \cellcolor{gray!6}{Tier 2}\\
Anterior cingulate area, ventral part, layer 1 & \cmark & \cmark & $\cdot$ & $\cdot$ & Tier 2\\
\cellcolor{gray!6}{Anterior cingulate area, ventral part, layer 2/3} & \cellcolor{gray!6}{\cmark} & \cellcolor{gray!6}{\cmark} & \cellcolor{gray!6}{$\cdot$} & \cellcolor{gray!6}{$\cdot$} & \cellcolor{gray!6}{Tier 2}\\
Anterior cingulate area, ventral part, layer 5 & \cmark & \cmark & $\cdot$ & $\cdot$ & Tier 2\\
 
\cellcolor{gray!6}{Anterior olfactory nucleus} & \cellcolor{gray!6}{\cmark} & \cellcolor{gray!6}{\cmark} & \cellcolor{gray!6}{$\cdot$} & \cellcolor{gray!6}{$\cdot$} & \cellcolor{gray!6}{Tier 2}\\
Anteromedial visual area, layer 1 & \cmark & \cmark & $\cdot$ & $\cdot$ & Tier 2\\
\cellcolor{gray!6}{Central amygdalar nucleus} & \cellcolor{gray!6}{\cmark} & \cellcolor{gray!6}{\cmark} & \cellcolor{gray!6}{$\cdot$} & \cellcolor{gray!6}{$\cdot$} & \cellcolor{gray!6}{Tier 2}\\
Ectorhinal area/Layer 1 & \cmark & \cmark & $\cdot$ & $\cdot$ & Tier 2\\
\cellcolor{gray!6}{Ectorhinal area/Layer 5} & \cellcolor{gray!6}{\cmark} & \cellcolor{gray!6}{\cmark} & \cellcolor{gray!6}{$\cdot$} & \cellcolor{gray!6}{$\cdot$} & \cellcolor{gray!6}{Tier 2}\\
 
Ectorhinal area/Layer 6a & \cmark & \cmark & $\cdot$ & $\cdot$ & Tier 2\\
\cellcolor{gray!6}{fiber tracts} & \cellcolor{gray!6}{\cmark} & \cellcolor{gray!6}{\cmark} & \cellcolor{gray!6}{$\cdot$} & \cellcolor{gray!6}{$\cdot$} & \cellcolor{gray!6}{Tier 2}\\
Hippocampal formation & \cmark & \cmark & $\cdot$ & $\cdot$ & Tier 2\\
\cellcolor{gray!6}{Infralimbic area, layer 1} & \cellcolor{gray!6}{\cmark} & \cellcolor{gray!6}{\cmark} & \cellcolor{gray!6}{$\cdot$} & \cellcolor{gray!6}{$\cdot$} & \cellcolor{gray!6}{Tier 2}\\
Infralimbic area, layer 5 & \cmark & \cmark & $\cdot$ & $\cdot$ & Tier 2\\
 
\cellcolor{gray!6}{Main olfactory bulb} & \cellcolor{gray!6}{\cmark} & \cellcolor{gray!6}{\cmark} & \cellcolor{gray!6}{\cmark} & \cellcolor{gray!6}{$\cdot$} & \cellcolor{gray!6}{Tier 2}\\
Nucleus accumbens & \cmark & \cmark & \cmark & $\cdot$ & Tier 2\\
\cellcolor{gray!6}{Orbital area, medial part, layer 1} & \cellcolor{gray!6}{\cmark} & \cellcolor{gray!6}{\cmark} & \cellcolor{gray!6}{$\cdot$} & \cellcolor{gray!6}{$\cdot$} & \cellcolor{gray!6}{Tier 2}\\
Orbital area, medial part, layer 5 & \cmark & \cmark & $\cdot$ & $\cdot$ & Tier 2\\
\cellcolor{gray!6}{Orbital area, ventrolateral part, layer 1} & \cellcolor{gray!6}{\cmark} & \cellcolor{gray!6}{\cmark} & \cellcolor{gray!6}{$\cdot$} & \cellcolor{gray!6}{$\cdot$} & \cellcolor{gray!6}{Tier 2}\\
 
Parasubiculum & \cmark & \cmark & $\cdot$ & $\cdot$ & Tier 2\\
\cellcolor{gray!6}{Perirhinal area, layer 1} & \cellcolor{gray!6}{\cmark} & \cellcolor{gray!6}{\cmark} & \cellcolor{gray!6}{$\cdot$} & \cellcolor{gray!6}{$\cdot$} & \cellcolor{gray!6}{Tier 2}\\
Perirhinal area, layer 2/3 & \cmark & \cmark & $\cdot$ & $\cdot$ & Tier 2\\
\cellcolor{gray!6}{Perirhinal area, layer 5} & \cellcolor{gray!6}{\cmark} & \cellcolor{gray!6}{\cmark} & \cellcolor{gray!6}{\cmark} & \cellcolor{gray!6}{$\cdot$} & \cellcolor{gray!6}{Tier 2}\\
Perirhinal area, layer 6a & \cmark & \cmark & $\cdot$ & $\cdot$ & Tier 2\\
 
\cellcolor{gray!6}{Piriform area} & \cellcolor{gray!6}{\cmark} & \cellcolor{gray!6}{\cmark} & \cellcolor{gray!6}{$\cdot$} & \cellcolor{gray!6}{$\cdot$} & \cellcolor{gray!6}{Tier 2}\\
Piriform-amygdalar area & \cmark & \cmark & \cmark & $\cdot$ & Tier 2\\
\cellcolor{gray!6}{Posterior auditory area, layer 2/3} & \cellcolor{gray!6}{\cmark} & \cellcolor{gray!6}{\cmark} & \cellcolor{gray!6}{$\cdot$} & \cellcolor{gray!6}{$\cdot$} & \cellcolor{gray!6}{Tier 2}\\
Posterolateral visual area, layer 2/3 & \cmark & \cmark & $\cdot$ & $\cdot$ & Tier 2\\
\cellcolor{gray!6}{Posterolateral visual area, layer 5} & \cellcolor{gray!6}{\cmark} & \cellcolor{gray!6}{\cmark} & \cellcolor{gray!6}{$\cdot$} & \cellcolor{gray!6}{$\cdot$} & \cellcolor{gray!6}{Tier 2}\\
 
Posteromedial visual area, layer 1 & \cmark & \cmark & $\cdot$ & $\cdot$ & Tier 2\\
\cellcolor{gray!6}{Postsubiculum} & \cellcolor{gray!6}{\cmark} & \cellcolor{gray!6}{\cmark} & \cellcolor{gray!6}{$\cdot$} & \cellcolor{gray!6}{$\cdot$} & \cellcolor{gray!6}{Tier 2}\\
Prelimbic area, layer 1 & \cmark & \cmark & $\cdot$ & $\cdot$ & Tier 2\\
\cellcolor{gray!6}{Prelimbic area, layer 5} & \cellcolor{gray!6}{\cmark} & \cellcolor{gray!6}{\cmark} & \cellcolor{gray!6}{$\cdot$} & \cellcolor{gray!6}{$\cdot$} & \cellcolor{gray!6}{Tier 2}\\
Primary auditory area, layer 1 & \cmark & \cmark & $\cdot$ & $\cdot$ & Tier 2\\
 
\cellcolor{gray!6}{Primary auditory area, layer 6a} & \cellcolor{gray!6}{\cmark} & \cellcolor{gray!6}{\cmark} & \cellcolor{gray!6}{$\cdot$} & \cellcolor{gray!6}{$\cdot$} & \cellcolor{gray!6}{Tier 2}\\
Primary motor area, Layer 1 & \cmark & \cmark & $\cdot$ & $\cdot$ & Tier 2\\
\cellcolor{gray!6}{Primary motor area, Layer 5} & \cellcolor{gray!6}{\cmark} & \cellcolor{gray!6}{\cmark} & \cellcolor{gray!6}{$\cdot$} & \cellcolor{gray!6}{$\cdot$} & \cellcolor{gray!6}{Tier 2}\\
Primary somatosensory area, barrel field, layer 1 & \cmark & \cmark & $\cdot$ & $\cdot$ & Tier 2\\
\cellcolor{gray!6}{Primary somatosensory area, barrel field, layer 2/3} & \cellcolor{gray!6}{\cmark} & \cellcolor{gray!6}{\cmark} & \cellcolor{gray!6}{\cmark} & \cellcolor{gray!6}{$\cdot$} & \cellcolor{gray!6}{Tier 2}\\
 
Primary somatosensory area, barrel field, layer 5  & \cmark & \cmark & $\cdot$ & $\cdot$ & Tier 2\\
\cellcolor{gray!6}{Primary somatosensory area, barrel field, layer 6a} & \cellcolor{gray!6}{\cmark} & \cellcolor{gray!6}{\cmark} & \cellcolor{gray!6}{$\cdot$} & \cellcolor{gray!6}{$\cdot$} & \cellcolor{gray!6}{Tier 2}\\
Primary somatosensory area, lower limb, layer 2/3 & \cmark & \cmark & $\cdot$ & $\cdot$ & Tier 2\\
\cellcolor{gray!6}{Primary somatosensory area, lower limb, layer 5} & \cellcolor{gray!6}{\cmark} & \cellcolor{gray!6}{\cmark} & \cellcolor{gray!6}{\cmark} & \cellcolor{gray!6}{$\cdot$} & \cellcolor{gray!6}{Tier 2}\\
Primary somatosensory area, mouth, layer 2/3 & \cmark & \cmark & $\cdot$ & $\cdot$ & Tier 2\\
\bottomrule
\end{tabular}}
\end{table}

\begin{table}[h!]
\centering
\resizebox{\linewidth}{!}{
\begin{tabular}[t]{lccccc}
\toprule
Brain region & HS & BH & lFDR & SnS & HSM\\
\midrule
\cellcolor{gray!6}{Primary somatosensory area, nose, layer 2/3} & \cellcolor{gray!6}{\cmark} & \cellcolor{gray!6}{\cmark} & \cellcolor{gray!6}{$\cdot$} & \cellcolor{gray!6}{$\cdot$} & \cellcolor{gray!6}{Tier 2}\\
Primary somatosensory area, trunk, layer 2/3 & \cmark & \cmark & $\cdot$ & $\cdot$ & Tier 2\\
\cellcolor{gray!6}{Primary somatosensory area, trunk, layer 4} & \cellcolor{gray!6}{\cmark} & \cellcolor{gray!6}{\cmark} & \cellcolor{gray!6}{$\cdot$} & \cellcolor{gray!6}{$\cdot$} & \cellcolor{gray!6}{Tier 2}\\
Primary somatosensory area, trunk, layer 5 & \cmark & \cmark & $\cdot$ & $\cdot$ & Tier 2\\
\cellcolor{gray!6}{Primary somatosensory area, upper limb, layer 1} & \cellcolor{gray!6}{\cmark} & \cellcolor{gray!6}{\cmark} & \cellcolor{gray!6}{$\cdot$} & \cellcolor{gray!6}{$\cdot$} & \cellcolor{gray!6}{Tier 2}\\
 
Primary somatosensory area, upper limb, layer 2/3 & \cmark & \cmark & $\cdot$ & $\cdot$ & Tier 2\\
\cellcolor{gray!6}{Primary visual area, layer 2/3} & \cellcolor{gray!6}{\cmark} & \cellcolor{gray!6}{\cmark} & \cellcolor{gray!6}{$\cdot$} & \cellcolor{gray!6}{$\cdot$} & \cellcolor{gray!6}{Tier 2}\\
Retrosplenial area, lateral agranular part, layer 1 & \cmark & \cmark & $\cdot$ & $\cdot$ & Tier 2\\
\cellcolor{gray!6}{Retrosplenial area, lateral agranular part, layer 5} & \cellcolor{gray!6}{\cmark} & \cellcolor{gray!6}{\cmark} & \cellcolor{gray!6}{\cmark} & \cellcolor{gray!6}{$\cdot$} & \cellcolor{gray!6}{Tier 2}\\
Retrosplenial area, ventral part, layer 1 & \cmark & \cmark & $\cdot$ & $\cdot$ & Tier 2\\
 
\cellcolor{gray!6}{Retrosplenial area, ventral part, layer 6a} & \cellcolor{gray!6}{\cmark} & \cellcolor{gray!6}{\cmark} & \cellcolor{gray!6}{$\cdot$} & \cellcolor{gray!6}{$\cdot$} & \cellcolor{gray!6}{Tier 2}\\
Secondary motor area, layer 2/3 & \cmark & \cmark & $\cdot$ & $\cdot$ & Tier 2\\
\cellcolor{gray!6}{Secondary motor area, layer 5} & \cellcolor{gray!6}{\cmark} & \cellcolor{gray!6}{\cmark} & \cellcolor{gray!6}{$\cdot$} & \cellcolor{gray!6}{$\cdot$} & \cellcolor{gray!6}{Tier 2}\\
Secondary motor area, layer 6a & \cmark & \cmark & \cmark & $\cdot$ & Tier 2\\
\cellcolor{gray!6}{Supplemental somatosensory area, layer 1} & \cellcolor{gray!6}{\cmark} & \cellcolor{gray!6}{\cmark} & \cellcolor{gray!6}{$\cdot$} & \cellcolor{gray!6}{$\cdot$} & \cellcolor{gray!6}{Tier 2}\\
 
Supplemental somatosensory area, layer 5 & \cmark & \cmark & $\cdot$ & $\cdot$ & Tier 2\\
\cellcolor{gray!6}{Supplemental somatosensory area, layer 6a} & \cellcolor{gray!6}{\cmark} & \cellcolor{gray!6}{\cmark} & \cellcolor{gray!6}{\cmark} & \cellcolor{gray!6}{$\cdot$} & \cellcolor{gray!6}{Tier 2}\\
Temporal association areas, layer 1 & \cmark & \cmark & $\cdot$ & $\cdot$ & Tier 2\\
\cellcolor{gray!6}{Temporal association areas, layer 5} & \cellcolor{gray!6}{\cmark} & \cellcolor{gray!6}{\cmark} & \cellcolor{gray!6}{\cmark} & \cellcolor{gray!6}{$\cdot$} & \cellcolor{gray!6}{Tier 2}\\
third ventricle & \cmark & \cmark & $\cdot$ & $\cdot$ & Tier 2\\
 
\cellcolor{gray!6}{Unlabeled} & \cellcolor{gray!6}{\cmark} & \cellcolor{gray!6}{\cmark} & \cellcolor{gray!6}{$\cdot$} & \cellcolor{gray!6}{$\cdot$} & \cellcolor{gray!6}{Tier 2}\\
Ventral auditory area, layer 5 & \cmark & \cmark & \cmark & $\cdot$ & Tier 2\\
\cellcolor{gray!6}{Ventral auditory area, layer 6a} & \cellcolor{gray!6}{\cmark} & \cellcolor{gray!6}{\cmark} & \cellcolor{gray!6}{\cmark} & \cellcolor{gray!6}{$\cdot$} & \cellcolor{gray!6}{Tier 2}\\
\bottomrule
\end{tabular}}
\end{table}

\begin{table}[h!]
\centering
\resizebox{\linewidth}{!}{
\begin{tabular}[t]{lccccc}
\toprule
Brain region & HS & BH & lFDR & SnS & HSM\\
\midrule
Agranular insular area, dorsal part, layer 6a & \cmark & $\cdot$ & $\cdot$ & $\cdot$ & Tier 3\\
\cellcolor{gray!6}{Agranular insular area, posterior part, layer 1} & \cellcolor{gray!6}{\cmark} & \cellcolor{gray!6}{\cmark} & \cellcolor{gray!6}{$\cdot$} & \cellcolor{gray!6}{$\cdot$} & \cellcolor{gray!6}{Tier 3}\\
 
Agranular insular area, posterior part, layer 2/3 & $\cdot$ & $\cdot$ & $\cdot$ & $\cdot$ & Tier 3\\
\cellcolor{gray!6}{Agranular insular area, posterior part, layer 6a} & \cellcolor{gray!6}{\cmark} & \cellcolor{gray!6}{\cmark} & \cellcolor{gray!6}{$\cdot$} & \cellcolor{gray!6}{$\cdot$} & \cellcolor{gray!6}{Tier 3}\\
Agranular insular area, ventral part, layer 2/3 & \cmark & \cmark & $\cdot$ & $\cdot$ & Tier 3\\
\cellcolor{gray!6}{Agranular insular area, ventral part, layer 5} & \cellcolor{gray!6}{\cmark} & \cellcolor{gray!6}{\cmark} & \cellcolor{gray!6}{$\cdot$} & \cellcolor{gray!6}{$\cdot$} & \cellcolor{gray!6}{Tier 3}\\
Anterolateral visual area, layer 2/3 & $\cdot$ & $\cdot$ & $\cdot$ & $\cdot$ & Tier 3\\
 
\cellcolor{gray!6}{Basolateral amygdalar nucleus, ventral part} & \cellcolor{gray!6}{\cmark} & \cellcolor{gray!6}{\cmark} & \cellcolor{gray!6}{$\cdot$} & \cellcolor{gray!6}{$\cdot$} & \cellcolor{gray!6}{Tier 3}\\
Basomedial amygdalar nucleus & \cmark & \cmark & $\cdot$ & $\cdot$ & Tier 3\\
\cellcolor{gray!6}{Central lobule} & \cellcolor{gray!6}{$\cdot$} & \cellcolor{gray!6}{$\cdot$} & \cellcolor{gray!6}{$\cdot$} & \cellcolor{gray!6}{$\cdot$} & \cellcolor{gray!6}{Tier 3}\\
cerebal peduncle & \cmark & \cmark & $\cdot$ & $\cdot$ & Tier 3\\
\cellcolor{gray!6}{Claustrum} & \cellcolor{gray!6}{$\cdot$} & \cellcolor{gray!6}{$\cdot$} & \cellcolor{gray!6}{$\cdot$} & \cellcolor{gray!6}{$\cdot$} & \cellcolor{gray!6}{Tier 3}\\
 
Cortical amygdalar area, anterior part & $\cdot$ & $\cdot$ & $\cdot$ & $\cdot$ & Tier 3\\
\cellcolor{gray!6}{Cortical amygdalar area, posterior part} & \cellcolor{gray!6}{$\cdot$} & \cellcolor{gray!6}{$\cdot$} & \cellcolor{gray!6}{$\cdot$} & \cellcolor{gray!6}{$\cdot$} & \cellcolor{gray!6}{Tier 3}\\
Culmen & $\cdot$ & $\cdot$ & $\cdot$ & $\cdot$ & Tier 3\\
\cellcolor{gray!6}{Dentate gyrus, granule cell layer} & \cellcolor{gray!6}{\cmark} & \cellcolor{gray!6}{$\cdot$} & \cellcolor{gray!6}{$\cdot$} & \cellcolor{gray!6}{$\cdot$} & \cellcolor{gray!6}{Tier 3}\\
Dentate gyrus, polymorph layer & $\cdot$ & $\cdot$ & $\cdot$ & $\cdot$ & Tier 3\\
 
\cellcolor{gray!6}{Dorsal auditory area, layer 4} & \cellcolor{gray!6}{\cmark} & \cellcolor{gray!6}{\cmark} & \cellcolor{gray!6}{$\cdot$} & \cellcolor{gray!6}{$\cdot$} & \cellcolor{gray!6}{Tier 3}\\
Dorsal cochlear nucleus & $\cdot$ & \cmark & $\cdot$ & $\cdot$ & Tier 3\\
\cellcolor{gray!6}{Entorhinal area, medial part, dorsal zone} & \cellcolor{gray!6}{\cmark} & \cellcolor{gray!6}{$\cdot$} & \cellcolor{gray!6}{$\cdot$} & \cellcolor{gray!6}{$\cdot$} & \cellcolor{gray!6}{Tier 3}\\
Entorhinal area, medial part, ventral zone & $\cdot$ & \cmark & $\cdot$ & $\cdot$ & Tier 3\\
\cellcolor{gray!6}{Field CA1} & \cellcolor{gray!6}{$\cdot$} & \cellcolor{gray!6}{$\cdot$} & \cellcolor{gray!6}{$\cdot$} & \cellcolor{gray!6}{$\cdot$} & \cellcolor{gray!6}{Tier 3}\\
 
Field CA2 & $\cdot$ & \cmark & $\cdot$ & $\cdot$ & Tier 3\\
\cellcolor{gray!6}{Flocculus} & \cellcolor{gray!6}{$\cdot$} & \cellcolor{gray!6}{$\cdot$} & \cellcolor{gray!6}{$\cdot$} & \cellcolor{gray!6}{$\cdot$} & \cellcolor{gray!6}{Tier 3}\\
Folium-tuber vermis (VII) & \cmark & \cmark & $\cdot$ & $\cdot$ & Tier 3\\
\cellcolor{gray!6}{Frontal pole, layer 1} & \cellcolor{gray!6}{$\cdot$} & \cellcolor{gray!6}{$\cdot$} & \cellcolor{gray!6}{$\cdot$} & \cellcolor{gray!6}{$\cdot$} & \cellcolor{gray!6}{Tier 3}\\
Frontal pole, layer 2/3 & $\cdot$ & $\cdot$ & $\cdot$ & $\cdot$ & Tier 3\\
 
\cellcolor{gray!6}{Gustatory areas, layer 1} & \cellcolor{gray!6}{\cmark} & \cellcolor{gray!6}{$\cdot$} & \cellcolor{gray!6}{$\cdot$} & \cellcolor{gray!6}{$\cdot$} & \cellcolor{gray!6}{Tier 3}\\
Gustatory areas, layer 6a & $\cdot$ & \cmark & $\cdot$ & $\cdot$ & Tier 3\\
\cellcolor{gray!6}{Hypothalamus} & \cellcolor{gray!6}{$\cdot$} & \cellcolor{gray!6}{$\cdot$} & \cellcolor{gray!6}{$\cdot$} & \cellcolor{gray!6}{$\cdot$} & \cellcolor{gray!6}{Tier 3}\\
inferior cerebellar peduncle & $\cdot$ & $\cdot$ & $\cdot$ & $\cdot$ & Tier 3\\
\cellcolor{gray!6}{Intermediate reticular nucleus} & \cellcolor{gray!6}{$\cdot$} & \cellcolor{gray!6}{\cmark} & \cellcolor{gray!6}{$\cdot$} & \cellcolor{gray!6}{$\cdot$} & \cellcolor{gray!6}{Tier 3}\\
 
Lateral amygdalar nucleus & $\cdot$ & $\cdot$ & $\cdot$ & $\cdot$ & Tier 3\\
\cellcolor{gray!6}{lateral recess} & \cellcolor{gray!6}{$\cdot$} & \cellcolor{gray!6}{$\cdot$} & \cellcolor{gray!6}{$\cdot$} & \cellcolor{gray!6}{$\cdot$} & \cellcolor{gray!6}{Tier 3}\\
Lateral reticular nucleus & $\cdot$ & $\cdot$ & $\cdot$ & $\cdot$ & Tier 3\\
\cellcolor{gray!6}{lateral ventricle} & \cellcolor{gray!6}{$\cdot$} & \cellcolor{gray!6}{$\cdot$} & \cellcolor{gray!6}{$\cdot$} & \cellcolor{gray!6}{$\cdot$} & \cellcolor{gray!6}{Tier 3}\\
Lateral visual area, layer 2/3 & $\cdot$ & $\cdot$ & $\cdot$ & $\cdot$ & Tier 3\\
 
\cellcolor{gray!6}{Magnocellular reticular nucleus} & \cellcolor{gray!6}{\cmark} & \cellcolor{gray!6}{\cmark} & \cellcolor{gray!6}{$\cdot$} & \cellcolor{gray!6}{$\cdot$} & \cellcolor{gray!6}{Tier 3}\\
Medial amygdalar nucleus & \cmark & $\cdot$ & $\cdot$ & $\cdot$ & Tier 3\\
\cellcolor{gray!6}{Medial vestibular nucleus} & \cellcolor{gray!6}{\cmark} & \cellcolor{gray!6}{$\cdot$} & \cellcolor{gray!6}{$\cdot$} & \cellcolor{gray!6}{$\cdot$} & \cellcolor{gray!6}{Tier 3}\\
Nodulus (X) & $\cdot$ & \cmark & $\cdot$ & $\cdot$ & Tier 3\\
\cellcolor{gray!6}{olfactory nerve layer of main olfactory bulb} & \cellcolor{gray!6}{\cmark} & \cellcolor{gray!6}{\cmark} & \cellcolor{gray!6}{$\cdot$} & \cellcolor{gray!6}{$\cdot$} & \cellcolor{gray!6}{Tier 3}\\
 
Olfactory tubercle & \cmark & \cmark & $\cdot$ & $\cdot$ & Tier 3\\
\cellcolor{gray!6}{optic tract} & \cellcolor{gray!6}{$\cdot$} & \cellcolor{gray!6}{$\cdot$} & \cellcolor{gray!6}{$\cdot$} & \cellcolor{gray!6}{$\cdot$} & \cellcolor{gray!6}{Tier 3}\\
Orbital area, lateral part, layer 1 & \cmark & $\cdot$ & $\cdot$ & $\cdot$ & Tier 3\\
\cellcolor{gray!6}{Orbital area, lateral part, layer 5} & \cellcolor{gray!6}{$\cdot$} & \cellcolor{gray!6}{\cmark} & \cellcolor{gray!6}{$\cdot$} & \cellcolor{gray!6}{$\cdot$} & \cellcolor{gray!6}{Tier 3}\\
Orbital area, medial part, layer 2/3 & $\cdot$ & $\cdot$ & $\cdot$ & $\cdot$ & Tier 3\\
 
\bottomrule
\end{tabular}}
\end{table}

\begin{table}[h!]
\centering
\resizebox{\linewidth}{!}{
\begin{tabular}[t]{lccccc}
\toprule
Brain region & HS & BH & lFDR & SnS & HSM\\
\midrule

\cellcolor{gray!6}{Orbital area, ventrolateral part, layer 2/3} & \cellcolor{gray!6}{$\cdot$} & \cellcolor{gray!6}{$\cdot$} & \cellcolor{gray!6}{$\cdot$} & \cellcolor{gray!6}{$\cdot$} & \cellcolor{gray!6}{Tier 3}\\
Pallidum & \cmark & \cmark & $\cdot$ & $\cdot$ & Tier 3\\
\cellcolor{gray!6}{Parabrachial nucleus} & \cellcolor{gray!6}{\cmark} & \cellcolor{gray!6}{\cmark} & \cellcolor{gray!6}{$\cdot$} & \cellcolor{gray!6}{$\cdot$} & \cellcolor{gray!6}{Tier 3}\\
Paragigantocellular reticular nucleus, dorsal part & $\cdot$ & $\cdot$ & $\cdot$ & $\cdot$ & Tier 3\\
\cellcolor{gray!6}{Paragigantocellular reticular nucleus, lateral part} & \cellcolor{gray!6}{\cmark} & \cellcolor{gray!6}{\cmark} & \cellcolor{gray!6}{$\cdot$} & \cellcolor{gray!6}{$\cdot$} & \cellcolor{gray!6}{Tier 3}\\
 
Paramedian lobule & $\cdot$ & $\cdot$ & $\cdot$ & $\cdot$ & Tier 3\\
\cellcolor{gray!6}{Posterior hypothalamic nucleus} & \cellcolor{gray!6}{$\cdot$} & \cellcolor{gray!6}{$\cdot$} & \cellcolor{gray!6}{$\cdot$} & \cellcolor{gray!6}{$\cdot$} & \cellcolor{gray!6}{Tier 3}\\
Posterolateral visual area, layer 1 & \cmark & \cmark & $\cdot$ & $\cdot$ & Tier 3\\
\cellcolor{gray!6}{Posterolateral visual area, layer 4} & \cellcolor{gray!6}{\cmark} & \cellcolor{gray!6}{\cmark} & \cellcolor{gray!6}{$\cdot$} & \cellcolor{gray!6}{$\cdot$} & \cellcolor{gray!6}{Tier 3}\\
Posteromedial visual area, layer 2/3 & $\cdot$ & $\cdot$ & $\cdot$ & $\cdot$ & Tier 3\\
 
\cellcolor{gray!6}{Prelimbic area, layer 2/3} & \cellcolor{gray!6}{$\cdot$} & \cellcolor{gray!6}{\cmark} & \cellcolor{gray!6}{$\cdot$} & \cellcolor{gray!6}{$\cdot$} & \cellcolor{gray!6}{Tier 3}\\
Prelimbic area, layer 6a & \cmark & \cmark & $\cdot$ & $\cdot$ & Tier 3\\
\cellcolor{gray!6}{Presubiculum} & \cellcolor{gray!6}{\cmark} & \cellcolor{gray!6}{\cmark} & \cellcolor{gray!6}{$\cdot$} & \cellcolor{gray!6}{$\cdot$} & \cellcolor{gray!6}{Tier 3}\\
Primary auditory area, layer 4 & \cmark & \cmark & $\cdot$ & $\cdot$ & Tier 3\\
\cellcolor{gray!6}{Primary motor area, Layer 2/3} & \cellcolor{gray!6}{\cmark} & \cellcolor{gray!6}{\cmark} & \cellcolor{gray!6}{$\cdot$} & \cellcolor{gray!6}{$\cdot$} & \cellcolor{gray!6}{Tier 3}\\
 
Primary motor area, Layer 6a & \cmark & \cmark & $\cdot$ & $\cdot$ & Tier 3\\
\cellcolor{gray!6}{Primary motor area, Layer 6b} & \cellcolor{gray!6}{$\cdot$} & \cellcolor{gray!6}{$\cdot$} & \cellcolor{gray!6}{$\cdot$} & \cellcolor{gray!6}{$\cdot$} & \cellcolor{gray!6}{Tier 3}\\
Primary somatosensory area, barrel field, layer 4 & \cmark & \cmark & $\cdot$ & $\cdot$ & Tier 3\\
\cellcolor{gray!6}{Primary somatosensory area, lower limb, layer 1} & \cellcolor{gray!6}{$\cdot$} & \cellcolor{gray!6}{$\cdot$} & \cellcolor{gray!6}{$\cdot$} & \cellcolor{gray!6}{$\cdot$} & \cellcolor{gray!6}{Tier 3}\\
Primary somatosensory area, lower limb, layer 4 & \cmark & \cmark & $\cdot$ & $\cdot$ & Tier 3\\
 
\cellcolor{gray!6}{Primary somatosensory area, mouth, layer 4} & \cellcolor{gray!6}{$\cdot$} & \cellcolor{gray!6}{$\cdot$} & \cellcolor{gray!6}{$\cdot$} & \cellcolor{gray!6}{$\cdot$} & \cellcolor{gray!6}{Tier 3}\\
Primary somatosensory area, mouth, layer 6a & $\cdot$ & $\cdot$ & $\cdot$ & $\cdot$ & Tier 3\\
\cellcolor{gray!6}{Primary somatosensory area, nose, layer 1} & \cellcolor{gray!6}{$\cdot$} & \cellcolor{gray!6}{$\cdot$} & \cellcolor{gray!6}{$\cdot$} & \cellcolor{gray!6}{$\cdot$} & \cellcolor{gray!6}{Tier 3}\\
Primary somatosensory area, nose, layer 5 & $\cdot$ & \cmark & $\cdot$ & $\cdot$ & Tier 3\\
\cellcolor{gray!6}{Primary somatosensory area, trunk, layer 1} & \cellcolor{gray!6}{\cmark} & \cellcolor{gray!6}{\cmark} & \cellcolor{gray!6}{$\cdot$} & \cellcolor{gray!6}{$\cdot$} & \cellcolor{gray!6}{Tier 3}\\
 
Primary somatosensory area, upper limb, layer 4 & $\cdot$ & $\cdot$ & $\cdot$ & $\cdot$ & Tier 3\\
\cellcolor{gray!6}{Primary somatosensory area, upper limb, layer 5} & \cellcolor{gray!6}{$\cdot$} & \cellcolor{gray!6}{\cmark} & \cellcolor{gray!6}{$\cdot$} & \cellcolor{gray!6}{$\cdot$} & \cellcolor{gray!6}{Tier 3}\\
Primary somatosensory area, upper limb, layer 6a & $\cdot$ & $\cdot$ & $\cdot$ & $\cdot$ & Tier 3\\
\cellcolor{gray!6}{Retrosplenial area, dorsal part, layer 1} & \cellcolor{gray!6}{\cmark} & \cellcolor{gray!6}{\cmark} & \cellcolor{gray!6}{$\cdot$} & \cellcolor{gray!6}{$\cdot$} & \cellcolor{gray!6}{Tier 3}\\
Retrosplenial area, dorsal part, layer 2/3 & \cmark & \cmark & $\cdot$ & $\cdot$ & Tier 3\\
 
\cellcolor{gray!6}{Retrosplenial area, dorsal part, layer 6a} & \cellcolor{gray!6}{\cmark} & \cellcolor{gray!6}{\cmark} & \cellcolor{gray!6}{$\cdot$} & \cellcolor{gray!6}{$\cdot$} & \cellcolor{gray!6}{Tier 3}\\
Retrosplenial area, ventral part, layer 2/3 & \cmark & \cmark & $\cdot$ & $\cdot$ & Tier 3\\
\cellcolor{gray!6}{Retrosplenial area, ventral part, layer 5} & \cellcolor{gray!6}{\cmark} & \cellcolor{gray!6}{\cmark} & \cellcolor{gray!6}{$\cdot$} & \cellcolor{gray!6}{$\cdot$} & \cellcolor{gray!6}{Tier 3}\\
Secondary motor area, layer 1 & \cmark & \cmark & $\cdot$ & $\cdot$ & Tier 3\\
\cellcolor{gray!6}{Simple lobule} & \cellcolor{gray!6}{\cmark} & \cellcolor{gray!6}{\cmark} & \cellcolor{gray!6}{$\cdot$} & \cellcolor{gray!6}{$\cdot$} & \cellcolor{gray!6}{Tier 3}\\
 
Spinal vestibular nucleus & \cmark & \cmark & $\cdot$ & $\cdot$ & Tier 3\\
\cellcolor{gray!6}{Substantia nigra, reticular part} & \cellcolor{gray!6}{$\cdot$} & \cellcolor{gray!6}{$\cdot$} & \cellcolor{gray!6}{$\cdot$} & \cellcolor{gray!6}{$\cdot$} & \cellcolor{gray!6}{Tier 3}\\
Superior colliculus, optic layer & \cmark & \cmark & $\cdot$ & $\cdot$ & Tier 3\\
\cellcolor{gray!6}{Supplemental somatosensory area, layer 2/3} & \cellcolor{gray!6}{\cmark} & \cellcolor{gray!6}{\cmark} & \cellcolor{gray!6}{$\cdot$} & \cellcolor{gray!6}{$\cdot$} & \cellcolor{gray!6}{Tier 3}\\
Supplemental somatosensory area, layer 4 & $\cdot$ & \cmark & $\cdot$ & $\cdot$ & Tier 3\\
 
\cellcolor{gray!6}{Thalamus} & \cellcolor{gray!6}{$\cdot$} & \cellcolor{gray!6}{$\cdot$} & \cellcolor{gray!6}{$\cdot$} & \cellcolor{gray!6}{$\cdot$} & \cellcolor{gray!6}{Tier 3}\\
Ventral auditory area, layer 1 & \cmark & \cmark & $\cdot$ & $\cdot$ & Tier 3\\
\cellcolor{gray!6}{Ventral cochlear nucleus} & \cellcolor{gray!6}{\cmark} & \cellcolor{gray!6}{\cmark} & \cellcolor{gray!6}{$\cdot$} & \cellcolor{gray!6}{$\cdot$} & \cellcolor{gray!6}{Tier 3}\\
Visceral area, layer 1 & $\cdot$ & $\cdot$ & $\cdot$ & $\cdot$ & Tier 3\\
\cellcolor{gray!6}{Visceral area, layer 2/3} & \cellcolor{gray!6}{\cmark} & \cellcolor{gray!6}{\cmark} & \cellcolor{gray!6}{$\cdot$} & \cellcolor{gray!6}{$\cdot$} & \cellcolor{gray!6}{Tier 3}\\

\bottomrule
\end{tabular}}
\end{table}

\begin{table}[h!]
\centering
\resizebox{\linewidth}{!}{
\begin{tabular}[t]{lccccc}
\toprule
Brain region & HS & BH & lFDR & SnS & HSM\\
\midrule
Agranular insular area, dorsal part, layer 2/3 & $\cdot$ & $\cdot$ & $\cdot$ & $\cdot$ & Tier 4\\
\cellcolor{gray!6}{Agranular insular area, dorsal part, layer 5} & \cellcolor{gray!6}{$\cdot$} & \cellcolor{gray!6}{$\cdot$} & \cellcolor{gray!6}{$\cdot$} & \cellcolor{gray!6}{$\cdot$} & \cellcolor{gray!6}{Tier 4}\\
Agranular insular area, posterior part, layer 5 & $\cdot$ & $\cdot$ & $\cdot$ & $\cdot$ & Tier 4\\
\cellcolor{gray!6}{alveus} & \cellcolor{gray!6}{$\cdot$} & \cellcolor{gray!6}{$\cdot$} & \cellcolor{gray!6}{$\cdot$} & \cellcolor{gray!6}{$\cdot$} & \cellcolor{gray!6}{Tier 4}\\
Ansiform lobule & $\cdot$ & $\cdot$ & $\cdot$ & $\cdot$ & Tier 4\\
 
\cellcolor{gray!6}{Anterior cingulate area, dorsal part, layer 2/3} & \cellcolor{gray!6}{$\cdot$} & \cellcolor{gray!6}{$\cdot$} & \cellcolor{gray!6}{$\cdot$} & \cellcolor{gray!6}{$\cdot$} & \cellcolor{gray!6}{Tier 4}\\
Anteromedial visual area, layer 2/3 & $\cdot$ & $\cdot$ & $\cdot$ & $\cdot$ & Tier 4\\
\cellcolor{gray!6}{arbor vitae} & \cellcolor{gray!6}{$\cdot$} & \cellcolor{gray!6}{$\cdot$} & \cellcolor{gray!6}{$\cdot$} & \cellcolor{gray!6}{$\cdot$} & \cellcolor{gray!6}{Tier 4}\\
Basolateral amygdalar nucleus, anterior part & $\cdot$ & $\cdot$ & $\cdot$ & $\cdot$ & Tier 4\\
\cellcolor{gray!6}{Basolateral amygdalar nucleus, posterior part} & \cellcolor{gray!6}{$\cdot$} & \cellcolor{gray!6}{$\cdot$} & \cellcolor{gray!6}{$\cdot$} & \cellcolor{gray!6}{$\cdot$} & \cellcolor{gray!6}{Tier 4}\\
 
Bed nuclei of the stria terminalis & $\cdot$ & $\cdot$ & $\cdot$ & $\cdot$ & Tier 4\\
\cellcolor{gray!6}{brachium of the inferior colliculus} & \cellcolor{gray!6}{$\cdot$} & \cellcolor{gray!6}{$\cdot$} & \cellcolor{gray!6}{$\cdot$} & \cellcolor{gray!6}{$\cdot$} & \cellcolor{gray!6}{Tier 4}\\
Caudoputamen & $\cdot$ & $\cdot$ & $\cdot$ & $\cdot$ & Tier 4\\
\cellcolor{gray!6}{cerebral aqueduct} & \cellcolor{gray!6}{$\cdot$} & \cellcolor{gray!6}{$\cdot$} & \cellcolor{gray!6}{$\cdot$} & \cellcolor{gray!6}{$\cdot$} & \cellcolor{gray!6}{Tier 4}\\
cingulum bundle & $\cdot$ & $\cdot$ & $\cdot$ & $\cdot$ & Tier 4\\
 
\cellcolor{gray!6}{Copula pyramidis} & \cellcolor{gray!6}{$\cdot$} & \cellcolor{gray!6}{$\cdot$} & \cellcolor{gray!6}{$\cdot$} & \cellcolor{gray!6}{$\cdot$} & \cellcolor{gray!6}{Tier 4}\\
corpus callosum & $\cdot$ & $\cdot$ & $\cdot$ & $\cdot$ & Tier 4\\
\cellcolor{gray!6}{Declive (VI)} & \cellcolor{gray!6}{$\cdot$} & \cellcolor{gray!6}{$\cdot$} & \cellcolor{gray!6}{$\cdot$} & \cellcolor{gray!6}{$\cdot$} & \cellcolor{gray!6}{Tier 4}\\
Dentate gyrus, molecular layer & $\cdot$ & $\cdot$ & $\cdot$ & $\cdot$ & Tier 4\\
\cellcolor{gray!6}{Diagonal band nucleus} & \cellcolor{gray!6}{$\cdot$} & \cellcolor{gray!6}{$\cdot$} & \cellcolor{gray!6}{$\cdot$} & \cellcolor{gray!6}{$\cdot$} & \cellcolor{gray!6}{Tier 4}\\
 
dorsal hippocampal commissure & $\cdot$ & $\cdot$ & $\cdot$ & $\cdot$ & Tier 4\\
\cellcolor{gray!6}{Dorsal peduncular area} & \cellcolor{gray!6}{$\cdot$} & \cellcolor{gray!6}{$\cdot$} & \cellcolor{gray!6}{$\cdot$} & \cellcolor{gray!6}{$\cdot$} & \cellcolor{gray!6}{Tier 4}\\
Dorsomedial nucleus of the hypothalamus & $\cdot$ & $\cdot$ & $\cdot$ & $\cdot$ & Tier 4\\
\cellcolor{gray!6}{Endopiriform nucleus, dorsal part} & \cellcolor{gray!6}{$\cdot$} & \cellcolor{gray!6}{$\cdot$} & \cellcolor{gray!6}{$\cdot$} & \cellcolor{gray!6}{$\cdot$} & \cellcolor{gray!6}{Tier 4}\\
Endopiriform nucleus, ventral part & $\cdot$ & $\cdot$ & $\cdot$ & $\cdot$ & Tier 4\\
 
\cellcolor{gray!6}{Entorhinal area, lateral part} & \cellcolor{gray!6}{$\cdot$} & \cellcolor{gray!6}{$\cdot$} & \cellcolor{gray!6}{$\cdot$} & \cellcolor{gray!6}{$\cdot$} & \cellcolor{gray!6}{Tier 4}\\
Facial motor nucleus & $\cdot$ & $\cdot$ & $\cdot$ & $\cdot$ & Tier 4\\
\cellcolor{gray!6}{Fastigial nucleus} & \cellcolor{gray!6}{$\cdot$} & \cellcolor{gray!6}{$\cdot$} & \cellcolor{gray!6}{$\cdot$} & \cellcolor{gray!6}{$\cdot$} & \cellcolor{gray!6}{Tier 4}\\
Field CA3 & $\cdot$ & $\cdot$ & $\cdot$ & $\cdot$ & Tier 4\\
\cellcolor{gray!6}{Gigantocellular reticular nucleus} & \cellcolor{gray!6}{$\cdot$} & \cellcolor{gray!6}{$\cdot$} & \cellcolor{gray!6}{$\cdot$} & \cellcolor{gray!6}{$\cdot$} & \cellcolor{gray!6}{Tier 4}\\
 
Gustatory areas, layer 2/3 & $\cdot$ & $\cdot$ & $\cdot$ & $\cdot$ & Tier 4\\
\cellcolor{gray!6}{Gustatory areas, layer 4} & \cellcolor{gray!6}{$\cdot$} & \cellcolor{gray!6}{$\cdot$} & \cellcolor{gray!6}{$\cdot$} & \cellcolor{gray!6}{$\cdot$} & \cellcolor{gray!6}{Tier 4}\\
Gustatory areas, layer 5 & $\cdot$ & $\cdot$ & $\cdot$ & $\cdot$ & Tier 4\\
\cellcolor{gray!6}{Inferior colliculus} & \cellcolor{gray!6}{$\cdot$} & \cellcolor{gray!6}{$\cdot$} & \cellcolor{gray!6}{$\cdot$} & \cellcolor{gray!6}{$\cdot$} & \cellcolor{gray!6}{Tier 4}\\
Inferior olivary complex & $\cdot$ & $\cdot$ & $\cdot$ & $\cdot$ & Tier 4\\
 
\cellcolor{gray!6}{Infralimbic area, layer 2/3} & \cellcolor{gray!6}{$\cdot$} & \cellcolor{gray!6}{$\cdot$} & \cellcolor{gray!6}{$\cdot$} & \cellcolor{gray!6}{$\cdot$} & \cellcolor{gray!6}{Tier 4}\\
internal capsule & $\cdot$ & $\cdot$ & $\cdot$ & $\cdot$ & Tier 4\\
\cellcolor{gray!6}{Interpeduncular nucleus} & \cellcolor{gray!6}{$\cdot$} & \cellcolor{gray!6}{$\cdot$} & \cellcolor{gray!6}{$\cdot$} & \cellcolor{gray!6}{$\cdot$} & \cellcolor{gray!6}{Tier 4}\\
Interposed nucleus & $\cdot$ & $\cdot$ & $\cdot$ & $\cdot$ & Tier 4\\
\cellcolor{gray!6}{Lateral dorsal nucleus of thalamus} & \cellcolor{gray!6}{$\cdot$} & \cellcolor{gray!6}{$\cdot$} & \cellcolor{gray!6}{$\cdot$} & \cellcolor{gray!6}{$\cdot$} & \cellcolor{gray!6}{Tier 4}\\
 
Lateral hypothalamic area & $\cdot$ & $\cdot$ & $\cdot$ & $\cdot$ & Tier 4\\
\cellcolor{gray!6}{lateral lemniscus} & \cellcolor{gray!6}{$\cdot$} & \cellcolor{gray!6}{$\cdot$} & \cellcolor{gray!6}{$\cdot$} & \cellcolor{gray!6}{$\cdot$} & \cellcolor{gray!6}{Tier 4}\\
lateral olfactory tract, body & $\cdot$ & $\cdot$ & $\cdot$ & $\cdot$ & Tier 4\\
\cellcolor{gray!6}{Lateral septal nucleus, rostral (rostroventral) part} & \cellcolor{gray!6}{$\cdot$} & \cellcolor{gray!6}{$\cdot$} & \cellcolor{gray!6}{$\cdot$} & \cellcolor{gray!6}{$\cdot$} & \cellcolor{gray!6}{Tier 4}\\
Medial habenula & $\cdot$ & $\cdot$ & $\cdot$ & $\cdot$ & Tier 4\\

\bottomrule
\end{tabular}}
\end{table}
\clearpage

\begin{table}[h!]
\centering
\resizebox{\linewidth}{!}{
\begin{tabular}[t]{lccccc}
\toprule
Brain region & HS & BH & lFDR & SnS & HSM\\
\midrule
\cellcolor{gray!6}{medial lemniscus} & \cellcolor{gray!6}{$\cdot$} & \cellcolor{gray!6}{$\cdot$} & \cellcolor{gray!6}{$\cdot$} & \cellcolor{gray!6}{$\cdot$} & \cellcolor{gray!6}{Tier 4}\\
medial longitudinal fascicle & $\cdot$ & $\cdot$ & $\cdot$ & $\cdot$ & Tier 4\\
\cellcolor{gray!6}{Medial preoptic area} & \cellcolor{gray!6}{$\cdot$} & \cellcolor{gray!6}{$\cdot$} & \cellcolor{gray!6}{$\cdot$} & \cellcolor{gray!6}{$\cdot$} & \cellcolor{gray!6}{Tier 4}\\
Mediodorsal nucleus of thalamus & $\cdot$ & $\cdot$ & $\cdot$ & $\cdot$ & Tier 4\\
\cellcolor{gray!6}{Medulla} & \cellcolor{gray!6}{$\cdot$} & \cellcolor{gray!6}{$\cdot$} & \cellcolor{gray!6}{$\cdot$} & \cellcolor{gray!6}{$\cdot$} & \cellcolor{gray!6}{Tier 4}\\
Medullary reticular nucleus, ventral part & $\cdot$ & $\cdot$ & $\cdot$ & $\cdot$ & Tier 4\\
\cellcolor{gray!6}{Midbrain} & \cellcolor{gray!6}{$\cdot$} & \cellcolor{gray!6}{$\cdot$} & \cellcolor{gray!6}{$\cdot$} & \cellcolor{gray!6}{$\cdot$} & \cellcolor{gray!6}{Tier 4}\\
Midbrain reticular nucleus & $\cdot$ & $\cdot$ & $\cdot$ & $\cdot$ & Tier 4\\
\cellcolor{gray!6}{middle cerebellar peduncle} & \cellcolor{gray!6}{$\cdot$} & \cellcolor{gray!6}{$\cdot$} & \cellcolor{gray!6}{$\cdot$} & \cellcolor{gray!6}{$\cdot$} & \cellcolor{gray!6}{Tier 4}\\
Nucleus of the lateral olfactory tract, layer 3 & $\cdot$ & $\cdot$ & $\cdot$ & $\cdot$ & Tier 4\\
\cellcolor{gray!6}{Nucleus of the solitary tract} & \cellcolor{gray!6}{$\cdot$} & \cellcolor{gray!6}{$\cdot$} & \cellcolor{gray!6}{$\cdot$} & \cellcolor{gray!6}{$\cdot$} & \cellcolor{gray!6}{Tier 4}\\
Nucleus prepositus & $\cdot$ & $\cdot$ & $\cdot$ & $\cdot$ & Tier 4\\
\cellcolor{gray!6}{Orbital area, lateral part, layer 2/3} & \cellcolor{gray!6}{$\cdot$} & \cellcolor{gray!6}{$\cdot$} & \cellcolor{gray!6}{$\cdot$} & \cellcolor{gray!6}{$\cdot$} & \cellcolor{gray!6}{Tier 4}\\
Orbital area, lateral part, layer 6a & $\cdot$ & $\cdot$ & $\cdot$ & $\cdot$ & Tier 4\\
\cellcolor{gray!6}{Orbital area, ventrolateral part, layer 5} & \cellcolor{gray!6}{$\cdot$} & \cellcolor{gray!6}{$\cdot$} & \cellcolor{gray!6}{$\cdot$} & \cellcolor{gray!6}{$\cdot$} & \cellcolor{gray!6}{Tier 4}\\
 
Parataenial nucleus & $\cdot$ & $\cdot$ & $\cdot$ & $\cdot$ & Tier 4\\
\cellcolor{gray!6}{Paraventricular hypothalamic nucleus} & \cellcolor{gray!6}{$\cdot$} & \cellcolor{gray!6}{$\cdot$} & \cellcolor{gray!6}{$\cdot$} & \cellcolor{gray!6}{$\cdot$} & \cellcolor{gray!6}{Tier 4}\\
Paraventricular nucleus of the thalamus & $\cdot$ & $\cdot$ & $\cdot$ & $\cdot$ & Tier 4\\
\cellcolor{gray!6}{Parvicellular reticular nucleus} & \cellcolor{gray!6}{$\cdot$} & \cellcolor{gray!6}{$\cdot$} & \cellcolor{gray!6}{$\cdot$} & \cellcolor{gray!6}{$\cdot$} & \cellcolor{gray!6}{Tier 4}\\
Periventricular hypothalamic nucleus, intermediate part & $\cdot$ & $\cdot$ & $\cdot$ & $\cdot$ & Tier 4\\
 
\cellcolor{gray!6}{Periventricular hypothalamic nucleus, preoptic part} & \cellcolor{gray!6}{$\cdot$} & \cellcolor{gray!6}{$\cdot$} & \cellcolor{gray!6}{$\cdot$} & \cellcolor{gray!6}{$\cdot$} & \cellcolor{gray!6}{Tier 4}\\
Pons & $\cdot$ & $\cdot$ & $\cdot$ & $\cdot$ & Tier 4\\
\cellcolor{gray!6}{Pontine central gray} & \cellcolor{gray!6}{$\cdot$} & \cellcolor{gray!6}{$\cdot$} & \cellcolor{gray!6}{$\cdot$} & \cellcolor{gray!6}{$\cdot$} & \cellcolor{gray!6}{Tier 4}\\
Pontine gray & $\cdot$ & $\cdot$ & $\cdot$ & $\cdot$ & Tier 4\\
\cellcolor{gray!6}{Pontine reticular nucleus} & \cellcolor{gray!6}{$\cdot$} & \cellcolor{gray!6}{$\cdot$} & \cellcolor{gray!6}{$\cdot$} & \cellcolor{gray!6}{$\cdot$} & \cellcolor{gray!6}{Tier 4}\\
 
Pontine reticular nucleus, caudal part & $\cdot$ & $\cdot$ & $\cdot$ & $\cdot$ & Tier 4\\
\cellcolor{gray!6}{Posterior complex of the thalamus} & \cellcolor{gray!6}{$\cdot$} & \cellcolor{gray!6}{$\cdot$} & \cellcolor{gray!6}{$\cdot$} & \cellcolor{gray!6}{$\cdot$} & \cellcolor{gray!6}{Tier 4}\\
Primary somatosensory area, mouth, layer 1 & $\cdot$ & $\cdot$ & $\cdot$ & $\cdot$ & Tier 4\\
\cellcolor{gray!6}{Primary somatosensory area, mouth, layer 5} & \cellcolor{gray!6}{$\cdot$} & \cellcolor{gray!6}{$\cdot$} & \cellcolor{gray!6}{$\cdot$} & \cellcolor{gray!6}{$\cdot$} & \cellcolor{gray!6}{Tier 4}\\
Primary somatosensory area, nose, layer 6a & $\cdot$ & $\cdot$ & $\cdot$ & $\cdot$ & Tier 4\\
 
\cellcolor{gray!6}{Primary somatosensory area, upper limb, layer 6b} & \cellcolor{gray!6}{$\cdot$} & \cellcolor{gray!6}{$\cdot$} & \cellcolor{gray!6}{$\cdot$} & \cellcolor{gray!6}{$\cdot$} & \cellcolor{gray!6}{Tier 4}\\
Principal sensory nucleus of the trigeminal & $\cdot$ & $\cdot$ & $\cdot$ & $\cdot$ & Tier 4\\
\cellcolor{gray!6}{Reticular nucleus of the thalamus} & \cellcolor{gray!6}{$\cdot$} & \cellcolor{gray!6}{$\cdot$} & \cellcolor{gray!6}{$\cdot$} & \cellcolor{gray!6}{$\cdot$} & \cellcolor{gray!6}{Tier 4}\\
Retrosplenial area, dorsal part, layer 5 & $\cdot$ & $\cdot$ & $\cdot$ & $\cdot$ & Tier 4\\
\cellcolor{gray!6}{Retrosplenial area, lateral agranular part, layer 2/3} & \cellcolor{gray!6}{$\cdot$} & \cellcolor{gray!6}{$\cdot$} & \cellcolor{gray!6}{$\cdot$} & \cellcolor{gray!6}{$\cdot$} & \cellcolor{gray!6}{Tier 4}\\
 
Spinal nucleus of the trigeminal, interpolar part & $\cdot$ & $\cdot$ & $\cdot$ & $\cdot$ & Tier 4\\
\cellcolor{gray!6}{spinal tract of the trigeminal nerve} & \cellcolor{gray!6}{$\cdot$} & \cellcolor{gray!6}{$\cdot$} & \cellcolor{gray!6}{$\cdot$} & \cellcolor{gray!6}{$\cdot$} & \cellcolor{gray!6}{Tier 4}\\
stria terminalis & $\cdot$ & $\cdot$ & $\cdot$ & $\cdot$ & Tier 4\\
\cellcolor{gray!6}{Substantia innominata} & \cellcolor{gray!6}{$\cdot$} & \cellcolor{gray!6}{$\cdot$} & \cellcolor{gray!6}{$\cdot$} & \cellcolor{gray!6}{$\cdot$} & \cellcolor{gray!6}{Tier 4}\\
superior cerebelar peduncles & $\cdot$ & $\cdot$ & $\cdot$ & $\cdot$ & Tier 4\\
 
\cellcolor{gray!6}{Superior colliculus, motor related, intermediate white layer} & \cellcolor{gray!6}{$\cdot$} & \cellcolor{gray!6}{$\cdot$} & \cellcolor{gray!6}{$\cdot$} & \cellcolor{gray!6}{$\cdot$} & \cellcolor{gray!6}{Tier 4}\\
Superior colliculus, superficial gray layer & $\cdot$ & $\cdot$ & $\cdot$ & $\cdot$ & Tier 4\\
\cellcolor{gray!6}{Superior vestibular nucleus} & \cellcolor{gray!6}{$\cdot$} & \cellcolor{gray!6}{$\cdot$} & \cellcolor{gray!6}{$\cdot$} & \cellcolor{gray!6}{$\cdot$} & \cellcolor{gray!6}{Tier 4}\\
Tegmental reticular nucleus & $\cdot$ & $\cdot$ & $\cdot$ & $\cdot$ & Tier 4\\
\cellcolor{gray!6}{Temporal association areas, layer 4} & \cellcolor{gray!6}{$\cdot$} & \cellcolor{gray!6}{$\cdot$} & \cellcolor{gray!6}{$\cdot$} & \cellcolor{gray!6}{$\cdot$} & \cellcolor{gray!6}{Tier 4}\\
 
Uvula (IX) & $\cdot$ & $\cdot$ & $\cdot$ & $\cdot$ & Tier 4\\
\cellcolor{gray!6}{Ventral auditory area, layer 4} & \cellcolor{gray!6}{$\cdot$} & \cellcolor{gray!6}{$\cdot$} & \cellcolor{gray!6}{$\cdot$} & \cellcolor{gray!6}{$\cdot$} & \cellcolor{gray!6}{Tier 4}\\
Ventral tegmental area & $\cdot$ & $\cdot$ & $\cdot$ & $\cdot$ & Tier 4\\
\cellcolor{gray!6}{Visceral area, layer 4} & \cellcolor{gray!6}{$\cdot$} & \cellcolor{gray!6}{$\cdot$} & \cellcolor{gray!6}{$\cdot$} & \cellcolor{gray!6}{$\cdot$} & \cellcolor{gray!6}{Tier 4}\\
Visceral area, layer 5 & $\cdot$ & $\cdot$ & $\cdot$ & $\cdot$ & Tier 4\\
 
\cellcolor{gray!6}{Visceral area, layer 6a} & \cellcolor{gray!6}{$\cdot$} & \cellcolor{gray!6}{$\cdot$} & \cellcolor{gray!6}{$\cdot$} & \cellcolor{gray!6}{$\cdot$} & \cellcolor{gray!6}{Tier 4}\\
\bottomrule
\end{tabular}}
\end{table}
\clearpage

\section{On the number of estimated tiers}

Estimating the exact number of relevance tiers is not a trivial nor an automated task for our model. Because the Horseshoe Mix is a mixture performed over scale parameters and, consequently, the mixture kernels are highly overlapping, the model's membership labels do not lead to clean results as the ones we would observe if we were to estimate a mixture over some location parameters. The overlap between the mixture kernels hinders the recovery of the number of clusters via theoretical-based methods such as minimization of loss functions (e.g., Variation of Information, \cite{Wade2018}). 

Therefore, we resorted to Medvedovic's approach and threshold the resulting dendrogram (obtained as described in our previous answer). 
We can detect the most meaningful number of clusters by monitoring how the observations are partitioned as the number of tiers $k$ increases. However, we observed in practice that after a few (usually 3-4) meaningful groups are detected, our post-processing approach dismembers the low-magnitude z-scores as $k$ increases. This behavior suggests that there is a limit to the number of eloquent tiers one could extract that we can exploit to fix $k$ in the absence of prior information.

\begin{figure}[th]
    \centering
    \includegraphics[width=\linewidth]{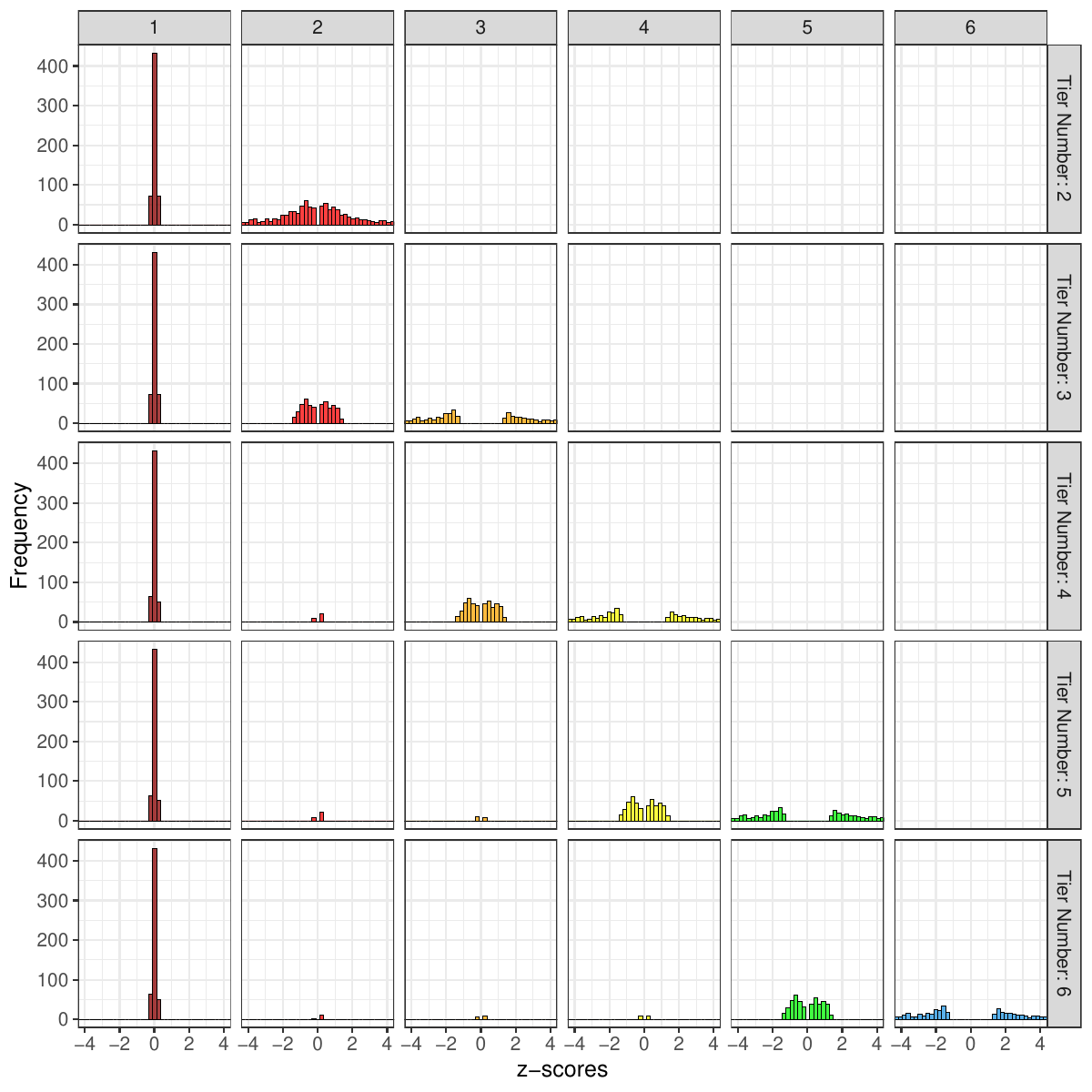}
    \caption{Histograms of the z-scores stratified into relevance tiers. The plots recovered by using an increasing number of clusters - retrieved by thresholding the dendrogram resulting from the posterior coclustering matrix - are displayed by rows. The columns indicate the tiers sorted by increasing magnitude order.}
    \label{fig:hist}
\end{figure}

\begin{figure}[th]
    \centering
    \includegraphics[width=\linewidth]{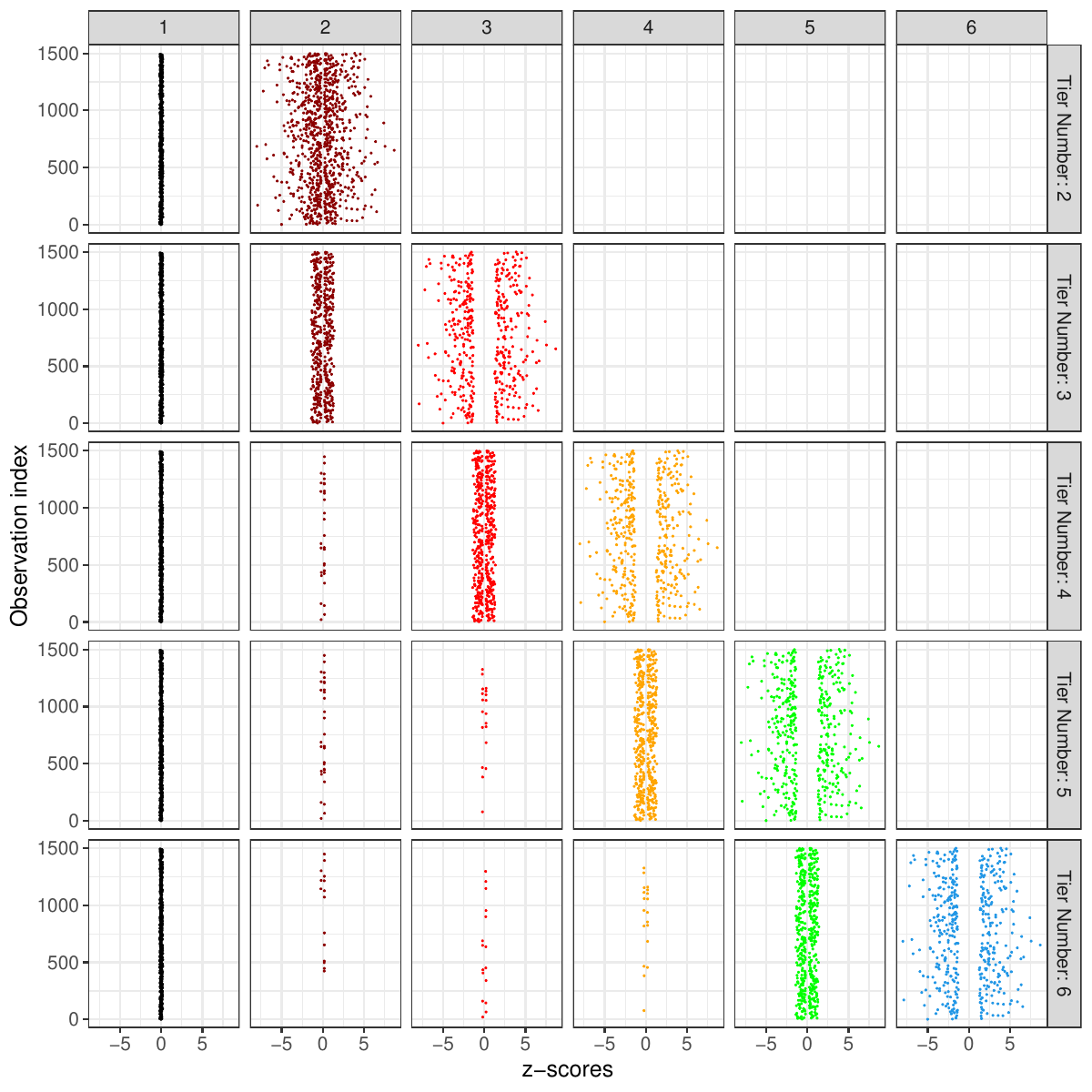}
    \caption{Scatterplots of the z-scores stratified into relevance tiers. The plots recovered by using an increasing number of clusters - retrieved by thresholding the dendrogram resulting from the posterior coclustering matrix - are displayed by rows. The columns indicate the tiers sorted by increasing magnitude order.}
    \label{fig:scat}
\end{figure}

Consider the following example. We simulate 1,500 observations from a mixture of three Normals centered in zero and with scale parameters equal to $\sigma_1=3$, $\sigma_1=1$, and $\sigma_1=0.01$. Then, we fit the HSM model and compute the resulting posterior probability of coclustering matrix $PPC$. Finally, we threshold the dendrogram obtained from the matrix considering $k^*\in\{2,3,4,5,6\}$.\\
We display a summary of the results in Figure \ref{fig:hist} (Figure \ref{fig:scat}), where we show a collection of histograms (scatterplots) of the generated z-scores, stratified by estimated tiers. Moreover, in Table \ref{tab:freq}, we also perform pairwise comparisons of the frequencies of the estimated groups for consecutive values of $k^*$.


 Each row of panels represents the results obtained for a specific value of $k^*$. The first line, trivially, reports how the model would partition the z-scores between relevant and irrelevant. As we investigate the results for moderate values of $k^*$ (3-4), more interesting partitions arise. However, when $k^*=5,6$, the clusters that are further partitioned are the ones that have low magnitude, resulting in tiers that contain only a few observations, becoming hardly interpretable. This application suggests that without any a-priori or field expert knowledge that one can leverage, we recommend exploring the results for a small collection of values $k$ and stop as soon as more than one tier contain only a small fraction of the observations.  

\begin{table}[th]
    \centering
    \begin{tabular}{cccc}
    \toprule
        Tier Number & 1 & 2 & 3 \\
        \midrule
        1 & 578 & 0& 0\\
        2 & 0 & 477& 445\\
    \bottomrule
    \end{tabular}
    
    \vspace{.5cm}
    \centering
    \begin{tabular}{ccccc}
        \toprule
        Tier Number & 1 & 2 & 3& 4 \\
        \midrule
        1 & 547 & 31& 0& 0\\
        2 & 0 & 0& 477& 0\\
        3 & 0 & 0& 0& 445\\
        \bottomrule
    \end{tabular}
    
    \vspace{.5cm}
    \centering
    \begin{tabular}{cccccc}
    \toprule
        Tier Number & 1 & 2 & 3 & 4& 5\\
        \midrule
        1 & 547 & 0 & 0 & 0& 0\\
        2 & 0 & 31 & 0 & 0& 0\\
        3 & 0 & 0 & 19 & 458& 0\\
        4 & 0 & 0 & 0 & 0& 445\\
        \bottomrule
    \end{tabular}
    
    \vspace{.5cm}
    \centering
    \begin{tabular}{ccccccc}
    \toprule
        Tier Number & 1 & 2 & 3& 4& 5& 6 \\
        \midrule
        1 & 547 & 0& 0& 0& 0& 0\\
        2 & 0 & 15& 16& 0& 0& 0\\
        3 & 0 & 0& 0& 19& 0& 0\\
        4 & 0 & 0& 0& 0& 458& 0\\
        5 & 0 & 0& 0& 0& 0& 445\\
        \bottomrule
    \end{tabular}
    \caption{Frequency tables comparing consecutive clustering solutions.}
    \label{tab:freq}
\end{table}